\documentclass[aps,groupedaddress,superscriptaddress,amsmath,amssymb,pra,twocolumn,longbibliography]{revtex4-2}
\usepackage{bm}
\usepackage{hyperref}
\usepackage{graphicx}
\usepackage{amsfonts}
\usepackage{verbatim}
\usepackage{bbm}
\usepackage{braket}
\usepackage{qcircuit}
\usepackage[normalem]{ulem}
\hypersetup{
    colorlinks=true,
    linkcolor=blue,
    filecolor=magenta,      
    urlcolor=cyan,
}
\newcommand{\tr}{\mathrm{Tr}}

\def\omq{\delta\omega_\mathrm{q}}
\def\sn{^{(n)}}
\def\sm{^{(m)}}

\def\th{\theta{}}
\def\ts{t_\mathrm{cyc}}
\def\ep{\varepsilon}

\def\ku{\ket{\uparrow}}
\def\mE{\mathbb{E}}
\def\mK{\mathbb{K}_{\{m\}_s}}
\def\kd{\ket{\downarrow}}

\newcommand{\sumprime}[1]{\sum_{#1}{\vphantom{\sum}}^{\!\!\prime}}
\def\RR{\mathfrak{R}}
\def\Rst{\mathfrak{R}_\mathrm{st}}
\def\Rcorr{\mathfrak{R}_\mathrm{corr}}
\def\DD{\mathcal{D}}

\def\hI{\hat{I}}
\def\h1{\mathds{1}}
\def\tT{\hat\tau}
 \def\NTLS{N_{\mathrm{TLS}}}
 
 \def\corr{\mathrm{corr}}
 \def\omegaMin{\omega_{\mathrm{min}}}

\begin{document}
\title{Characterizing low-frequency qubit noise}
\author{Filip Wudarski}
\affiliation{Quantum Artificial Intelligence Laboratory, Exploration Technology Directorate,
NASA Ames Research Center, Moffett Field, CA 94035, USA}
\affiliation{USRA Research Institute for Advanced Computer Science (RIACS), Mountain View, CA 94043, USA}
\author{Yaxing Zhang}
\affiliation{Google Research, Mountain View, CA 94043, USA}
\author{Alexander Korotkov}
\affiliation{Google Research, Mountain View, CA 94043, USA}
\author{A. G. Petukhov}
\affiliation{Google Research, Mountain View, CA 94043, USA}
\author{M. I. Dykman}
\affiliation{Department of Physics and Astronomy, Michigan State University, East Lansing, MI 48824, USA}

\begin{abstract} 
Fluctuations of the qubit frequencies are one of the major problems to overcome on the way to scalable quantum computers. Of particular importance are fluctuations with the correlation time that exceeds the decoherence time due to decay and dephasing by fast processes. The statistics of the fluctuations can be characterized by measuring the correlators of the outcomes of periodically repeated Ramsey measurements. This work suggests a method that allows describing qubit dynamics during repeated measurements in the presence of evolving noise. It made it possible, in particular, to evaluate the two-time correlator for the noise from two-level systems and obtain two- and three-time correlators for a Gaussian noise. The explicit expressions for the correlators are compared with simulations. A significant difference of the three-time correlators for the noise from two-level systems and for a Gaussian noise is demonstrated. Strong broadening of the distribution of the outcomes of Ramsey measurements, with a possible fine structure, is found for the data acquisition time comparable to the noise correlation time.
\end{abstract}

\date{\today}

\maketitle

\section{Introduction}
\label{sec:Intro}

Decoherence, and in particular fluctuations of  qubit frequencies, are one of the major obstacles faced by quantum computing. Understanding the mechanisms of these  fluctuations has been attracting much attention. To this end, much work has focused on the analysis of the fluctuation spectra, cf. \cite{Alvarez2011,Bylander2011,Sank2012,Yan2012,Paz-Silva2014,Yoshihara2014,Kim2015,Brownnutt2015,O'Malley2015,Szankowski2016,Yan2016,Myers2017,Quintana2017,Ferrie2018,Noel2019,vonLupke2020,Wolfowicz2021,Wang2021}
and references therein. For condensed-matter based qubits, the analysis is frequently based on the assumption that the fluctuations are Gaussian noise that comes from many independent sources each of which is weakly coupled to a qubit. However, this assumption does not necessarily apply, particularly for low-frequency noise,  \cite{Riste2013,Serniak2018,Christensen2019,Schlor2019}. Such noise is often thought to come from random hops between the states of two-level systems (TLSs) coupled to a qubit. However, qubit noise from the TLSs is generally non-Gaussian \cite{Paladino2002,Galperin2004,Galperin2006a,Paladino2014,Muller2019}, reminiscent of the problem of spin decoherence in  nuclear magnetic resonance associated with the spectral diffusion, see \cite{Herzog1956,Klauder1962}.

To characterize noise that causes qubit dephasing it is important to know not just the noise spectrum, but also its statistics. The problem has attracted considerable interest \cite{Li2013a,Norris2016,Szankowski2017,Sung2019}. A natural approach is to characterize higher-order spectral moments of noise of the qubit frequency. For zero-mean Gaussian noise, all moments are expressed in terms of the second moment. A deviation from the corresponding interrelation between the moments is a signature of  the noise being non-Gaussian. For frequency noise of a mesoscopic vibrational system higher-order moments were studied in  \cite{Maizelis2011,Sun2015}. The possibility to measure the third spectral moment of qubit noise (the bispectrum) was demonstrated in \cite{Sung2019}  by building up on the approach \cite{Norris2016}.  This approach is based, ultimately, on using distinct sequences of refocusing pulses  during Ramsey measurements. Therefore it is limited to the noise frequencies that are higher than the reciprocal lifetime of a qubit.

In the present paper we develop means for studying  the statistics of a qubit noise  with frequencies smaller than the reciprocal qubit lifetime. Such noise plays a critical role in the operation of a quantum computer as it limits the time over which repeated gate operations can be performed on a qubit without recalibrating it. However, we are not studying the range of extremely low frequencies, where the qubit frequency can be effectively measured in real time \cite{Sank2012}. The stability of qubits over very long times is affected by several factors, such as cosmic rays or high-energy photons for superconducting qubits \cite{Wilen2021,Liu2022a}. Our approach can be extended to this time range, but here we concentrate on shorter times.

Our goal is to develop an analytical theory of the effects of noise statistics. The theory should be sufficiently general to account for different types of noise, including both Gaussian and non-Gaussian noise, and for the qubit dynamics involved in the measurements. We also aim at performing numerical simulations, which can be compared with the theory and in some cases  can go beyond the range where analytical results can be obtained or become too cumbersome.

We study the first three moments of the qubit frequency noise. Such a study can be conveniently done by periodically repeating Ramsey measurements, cf. Ref.~\cite{Yan2012}, but going beyond noise spectroscopy. For the low-frequency noise it is advantageous to analyze the results primarily in the time rather than the frequency domain, particularly where we are not limited to the effects of the lowest order in the noise intensity.

The interplay of a large correlation time and the noise statistics should lead to a number of consequences, and we identify some of them. An example is a potentially large  change of the variance of the qubit measurement outcomes beyond the binomial (Bernoulli) limit. Of significant interest is the  occurrence of ``anomalous'' measurement outcomes, that is, of having an unlikely outcome with the probability much higher than what is expected from the Gaussian distribution of the outcome probabilities. The effect is particularly pronounced for TLSs, but it is different from the familiar  mechanism associated with a strong coupling to a group of TLSs \cite{Galperin2006a,Paladino2014}.  It emerges where the number of measurements is large, but not too large so that the overall duration of the data acquisition is not too long. 

Identifying the mechanism of classical non-Gaussian qubit noise based on its moments is a hard problem, generally: such noise is often  a result of ``processing'' of  a Gaussian noise by nonlinear systems coupled to a qubit. A familiar example is the telegraph noise that comes from TLSs. Even in a simple model of two-level states in glasses \cite{Anderson1972,Phillips1972,Phillips1987} this noise results ultimately from the interstate switching of a strongly-nonlinear (two-state) system due to its coupling to a bosonic reservoir. Noise identification is further exacerbated by the quantum uncertainty of the measurement outcomes. 

Therefore it is helpful to have a ``map'' of the outcomes of the measurements depending on the noise correlation and statistics for different types of noise sources. We aim to develop such a map for TLSs, analytically and via simulations. We also study correlation functions  for three important types of a Gaussian noise with a large correlation time, the exponentially correlated  noise, the noise with a definite ``color'', i.e., with a comparatively narrow peak in the power spectrum, and $1/f$ noise.

The analytical calculations are fairly cumbersome. Therefore we separate the paper into three  parts. One part present the results of the analytical calculations and describes the simulations and the comparison of the theory and simulations. The second part describes the general theoretical approach. The details of the theoretical calculations and some auxiliary results of the simulations are presented in the Appendices.

In Sec.~\ref{sec:correlation_general} we describe the scheme of periodically repeated Ramsey measurements and define the two- and three-time correlation functions of the measurement outcomes. 
Section~\ref{sec:summary_TLS} summarizes, without a derivation,  the major analytical results on the effect of coupling to TLSs. It gives the explicit general expression for the two-time correlator and discusses several important limiting cases. Section~\ref{sec:class_noise_summary} provides, also without a derivation, explicit  expressions for the two- and three-time correlators in the case of a Gaussian noise. The parameters in these expressions are evaluated for several important types of the noise.  Section~\ref{sec:theory_simulations} presents the results of the simulations and a detailed comparison of the theory and simulations for different types of noise. Section~\ref{sec:short_acquisition} presents analytical results and the results of simulations for a moderately large acquisition time, comparable to the noise correlation time, where the properties of the noise are pronounced particularly strongly.  Section~\ref{sec:periodic}  shows how periodic modulation of the qubit frequency affects the power spectrum of the measurement outcomes. In Sec.~\ref{sec:master_equation} we derive a master equation for a qubit coupled to TLSs and consider the qubit dynamics during the Ramsey measurement and the probability of the measurement outcome. In Sec.~\ref{sec:r_2_TLS} the analysis is extended to find the two-time correlator of the outcomes. The approach is further developed in Sec.~\ref{sec:Gaussian_general} to analyze the effect of a Gaussian noise on the one-, two- and three-time correlators. Section~\ref{sec:conclusions} provides a summary of the results.


\section{The correlation function of the measurement outcomes}
\label{sec:correlation_general}

\begin{figure}[h]
\includegraphics[width=0.47\textwidth]{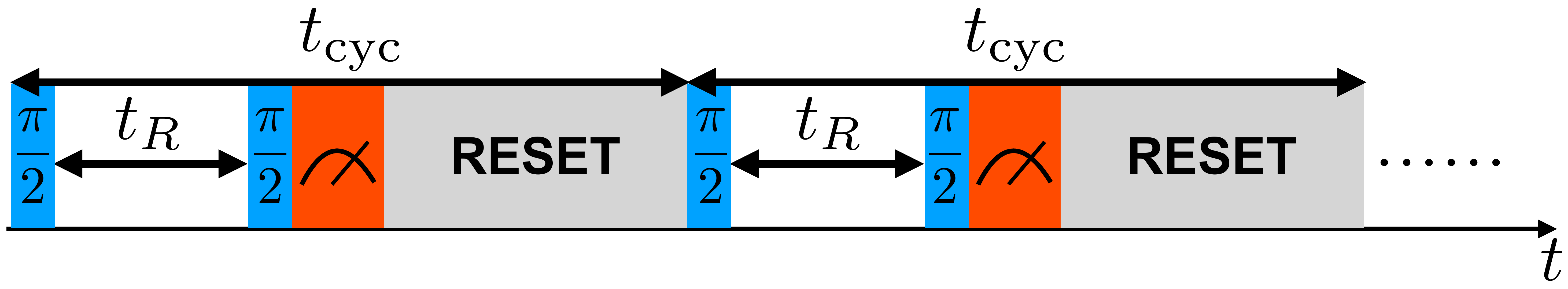} \\
\caption{Schematics of the Ramsey measurements. The measurements of duration $t_R$ are repeated with period $\ts$. After each measurement the qubit is reset to the ground state. Before each $\pi/2$ pulse that precedes the measurement the phase of the qubit is incremented by $\phi_R$ (not shown). The phase $\phi_R$ mimics the phase accumulation due to the detuning of the qubit transition frequency from the frequency of the reference signal. }
\label{fig:pulse_sequence}
\end{figure}

We associate the operators acting on the qubit states with the Pauli operators $\sigma_{x,y,z}$ and the unit operator $\hI_q$.  The ground and excited states of the qubit are the eigenstates of $\sigma_z$. In the Bloch sphere representation they are $\ket{0}\equiv \ku$ and $\ket{1}\equiv \kd$, respectively. We consider a periodic sequence of Ramsey measurements sketched in Fig.~\ref{fig:pulse_sequence} \cite{Sank2012,Yan2012}.   In the first Ramsey measurement  the qubit, initially in the state $\ket{0}$, is  rotated at time $t=0$ by $\pi/2$ around the $y$-axis into the state $(\ket{0}+\ket{1})/\sqrt{2}$. At $t_R$ it is rotated by $\pi/2$ around the $y$-axis again and the occupation of the state $\ket{1}$ is measured.  After the measurement the qubit is reset to the ground state. The Ramsey measurements  are then repeated with period $\ts$, which we call the cycle period. For simplicity we disregard the duration of the gate operations and the measurement as well as the gate and measurement errors.

In a Ramsey measurement, the phase accumulated by the qubit over time $t_R$ is compared with the phase accumulated over this time by a reference resonant signal. The accumulation of the phase difference $\th$ is thus determined by fluctuations $\omq(t)$ of the qubit frequency. These fluctuations are described by the Hamiltonian 
\begin{align}
\label{eq:just_delta_omega}
H_\mathrm{fl} = -\frac{1}{2}\omq (t)\sigma_z \qquad (\hbar = 1).
\end{align}
The random phase $\th_k$ accumulated over a $k$th  cycle, i.e., over the time interval $(k\ts, k\ts+t_R)$, is 
\begin{align}
\label{eq:phase_k_classical}
\theta_k = \int_{k\ts}^{k\ts+t_R}\omq (t)dt\ .
\end{align}
As seen from this equation, fluctuations $\omq(t)$ with a typical correlation time much shorter than $t_R$ are largely averaged out. Periodic repetition of the Ramsey measurements allows revealing fluctuations with correlation times not only on the scale of $t_R$, but also on the scale determined by the cycle period $\ts$.

We assume that the system, including noise, is stationary. Then  measurement outcomes depend only on the time interval between the measurements. We consider the expectation values of the outcomes $r_1, r_2(k)$, and $r_3(k,l)$  of obtaining ``1'' in a single measurement, obtaining ``1'' and ``1'' in  two measurements separated by $k$ cycles, and obtaining  three ``1''s in three measurements in which two measurements are separated from the first one by $k$ and $l$ cycles, respectively; for concreteness, we assume $l>k\geq 1$. These expectation values are time correlators and are given by the correlation functions of the projection operator $\hat\pi=(\hI_q - \sigma_z)/2$,
\begin{align}
\label{eq:correlators1_defined}
&r_1=\tr[\hat\pi(t_R^+)\rho(0^-)], \nonumber
\\
&r_2(k) = \tr[\hat\pi(k\ts+ t_R^+)\hat\pi(t_R^+)\rho(0^-)],
\end{align}
and 
\begin{align}
\label{eq:correlators2_defined}
&r_3(k,l) = \tr[\hat\pi(l\ts+t_R^+)\hat\pi(k\ts+ t_R^+) \nonumber\\
&\times\hat\pi(t_R^+)\rho(0^-)],
\end{align}
where  $\rho$ is the density matrix of the system; $\hI_q$ is the unit operator in the qubit space. The superscripts ``+'' and ``-'' of the time arguments indicate that the operator is evaluated, respectively, right after or right before the corresponding instant of time, i.e., right after or right before the  gate operation performed at this time, $t^\pm \equiv t\pm \ep$ with $\ep\to +0$; 
in particular, $0^-$ is the time right before the rotation around the $y$-axis at $t=0$, whereas $t_R^+$ is the time right after the rotation at $t=t_R$.  Clearly, $r_1$ is just the probability of obtaining ``1'' in a Ramsey measurement. 

The traces in Eqs.~(\ref{eq:correlators1_defined}) and (\ref{eq:correlators2_defined}) are calculated over the states of the qubit and the thermal reservoir coupled to the qubit, including the states of the TLSs if the TLSs play a role. The traces also imply averaging over the realizations of noise in the case where noise is a classical random force that modulates the qubit frequency. 

The outcome of an $n$th Ramsey measurement $x_n$ takes on values 1 or 0. The correlators $r_{1,2,3}$ are determined by the expectation values of these outcomes and their products,  
\begin{align}
\label{eq:expectation_values}
&r_1= \mE[x_n],\quad r_2(k) = \mE[x_{n+k}x_n], \nonumber\\
 &r_3(k,l) = \mE[x_{n+l}x_{n+k}x_n]
\end{align}
For classical Gaussian noise the correlators $r_3(k,l)$ are fully determined by $r_1$ and $r_2(k)$. We find the corresponding relations. If they do not hold, this indicates that noise is non-Gaussian. 

Along with $r_{2,3}$ we will also consider centered correlators
\begin{align}
\label{eq:centered correlators}
&\tilde r_2(k) = \mE[(x_{n+k}-r_1)(x_n-r_1)]\ ,\nonumber\\
&\tilde r_3(k,l) = \mE[(x_{n+l}-r_1)(x_{n+k}-r_1)(x_n-r_1)]\ .
\end{align}

In what follows we analytically calculate the correlators $r_{1,2,3}$ and compare them with the results of simulating the sequences $x_n$ for several types of fluctuations. We typically simulate $N=10^5$ cycles and average the results over 300 repetitions. This limits noise correlation time we can reliably simulate to $\lesssim 10^4 \ts$. We also do simulations where the number of cycles  $N$ is much smaller,$\lesssim 10^2$, so as to reveal the tail of the distribution of the outcomes that emerges in this case.

A note is due on the difference of the effects of TLSs on the correlators of different order. The Ramsey measurement probability $r_1$ depends on the phase $\th$ accumulated between the Ramsey pulses due to noise. For a given $\theta$, the probability  of obtaining ``1'' in a measurement is \cite{Nielsen2011}
\begin{align}
\label{eq:standard_probability}
p(\theta) = \frac{1}{2}\left[1+e^{-t_R/T_2}\cos(\phi_R + \theta)\right]\ .
\end{align}
For a random $\theta$, the probability $r_1$ is given by the mean value of $p(\theta)$.
In Eq.~(\ref{eq:standard_probability}),  $T_2^{-1}$ is the qubit decoherence  rate due to fast decay and dephasing processes. The phase $\phi_R$ mimics the phase accumulated due to the difference between the qubit frequency $\omega_\mathrm{q}$ and the frequency of the reference signal $\omega_\mathrm{ref}$,
\begin{align}
\label{eq:phi_R_as_detuning}
\phi_R = (\omega_\mathrm{q} - \omega_\mathrm{ref})t_R.
\end{align}
This phase can be (and frequently is) also added in a controlled way by a gate operation, see Eq.~(\ref{eq:H_R}) below. 

In the model of noninteracting TLSs that we consider, the TLSs contribute to the phase $\th$ independently. The overall phase is a sum of these contributions. From Eq.~(\ref{eq:standard_probability}), the value of $r_1$ is determined by $\langle \exp(i\th)\rangle$ and can be calculated by multiplying the contributions of individual TLSs \cite{Paladino2002,Galperin2004,Galperin2006a}. In contrast, the correlators $r_{2}$ and higher-order correlators should contain terms that decay as individual TLSs, their pairs, triples, etc. Therefore these correlators may not be calculated as products of the contributions of individual TLSs. 


\section{Analytical results on dispersive  coupling to two-level systems}
\label{sec:summary_TLS}

We study two major mechanisms of qubit decoherence, classical fluctuations of the qubit frequency and the effect of coupling to two-level systems. The level spacing of the TLSs is assumed to be much smaller than the qubit level spacing. In this case the major effect of the TLSs is to modulate the qubit frequency as the TLSs switch between their states. We also discuss the effect of modulating the qubit frequency by classical Gaussian noise. The analysis is somewhat involved. Therefore we first provide the results, whereas their derivation is postponed till  Secs.~\ref{sec:master_equation} - \ref{sec:Gaussian_general}. Here we start with the results on the TLSs.

\subsection{Explicit general expressions}
\label{subsec:general_TLS}
Qubit decoherence due to  the dispersive qubit-to-TLSs coupling is described by the Hamiltonian
\begin{align}
\label{eq:q_TLS_dispersive}
H_\mathrm{q-TLS} =- \frac{1}{2}\sigma_z\sum_nV\sn{} \tT_z\sn
\end{align}
Here $n$ enumerates the TLSs, $\tT_z\sn$ is the Pauli operator  of the $n$th TLS, and $V\sn{} $ is the coupling parameter; the states of an $n$th TLS are $\ket{0}\sn$ and $\ket{1}\sn$,
and $\tT_z\sn\ket{i}\sn = (-1)^i\ket{i}\sn$ with $i=0,1$. The Hamiltonian $H_\mathrm{q-TLS}$ has the same form as the Hamiltonian of the qubit frequency fluctuations $H_\mathrm{fl}$ except that the fluctuations are described by  operators in the TLSs' space. Such treatment is advantageous in view of the formulation (\ref{eq:correlators1_defined}), as it allows describing the qubit and the TLSs in a single framework.

The effect of the TLSs on the qubit dynamics depends on the relation between the coupling $V\sn{} $ and the rates $W_{ij}\sn$ of the interstate switching $\ket{i}\sn \to \ket{j}\sn$ (we remind that $i$ and $j$ take on the values $0$ and $1$). Our analysis  gives the explicit expressions for the probability $r_1$ of having ``1'' as an outcome of the Ramsey measurement and for the pair correlation function $r_2(k)$, for an arbitrary ratio $V\sn{} /W_{ij}\sn$. In particular, we find
\begin{align}
\label{eq:Pi_1_multi_TLS_explicit}
&r_1 =  \frac{1}{2} + \frac{1}{2} e^{-t_R/T_2}\mathrm{Re}\, \Bigl[e^{i\tilde\phi_R}\,
\prod_n \Xi\sn(t_R)\Bigr]
\end{align}
where
\begin{align}
\label{eq:Xi_n}
&\Xi\sn(t_R)=\Bigl[\left(\frac{W\sn{}}{2\gamma\sn} +iV\sn{} \frac{\Delta W\sn{}}
{\gamma\sn W\sn{}}\right)\sinh \gamma\sn t_R
\nonumber\\
&  +\cosh\gamma\sn t_R\Bigr]\exp(-W\sn{}t_R/2)
\end{align}
and
\begin{align}
\label{eq:phi_R_TLS}
\tilde\phi_R = \bigl( \omega_\mathrm{q} - \omega_\mathrm{ref} -\sum_nV\sn{} \langle \tT_z\sn\rangle\bigr) t_R.
\end{align}
[we emphasize that here $\omega_\mathrm{q}$ is the observable qubit frequency; it incorporates the renormalization that comes from the coupling to the TLSs described by the Hamiltonian (\ref{eq:q_TLS_dispersive})].

The effect of the TLSs is described by the factor $\Xi\sn$. The expression for $\Xi\sn$ coincides with the previously obtained expression for the factor that describes decay of $\langle \sigma_\pm(t)\rangle$ due to the coupling to TLSs \cite{Paladino2002}. The form of $\Xi\sn$ is determined by the parameter $\gamma\sn$,
\begin{align}
\label{eq:mu_parameter}
&\gamma\sn=\frac{1}{2}\left[W\sn{}{}^2 +4iV\sn{} (\Delta W\sn{}+iV\sn{} )\right]^{1/2}, \nonumber\\
&W\sn = W_{01}\sn + W_{10}\sn, \quad \Delta W\sn = W_{10}\sn - W_{01}\sn 
\end{align}
which depends on the interrelation between $V\sn{} $ and the TLS relaxation rate $W\sn$. It also depends on the TLS asymmetry $\Delta W\sn$. Equation (\ref{eq:Pi_1_multi_TLS_explicit}) shows that, as expected from the qualitative arguments, the TLS-induced change of the measurement probability is determined by the product of the contributions from individual TLSs. 

\begin{widetext}

The correlator of the measurement outcomes $r_2(k)$ due to the coupling to TLSs is described by a more complicated expression. As mentioned above, it should contain contributions from the decay of groups of different numbers of TLSs. The expression below shows that $r_2(k)$ is actually expressed in terms of a sum over such groups. The number of TLSs in each group is given by $s$, with $N_\mathrm{TLS}\geq s\geq 1$, where $N_\mathrm{TLS}$ is the total number of TLSs. A group with a given $s$ includes all possible subsets of $s$ TLSs $\{m\}_s = m_1,m_2,...,m_s$ with $N_\mathrm{TLS}\geq m_i\geq 1$ for $i=1,\ldots,s$. The centered correlator has the form
\begin{align}
\label{eq:pair_full_1}
&\tilde r_2(k) = \frac{1}{4}e^{-2t_R/T_2}\sum_{s=1}^{N_\mathrm{TLS}}\sum_{\{m\}_s}
\left\{\mathrm{Re}\left( e^{i\tilde\phi_R}\Bigl[\prod_{j=1}^s \xi_k^{(m_j)}(t_R)\Bigr]
\Bigl[\prod_{n\notin\{m\}_s}\Xi\sn(t_R)\Bigr]\right)\right\}^2\nonumber\\
&\xi_k\sn(t_R) = i{}w\sn (V\sn{} /\gamma\sn)e^{-kW\sn \ts/2}\sinh \gamma\sn t_R, \quad 
{}w\sn = 2(W_{01}\sn W_{10}\sn)^{1/2}/W\sn
\end{align}
where $\Xi\sn(t_R)$ and the characteristic coupling parameter $\gamma\sn$ are given by Eq.~\eqref{eq:Xi_n} and \eqref{eq:mu_parameter}, respectively.   
The sum $\sum_{\{m\}_s}$ implies summation over all TLSs in the set $\{m\}_s$, i.e., over all $m_1,m_2,...,m_s$, which can be arranged as  $N_\mathrm{TLS}\geq m_1>m_2>...>m_s\geq 1$. We note that the number of terms is exponential in the number of the TLSs $N_\mathrm{TLS}$.
\end{widetext}

The derivation of the above results based on the master equation for the coupled qubit and TLSs is given in Secs.~\ref{sec:master_equation} - \ref{sec:r_2_TLS}. In Appendix~\ref{sec:Yaxing_method} we provide an alternative derivation where the effect of the TLSs is modeled by telegraph noise, i.e.,  $\omq(t)$ in Eq.~(\ref{eq:just_delta_omega}) is assumed to be telegraph noise.

The general expressions for the effect of the coupling to TLSs on the outcomes of Ramsey measurements simplify in the limiting cases of strong and weak coupling, as determined by the ratios of the coupling parameters $|V\sn{} |$ and the decay rates of the TLSs $W\sn$. They also simplify in the limit of comparatively short duration $t_R$ of the Ramsey measurements. The corresponding limiting cases are discussed in the following subsections. For concreteness, we assume $V\sn{} >0$, as the change of the sign of $V\sn{} $ corresponds to swapping the TLS states $\ket{0}\sn$ and $\ket{1}\sn$.

\subsection{Weak coupling}
\label{subsec:weak_cplng}
 
The case of weak coupling, where $V\sn{} \ll W\sn$ is interesting not only on its own, but also because it extends to the case where noise of the TLSs is Gaussian. The Gaussian limit corresponds to $V\sn{} \propto N_\mathrm{TLS}^{-1/2}$, with the number of the TLSs $N_\mathrm{TLS}\gg 1$, but we emphasize that the expressions in this subsection are not based on the assumption of Gaussianity. We start with a formal expansion of the general expressions for $r_1$ and $r_2(k)$ and then provide  a physical insight into the results.

Formally, for $V\sn{} \ll W\sn$ we can expand the parameter $\gamma\sn$ in Eq.~(\ref{eq:mu_parameter}) to the second order in $V\sn{} /W\sn$. Substituting the expansion into the expression (\ref{eq:Pi_1_multi_TLS_explicit}) for the probability  $r_1 $ of having ``1'' as the Ramsey measurement outcome, we obtain
 \begin{align}
 \label{eq:weak_cplng_Pi_1}
&r_1 \approx \frac{1}{2} + \frac{1}{2}\exp\left(-\frac{t_R}{T_2} - \frac{1}{2}\overline{\theta^2}\right)
\cos\phi_R
,\nonumber\\
& \overline{\theta^2}= \sum_n\overline{ \theta\sn{}^2}, \quad \overline{\theta\sn{}^2}= 2 \left({}w\sn V\sn{} /W\sn\right)^2\nonumber\\
&\qquad \times\left( W\sn{}t_R + e^{-W\sn{}t_R} -1\right).
\end{align}
Here we used that the observable control phase is $\phi_R = \tilde\phi_R + \overline{\delta\omega}_\mathrm{TLS}t_R$,
%
\begin{align}
\label{eq:frequency_shifts_TLS}
\overline{\delta\omega}_\mathrm{TLS} = \sum_n \overline{\delta\omega}{}\sn, \quad \overline{\delta\omega}{}\sn=
V\sn{} \Delta W\sn/W\sn.
\end{align}
%
The parameter $\overline{\delta\omega}_\mathrm{TLS}=\sum_n V\sn \langle\tT_z\sn\rangle$ is the mean shift of the qubit frequency due to the asymmetry of the TLSs. Indeed, by construction of the operator $\tT\sn$, $\langle\tT_z\sn\rangle$ is the difference of the stationary populations  $w_0\sn$ and $w_1\sn$ of the TLS states $\ket{0}\sn$ and $\ket{1}\sn$. These populations are  expressed in terms of the transition rates using the balance equation, 
\begin{align}
\label{eq:stationary_populations}
 w_0\sn = W_{10}\sn/W\sn,\quad w_1\sn = W_{01}\sn/W\sn,
 \end{align}
which leads to the above expression for $\overline{\delta\omega}_\mathrm{TLS}$.

The term $\overline{\theta^2}$  in Eq.~(\ref{eq:weak_cplng_Pi_1}) is related to the variance of the phase accumulated by the qubit during a Ramsey measurement. It is seen from the Hamiltonian (\ref{eq:q_TLS_dispersive}) of the qubit-to-TLS coupling that the random part of the qubit phase accumulated over time $t_R$ in the $k$th Ramsey measurement is determined by the operator 
\begin{align}
\label{eq:hat_theta_n}
\hat\theta_k\sn{} = V\sn{}  \int_{k\ts}^{k\ts+t_R} dt (\tT_z\sn(t) - \langle \tT_z\sn\rangle\ .
\end{align}
The variance of the qubit phase due to the coupling to an $n$th TLS is $\overline{\theta\sn{}^2} = \langle \hat\theta_k\sn{}^2\rangle$ (clearly, $\overline{\theta\sn{}^2}$ is independent of $k$). It is easy to obtain Eq.~(\ref{eq:weak_cplng_Pi_1}) for $\overline{\theta^2}$ by taking into account that, from the Bloch equations, 
\begin{align}
\label{eq:TLS_spectrum}
\langle \tT_z\sn(t)\tT_z\sn(0)\rangle -  \langle \tT_z\sn\rangle^2 = w\sn{}^2 \exp(-W\sn t)\ ,
\end{align}
see also Sec.~\ref{sec:master_equation}.

For small $W\sn t_R$ we have 
\[\overline{\theta^2}\approx \sum_n {}(w\sn V\sn{}  t_R)^2, \quad W\sn t_R\ll 1\ ,
\]
i.e., $r_1-1/2$ displays a ``Gaussian'' decay  with $t_R$, no assumption about the TLS spectrum is needed. On the other hand, for longer Ramsey time, $W\sn t_R >1$, the decay of  $r_1-1/2$ with the increasing $t_r$ is close to exponential.

In the weak-coupling limit,  to the leading order in $V\sn{} /W\sn$ Eq.~(\ref{eq:pair_full_1}) for the centered pair correlator $\tilde r_2(k)$ takes the form
\begin{align}
\label{eq:weak_cplng_Pi11}
&\tilde r_2(k)\approx \frac{1}{4}e^{-2t_R/T_2}\sum_{n=1}^{N_\mathrm{TLS}}\Bigl[
\langle \hat\theta_0\sn{}\hat\theta_k\sn{}\rangle
\exp\Bigl(-\sum_{m\neq n}\overline{\theta^{(m)}{}^2}\Bigr)
\nonumber\\
&\times \sin^2(\phi_R - \overline{\delta\omega^{(n)}} t_R)\Bigr]\ ,
\end{align}
where 
\begin{align}
\label{eq:phase_TLS_correlator}
    &\langle \hat\theta_0\sn{}\hat\theta_k\sn{}\rangle
=(2V\sn w\sn/W\sn)^2
\exp(-kW\sn \ts)\nonumber\\
&\times 
\sinh^2 (W\sn t_R/2), \qquad k\geq 1.
\end{align}
Equation (\ref{eq:weak_cplng_Pi11}) is written in the form that relates it to the expression  for the phase correlator $\langle \hat\theta_0\sn{}\hat\theta_k\sn{}\rangle$ that follows from  Eqs.~(\ref{eq:hat_theta_n}) and (\ref{eq:TLS_spectrum}).

An important feature of the correlator $\tilde r_2(k)$ is that it does not exponentially decay with $k$. The decay is described by a superposition of the exponential factors that come from the decay of correlations of individual TLSs, with the decay rates $W\sn$. We have also kept in Eq.~(\ref{eq:weak_cplng_Pi11}) the factor $\exp(-\sum_{m\neq n}\overline{\theta^{(m)}{}^2})$ that describes the collective effect of the system of the TLSs on the contribution of an individual TLS to the decay. Even though $\overline{\theta\sn{}^2}\ll 1$ for each TLS, the sum does not have to be small, if the number of the TLSs is large. 


\subsection{Short Ramsey measurement time}
\label{subsec: short_Ramsey}

The expressions for the probability $r_1$ and the centered pair correlator $r_2(k)$ are simplified also in the  case where the duration of the Ramsey measurement is small, so that $V\sn t_R, W\sn t_R\ll 1$. The general expressions (\ref{eq:Pi_1_multi_TLS_explicit}) and (\ref{eq:pair_full_1}) give in this case,
\begin{align}
\label{eq:short_tR_r1}
&r_1\approx \frac{1}{2} + \frac{1}{2}\exp\left[-(t_R/T_2) - \sum_n (w\sn V\sn t_R)^2/2\right]
\nonumber\\
&\times \cos\phi_R 
\end{align}
and 
\begin{align}
\label{eq:short_tR_r2}
&\tilde r_2(k) \approx \frac{1}{4}e^{-2t_R/T_2}t_R^2\sum_n V\sn{}^2 e^{-kW\sn \ts}
\nonumber\\
 & \times \exp\Bigl[-\sum_{m\neq n}(w^{(m)}V^{(m)} t_R)^2\Bigr]
 \sin^2(\phi_R -\overline{\delta\omega}^{(n)}t_R) \ .
 \end{align}

Equations (\ref{eq:short_tR_r1}) and (\ref{eq:short_tR_r2}) hold for  arbitrary ratios $V\sn/W\sn$. The considered limit is important. Indeed,  TLSs with  $W\sn t_R\gg 1$ do not contribute to the centered correlators $\tilde r_2(k)$ and higher-order centered correlators: fast switching averages out their effect. The correlators $\tilde r_2(k)$ are formed by the TLSs with the switching rates smaller than $\ts^{-1} < t_R^{-1}$, i.e., $W\sn t_R <1$. The condition $V\sn t_R\ll 1$ is the weak-coupling condition, where the fluctuations weakly affect the quantum correlations responsible for the probability of observing ``1'' in a Ramsey measurement being $r_1 >1/2$. 


\subsection{Strong coupling}
\label{subsec:strong_cplng}

In the  limit of strong coupling to the TLSs, $V\sn{} \gg W\sn$, to the leading order
\begin{align}
\label{eq:r1_strong_cplng}
&r_1 \approx \frac{1}{2} + \frac{1}{2}e^{-t_R/T_2}\mathrm{Re}\Bigl\{e^{i\tilde\phi_R} \prod_n e^{-W\sn t_R/2} 
\nonumber\\
&
\times[\cos V\sn{} t_R + i (\Delta W\sn/W\sn)\sin V\sn{} t_R]\Bigr\}\ .
\end{align}
The dependence of $r_1$ on the duration of the Ramsey measurement $t_R$ is determined by the product of the oscillating factors. For $V\sn{} t_R\ll 1$  we have $r_1-1/2 \propto \exp(-\sum_nV\sn{} ^2t_R^2/2)$. As the characteristic $V\sn{} t_R$ increases, $r_1-1/2$ quickly falls off for a large number of TLSs with different $V\sn$. 

The effect of the oscillating factor in Eq.~(\ref{eq:r1_strong_cplng}) is particularly clear if the values of $V\sn$ for different TLSs are close. Here, for symmetric TLSs ($\Delta W\sn =0$), we have $r_1-1/2 \propto( \cos V\sn t_R)^{N_\mathrm{TLS}}$, which is a sharp periodic function of $V\sn t_R$ for a large number  of TLSs. 

Strong coupling to the qubit is pronounced provided the condition $V\sn t_R\gtrsim 1$ holds for $t_R \ll T_2$. It is plausible therefore that only one TLS will display a sufficiently strong coupling. In this case $r_1$ oscillates with $t_R$ with period $2\pi/V\sn$. The correlator $\tilde r_2(k)$ is oscillating with half of this  period. For a symmetric TLS,
\begin{align}
\label{eq: one_TLS_strong_r2}
&\tilde r_2(k)\approx   \frac{1}{4}e^{-2t_R/T_2}\sin^2\phi_R \sin^2(V\sn t_R)\nonumber\\
&\times \exp(-kW\sn \ts) \ .
\end{align}
It is seen from a comparison of this expression with Eq.~(\ref{eq:class_noise_results_1}) below that the dependence of $\tilde r_2(k)$ on the angle $\phi_R$ is different for a TLS and for Gaussian frequency noise, if the coupling is not weak.

Oscillations with the varying $t_R$ is a characteristic feature of a strong coupling of TLSs to the qubit. However, for slowly decaying TLSs it might be easier to observe oscillations of the probability distribution of the measurement outcomes for a comparatively short acquisition time discussed in Sec.~\ref{subsec:slow_TLSs_1}.  The oscillations discussed in that section have the same physical origin but can be pronounced already for comparatively small $V\sn t_R$, and observing them does not require varying the duration of the Ramsey measurement.

\section{Analytical results on Gaussian fluctuations of the qubit frequency}
\label{sec:class_noise_summary}

An important cause of the qubit frequency fluctuations is external classical noise with frequencies much lower than the qubit transition frequency. The effect of such noise is described by the term $-\frac{1}{2}\omq(t)\sigma_z$ in the qubit Hamiltonian, cf. Eq.~(\ref{eq:just_delta_omega}). We consider zero-mean stationary Gaussian frequency fluctuations $\omq(t)$. Such fluctuations are fully characterized by their power spectrum 
\begin{align}
\label{eq:freq_noise_spectrum} 
S_q(\omega) = \int_{-\infty}^\infty dt e^{i\omega t} \langle \omq(t)\omq(0)\rangle.
\end{align}
For classical fluctuations $S_q(\omega) = S_q(-\omega)$. The technique we develop can be extended to quantum noise as well.

\subsection{Explicit general expressions}
\label{subsec:explicit_Gaussian}

Of relevance for the qubit is the accumulation of its phase due to the frequency fluctuations. For a classical noise of the qubit frequency, the phase accumulated over the time interval $(k\ts, k\ts+t_R)$, is 
$
\theta_k = \int_{k\ts}^{k\ts+t_R}\omq (t)dt
$,
%
cf. Eq.~(\ref{eq:phase_k_classical}). This expression is the classical analog of the operator $\sum_n\hat\theta_k\sn$, see Eq.~(\ref{eq:hat_theta_n}). For  a zero-mean Gaussian noise
\[\langle\theta_k\rangle = 0.\]

The phases accumulated during subsequent measurements are related via noise power spectrum,
\begin{align}
\label{eq:phase_correlation}
&f_{k}\equiv \langle\theta_n\theta_{n+k}\rangle =\frac{1}{\pi}\int d\omega 
S_q(\omega)\exp(i\omega k\ts)\ ,\nonumber\\
&\times (1-\cos \omega t_R)/\omega^2 \ .
\end{align} 
We use here that, because of the stationarity of noise $\omq(t)$, the correlator $\langle\theta_n\theta_{m}\rangle$ depends only on $|n-m|$.  The probability distribution of the phases $\theta_n$ is Gaussian for the Gaussian distribution of $\omq(t)$. It is thus fully determined by the parameters $f_k$. As seen from Eq.~(\ref{eq:phase_correlation}) $f_0>0$ and $f_0>|f_k|$ for $k\neq 0$. While $f_k=f_{-k}$, the correlators $f_k$ can be negative, in general.

The phase correlators should be directly related to the probability $r_1$  and the correlators $r_2(k), r_3(k,l)$ of the Ramsey measurements. Intuitively, one can express these parameters in terms of the probability $p(\theta)$, Eq.~(\ref{eq:standard_probability}),
of having ``1'' as an outcome of the Ramsey measurement for a given $\theta$,
\begin{align}
\label{eq:trivial_probability}
&r_1 = \langle p(\theta_n)\rangle, \quad r_2(k) = \langle p(\theta_n)p(\theta_{n+k})\rangle
\nonumber\\
&r_3(k,l) = \langle p(\theta_n) p(\theta_{n+k}) p(\theta_{n+l})\rangle
\end{align}
The averaging here is done over the distribution of the phases $\{\theta\}$.  For the stationary distribution the result is  independent of $n$. 

Equation (\ref{eq:trivial_probability}) applies independent of the noise statistics. It is substantiated by the analysis of Sec.~\ref{sec:Gaussian_general}, which is based on solving the master equation for a qubit in the presence of noise.

For Gaussian noise, $r_1, r_2(k)$, and $r_3(k,l)$ can be explicitly expressed in terms of the correlators $f_k$, and thus in terms of the noise power spectrum,
\begin{align}
\label{eq:class_noise_results_1}
&r_1 = \frac{1}{2}\left[1+ e^{-t_R/T_2}\exp(-f_0/2)\cos\phi_R\right], \nonumber\\
&\tilde r_2(k)= \frac{1}{8}e^{-2t_R/T_2}\exp(-f_0)\left[e^{f_k}-1  \right.\nonumber\\
&\left. -\cos(2\phi_R)(1-e^{-f_k})\right] \ ,
\end{align}
and
\begin{widetext}
\begin{align}
\label{eq:class_noise_results_2}
&\tilde r_3(k,l)=\frac{1}{32}\exp[-3(f_0/2) - 3(t_R/T_2)]
\left\{\cos 3\phi_R\left[\exp\Bigl(-\sum_i f_i\Bigr) + 2-\sum_i\exp(-f_i)\right] \right.
\nonumber\\
&\left.  + \cos\phi_R\left[\sum_i \exp \Bigl(-f_i+\sumprime{j} f_j \Bigr) + 6
-\sum_i \Bigl(2e^{f_i} + e^{-f_i}\Bigr)
\right]
\right\}.
\end{align}
In Eq.~(\ref{eq:class_noise_results_2}) the summation goes over $i,j =k,l,l-k$ and the prime over the sum means that $j\neq i$.
\end{widetext} 

It is seen from Eq.~(\ref{eq:class_noise_results_1}) that the correlators $f_k$ can be found for all $k$ by measuring $r_1$ and $\tilde r_2(k)$. They define $\tilde r_3(k,l)$, and therefore by measuring $\tilde r_3(k,l)$ one can tell whether frequency noise is Gaussian. 

It follows from Eq.~(\ref{eq:class_noise_results_2}) that the centered correlator $\tilde r_3$ goes to zero as $\phi_R$ approaches $\pi/2$. On the other hand, for weak Gaussian noise, where $f_k\ll 1$, we have from Eq.~(\ref{eq:class_noise_results_1}) $\tilde r_2(k)\propto f_k\sin^2\phi_R$, which means that in this limit $\tilde r_2\to 0$ for $\phi_R\to 0$. Therefore one may be interested in measuring the pair and triple correlators for different values of $\phi_R$ and comparing the results.

For weak noise, $f_k\ll 1$, the centered pair and triple correlators are
\begin{align}
\label{eq:weak_correlated}
&\tilde r_2(k) \approx \frac{1}{4}e^{-2t_R/T_2} \exp(-f_0) f_k\sin^2\phi_R\nonumber\\
&\tilde r_3(k,l) \approx -\frac{1}{8}e^{-3t_R/T_2}\exp(-3f_0/2)\nonumber\\
&\times (f_k f_l + f_k f_{l-k} + f_lf_{l-k})\cos\phi_R \sin^2\phi_R
\end{align}
We keep the term $\propto f_0$ in the exponent to account for the case where the parameters $f_{n>0}$ are small because $\ts \gg \tau_\mathrm{corr}$.  Respectively, we do not simplify the expression for $r_1$. The correlator $\tilde r_2(k)$ is of the first order in $f_k$ whereas the correlator $\tilde r_3(k,l)$ is bilinear in $f_k, f_l$. In many cases of interest $f_n>0$ for all $n$; then $\tilde r_2(k)>0$ whereas $\tilde r_3(k,l)<0$ for $\cos\phi_R>0$. Then, if in the experiment it is found that, for a weak noise, both $\tilde r_2>0$ and $\tilde r_3 >0$, this indicates that the noise is non-Gaussian. 


 \subsection{Exponentially correlated frequency noise}
\label{subsec:exponentially_correlated_theory}

We now provide explicit expressions for the correlators $f_k=\langle \theta_n\theta_{n+k}\rangle$ of the phases accumulated during Ramsey measurement for two explicit types of Gaussian frequency noise. These expressions are used in Sec.~\ref{subsec:simulations_Gaussian} to compare the analytical expressions with the results of simulations. The effect of noise correlations comes into play if the characteristic correlation time is comparable to $\ts$. Otherwise one can think of frequency noise as white ($\delta$-correlated). For $\langle\omq(t)\omq(t')\rangle = D_\mathrm{w}\delta(t-t')$ we have 
\[f_k = D_\mathrm{w} t_R \delta_{k,0},\]
where $D_\mathrm{w}$ is white noise intensity. For such noise the centered measurement correlators $\tilde r_2(k), \tilde r_3(k,l)$ should vanish, as indeed seen from Eqs.~(\ref{eq:class_noise_results_1}) and (\ref{eq:class_noise_results_2}). 
 
An important type of correlated Gaussian noise is exponentially correlated noise,
\begin{align}
\label{eq:expon_correlated_noise}
& \langle\omq(t)\omq(t')\rangle =\frac{1}{2} D_\mathrm{corr}\tau_\mathrm{corr}^{-1}e^{-|t-t'|/\tau_\mathrm{corr}}.  
\end{align}
The power spectrum of such noise is Lorentzian,
\begin{align}
\label{eq:Lorentzian_spectrum}
&S_q(\omega) \equiv \int_{-\infty}^\infty dt e^{i\omega t}\langle\omq(t)\omq(0)\rangle \nonumber\\
&=D_\mathrm{corr}/(1+\omega^2\tau_\mathrm{cor}^{2}).
\end{align}
The exponentially correlated noise may come from a filtered white Gaussian noise, a simple example being broad-band random voltage filtered by an $RC$-circuit. Another example is noise from dispersive coupling to thermal photons in a multi-mode cavity with the modes decay times being close to each other, so that these times can be approximated by $\tau_\mathrm{corr}$. In Eq.~(\ref{eq:expon_correlated_noise}) $D_\mathrm{corr}$ characterizes noise intensity, whereas $\tau_\mathrm{corr}$ is noise correlation time.

Using   Eq.~(\ref{eq:phase_correlation}), which expresses the phase correlators $f_k$ in terms of the power spectrum $S_q(\omega)$,  one obtains 
\begin{align}
\label{eq:Ornstein_f_function}
&f_{0}=D_\mathrm{corr}\left[t_R-\tau_\mathrm{corr}\left(1-e^{-t_R/\tau_\mathrm{corr}}\right)\right], \nonumber\\
&f_{k} = D_\mathrm{corr}\tau_\mathrm{corr}\exp(-|k|\ts/\tau_\mathrm{corr})\nonumber\\
&\times [\cosh(t_R/\tau_\mathrm{corr})-1] \quad (|k|>0)
\end{align}
As seen from Eq.~(\ref{eq:Ornstein_f_function}), the correlators $f_k$ fall off exponentially with the increasing $|k|$ for exponentially correlated noise. The rate of the decay is determined by the relation between the duration of the cycle $\ts$ and noise correlation time $\tau_\mathrm{corr}$.

From Eq. ~(\ref{eq:weak_correlated}), for weak noise, $f_n\ll 1$, the correlator $\tilde r_2(k)$ falls off exponentially with the increasing $k$. The correlator $\tilde r_3(k,l)$, on the other hand, which is quadratic in $f_n$, shows non-exponential decay even where the decay of $\tilde r_2(k)$ is close to exponential. This is in agreement with the simulations discussed in Sec.~\ref{subsec:simulations_Gaussian}. 

For stronger noise the decay of the correlators $\tilde r_2(k), \tilde r_3(k,l)$ becomes nonexponential, even though $f_k$ exponentially falls off with the increasing $k$. 

\subsection{Noise with ``color''}
\label{subsec:quasimonochrom}

To illustrate the possibility of a nonmonotonic behavior of the correlators $f_k$ as functions of $k$, we briefly describe the effect of noise with ``color'', that is noise with a spectrum $S_q(\omega)$ that has a peak at a nonzero frequency. A simple example is Johnson-Nyquist noise filtered by an RCL circuit. The power spectrum of such noise is  
\begin{align}
\label{eq:monochrom}
S_q(\omega) = D_\mathrm{clr}[(\omega^2-\omega_\mathrm{clr}^2)^2 + 4\Gamma_\mathrm{clr}^{2}\omega^2]^{-1}
\end{align}
This spectrum has a peak at $(\omega_\mathrm{clr}^2 + 2\Gamma_\mathrm{clr}^2)^{1/2}$. For $\omega_\mathrm{clr}\gg \Gamma_\mathrm{clr}$ this peak is Lorentzian with halfwidth $\Gamma_\mathrm{clr}$.

It is straightforward to see that, for $ \sqrt{2}\Gamma_\mathrm{clr}t_R < \omega_\mathrm{clr} t_R \ll 1$ we have 
\begin{align}
\label{eq:colored_f_k}
&f_0=D_\mathrm{clr}t_R^2/4\omega_\mathrm{clr}^3\sin\phi_\mathrm{clr},\nonumber\\
&f_k=\frac{D_\mathrm{clr}t_R^2}{2\omega_\mathrm{clr}^3\sin(2\phi_\mathrm{clr})}
\exp(-k\ts\omega_\mathrm{clr}\sin\phi_\mathrm{clr})\nonumber\\
&
\times\cos(k\ts \omega_\mathrm{clr}\cos\phi_\mathrm{clr} - \phi_\mathrm{clr}) \quad (k>0),
\end{align}
where 
\[\phi_\mathrm{clr} = \frac{1}{2}\arctan\frac{2\Gamma_\mathrm{clr}\sqrt{\omega_\mathrm{clr}^2 - \Gamma_\mathrm{clr}^2}}{\omega_\mathrm{clr}^2 - 2\Gamma_\mathrm{clr}^2}.\]

From Eq.~\eqref{eq:colored_f_k}, the correlators $f_k$ display exponentially decaying oscillations as functions of $k$. For weak noise, such oscillations will be immediately seen in the correlators $r_2(k)$ and will be also seen as broadened peaks in the power spectrum of the measurement outcomes discussed in Sec.~\ref{sec:periodic}.


\subsection{$1/f$ noise }
\label{subsec:1/f_theory}

Very often qubit decoherence is caused by $1/f$ frequency noise, i.e., by noise with the power spectrum $S_q(\omega) \propto 1/\omega\equiv 1/2\pi f$, cf.  \cite{Yan2012,Paladino2014,Quintana2017,You2021} and references therein. If such noise comes from a large number of fluctuators, it becomes Gaussian, and the assumption that $1/f$ noise is Gaussian is often made. Since the integral intensity of $1/f$ noise diverges, the spectrum has to be cut both at low and high frequencies. A high-frequency cutoff is irrelevant for the problem of frequency noise that we consider, since the integration over the interval $t_R$ between the Ramsey pulses filters out high-frequency noise components. 

The low-frequency cutoff is model-dependent. We will present results for a simple physically motivated model in which the spectrum is smooth. This model is also  related to the model of the noise from TLSs used in the simulations. In contrast to the simulated TLS models, it corresponds to coupling to a very large number of  TLSs with the coupling constant being the same for all TLSs (cf. Refs.~\cite{Ithier2005,Yoshihara2014}) and with log-normal distribution of the switching rates $W\sn$; the rates are assumed to be limited from below by $\omega_{\min}$.  From Eq.~(\ref{eq:TLS_spectrum}), the power  spectrum of such noise has the form
\begin{align}
\label{eq:1_f_spectrum}
&S_q(\omega) = \frac{2}{\pi}D_\mathrm{fl} \int_{\omega_{\min}}^\infty \frac{dW}{W^2+\omega^2}\nonumber\\
&=D_\mathrm{fl}\omega^{-1}[1-(2/\pi)\tan^{-1}(\omega_{\min}/\omega)]
\end{align}
In the range $\omega\gg \omega_{\min}$ we have $S_q(\omega) \approx D_\mathrm{fl}/\omega$, whereas $S_q(\omega)\approx 2D_\mathrm{fl}/\pi\omega_{\min}$ for $\omega\ll \omega_{\min}$.
 
Calculating the integral in Eq.~(\ref{eq:phase_correlation}) by closing the contour in the $\omega$-plane and using Eq.~(\ref{eq:1_f_spectrum}), one can write  the correlators of the qubit phase accumulated during remote measurements as
\begin{align}
\label{eq:1_f_correlators_integral_k}
&f_{k} = \frac{2}{\pi}D_\mathrm{fl}\int_{\omega_{\min}}^\infty \frac{dW}{W^3}[\cosh(Wt_R)-1]
\nonumber\\
&\times \exp(-kW\ts),\qquad k>0,
\end{align}
 whereas the mean square phase is 
 \begin{align}
 \label{eq:1_f_correlators_integral_0}
 &f_0=\frac{2}{\pi}D_\mathrm{fl}\left[\frac{t_R}{\omega_{\min}} + \int_{\omega_{\min}}^\infty \frac{dW}{W^3}\Bigl(e^{-Wt_R}-1\Bigr)\right]
 \end{align}
For small $\omega_{\min}t_R \ll 1$ the leading-order terms in $f_k$ and $f_0$ are logarithmic in $\omega_{\min}t_R$.  The decay of the correlators $f_k$ with the increasing $k$ is nonexponential, in contrast to the case of exponentially correlated Gaussian noise. However, it becomes close to exponential in the limit of large $k\omega_{\min}\ts \gg 1$, 
\[f_k\approx D_\mathrm{fl}(t_R^2/\pi  k\ts\omega_{\min})\exp(-k\omega_{\min}\ts).\]
On the other hand, for $k\ts \gg t_R$ but $k\omega_{\min}\ts \ll 1$ a reasonable numerical approximation is 
\begin{align}
\label{eq:1_f_intermediate}
f_k\approx D_\mathrm{fl}(t_R^2/\pi)[-\gamma_E-\log(k\omega_{\min}\ts)]
\end{align}
where $\gamma_E\approx 0.58$ is the Euler constant.

 The integrals in Eqs.~(\ref{eq:1_f_correlators_integral_k}) and (\ref{eq:1_f_correlators_integral_0}) can be expressed in terms of exponential integral and hyperbolic integral functions. These expressions are given in Appendix~\ref{sec:1_f_integrals}.


\section{Comparison of the theory and simulations}
\label{sec:theory_simulations}

In this section we present the results of  simulations of periodically repeated Ramsey measurements and compare them  with the theoretical predictions. The results are aimed at illustrating major qualitative aspects of the effect of low-frequency fluctuations of the qubit frequency on the measurements. We chose the ratio of the period $\ts$ of the repeated operation ``measurement+reset''  to the accumulation time of the  measurement itself $t_R$ (see Fig.~\ref{fig:pulse_sequence}) to be equal to $\ts/t_R=3$. We present the results for $\phi_R=\pi/4$, where all centered correlators are ``visible''. With other simple choices of $\phi_R$, as seen from Eqs.~(\ref{eq:class_noise_results_1}) and (\ref{eq:class_noise_results_2}), we have $\tilde r_2=0$ for $\phi_R=0$ in the case of  weak coupling, whereas $\tilde r_3=0$ for $\phi_R=\pi/2$. Also we assume that $t_R$ is much smaller than $T_2$ and disregard corrections $\propto t_R/T_2$. In this section we describe the correlators $r_2(k)$ and $r_3(k,l)$ found  in sufficiently long simulations to obtain good statistics. The simulation algorithms are outlined in Appendix~\ref{sec:simulation_algorithms}.

\subsection{Effect of the coupling to TLSs}
\label{subsec:simulations_TLSs}

One of the goals of the paper is to compare the effects of qubit frequency fluctuations $\omq(t)$ induced by  coupling to the TLSs and by Gaussian noise. Important for such comparison is the power spectrum $S_q(\omega)$ of the TLSs-induced fluctuations $\omq(t)$. From Eq.~(\ref{eq:q_TLS_dispersive}), $\omq(t)$ is an operator in the space of the TLSs' states, $\omq(t)\Rightarrow \delta\hat\omega_\mathrm{q}(t) = \sum_nV\sn\tT_z\sn$, and then
\[S_q(\omega) = \int_{-\infty}^\infty dt e^{i\omega t}\langle \delta\hat\omega_\mathrm{q}(t)[\delta\hat\omega_\mathrm{q}(0)-\langle\delta\hat\omega_\mathrm{q}\rangle)]\rangle\ .
\]
Using the explicit expression (\ref{eq:TLS_spectrum}) for the correlator of $\tT_z\sn$, we obtain
\begin{align}
\label{eq:TLS_power_spectrum}
&S_q(\omega)=2\sum_{n=1}^{N_\mathrm{TLS}} (w\sn V\sn)^2 W\sn \nonumber\\
&\times(W\sn{}^2+ \omega^2)^{-1}\ .
\end{align}
We present results of the simulations in which the coupling parameters are the same for all TLSs, $V\sn=V$. We also assumed that the switching rates $W_{ij}\sn$ depend on $n$ exponentially. In particular, for symmetric TLSs, where $W_{01}\sn = W_{10}\sn = W\sn/2$, we assumed that
\begin{align*}
V\sn=V, \qquad  W\sn t_R= \exp[-\alpha (n+n_0)].
\end{align*}
This leads to the spectrum $S_q(\omega)$ being of the $1/f$ form in a broad frequency range already for a small number of TLSs, see Appendix~\ref{sec:TLS_power_spectrum}. For a given $N_\mathrm{TLS}$, this range depends on the values of $n_0$ and $\alpha$.

\begin{widetext}

\begin{figure}[t]
    \centering
    \includegraphics[width=0.48\textwidth]{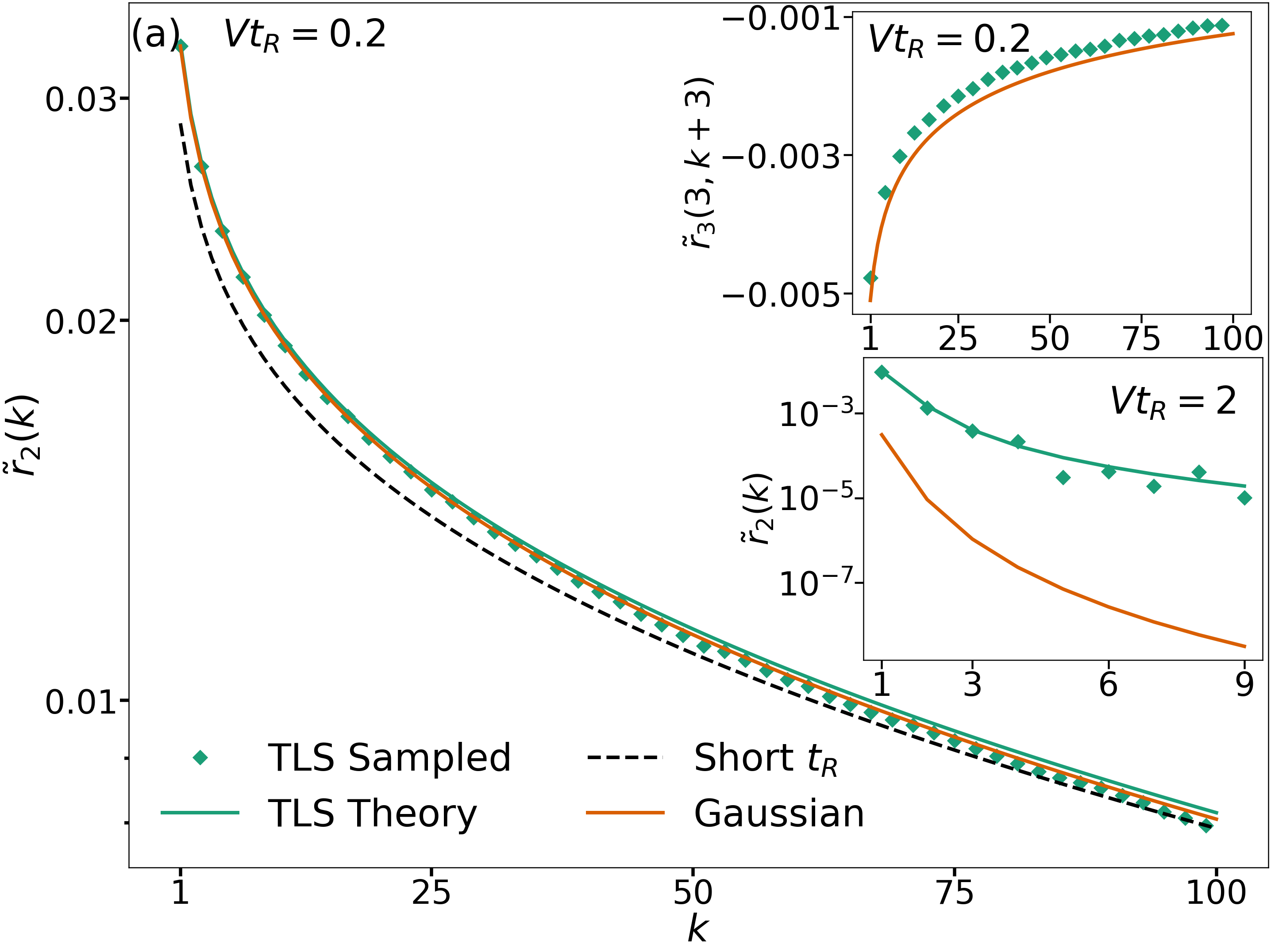}
    \includegraphics[width=0.48\textwidth]{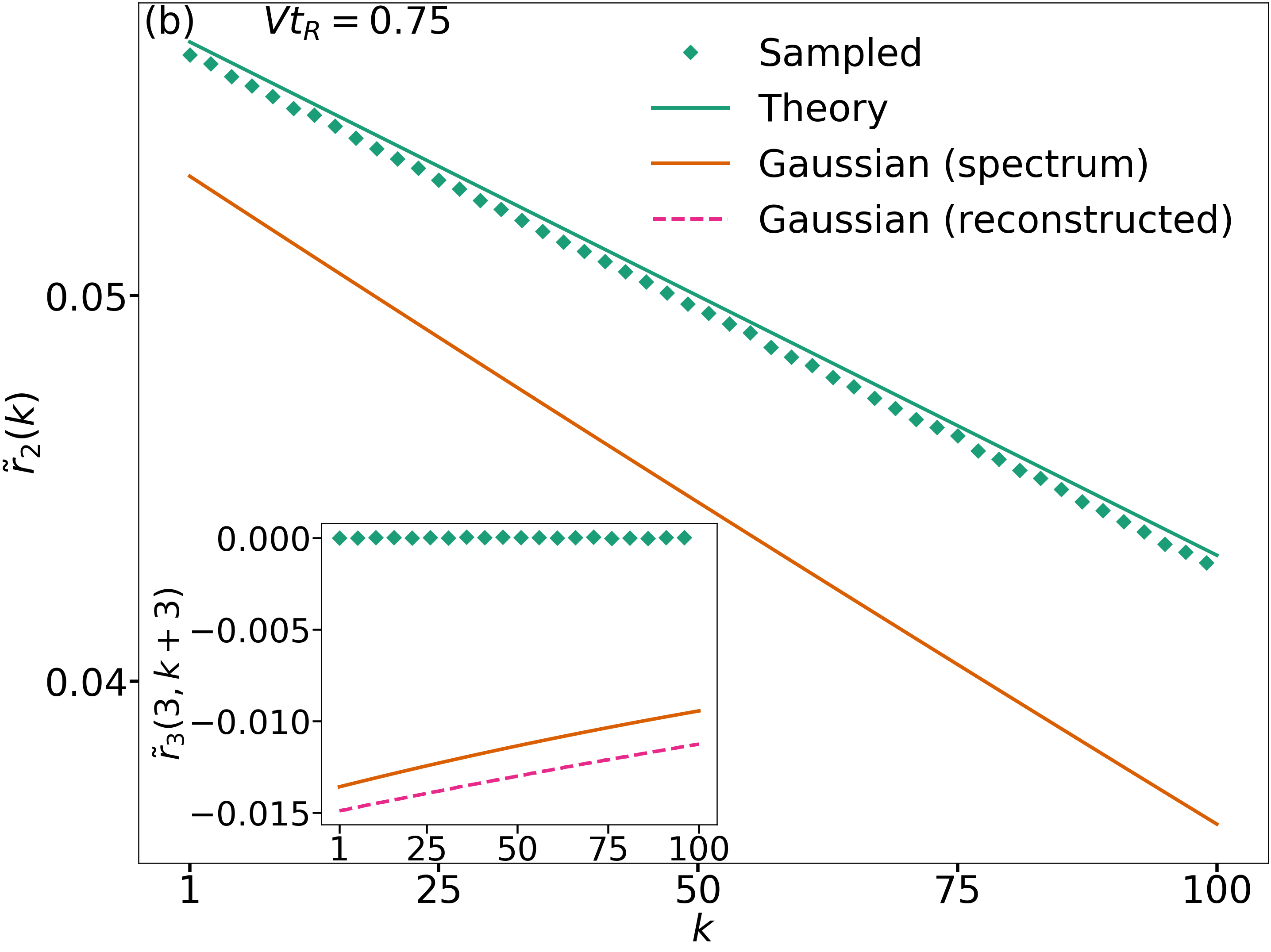}
    \includegraphics[width=0.48\textwidth]{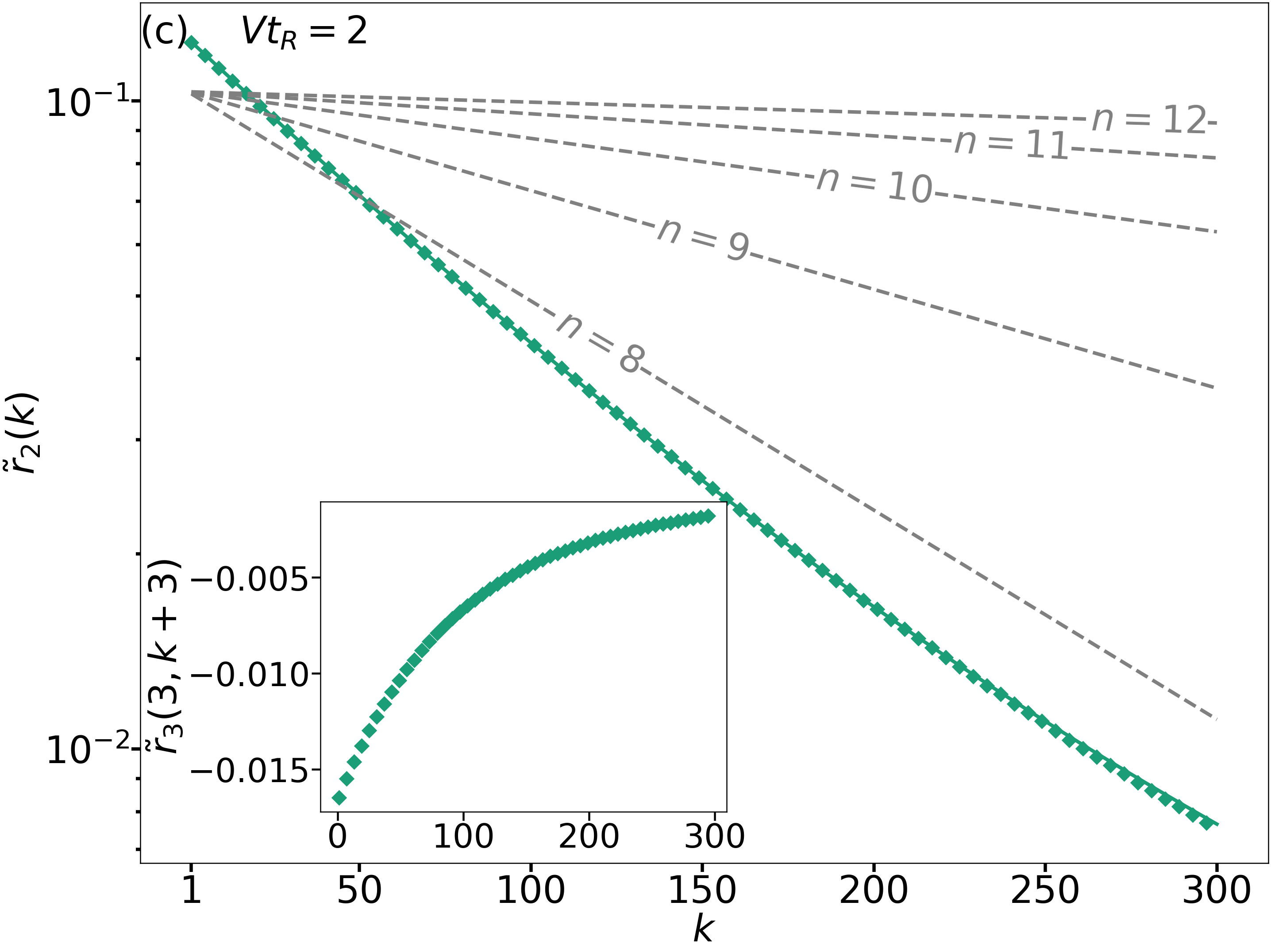}
    \includegraphics[width=0.48\textwidth]{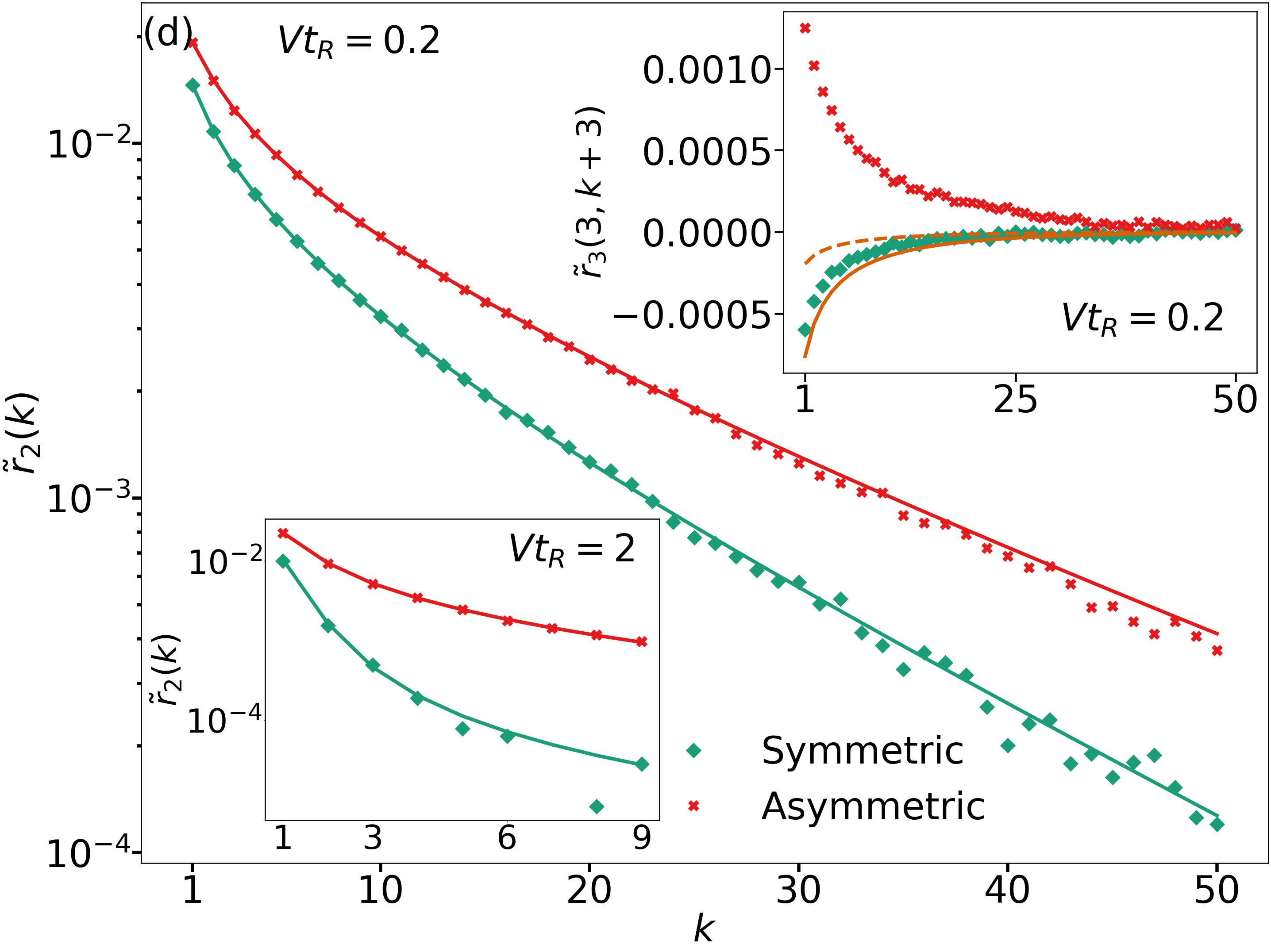}
    \caption{Centered correlators $\tilde r_2(k)$ and $\tilde r_3(k,l)$  for a qubit dispersively coupled to a system of TLSs. The coupling is the same for all TLS, $V\sn =V$. Diamonds (symmetric TLSs) and crosses (asymmetric TLSs) show the sampling data. (a) Coupling to 10 symmetric TLSs, $W_{01}\sn = W_{10}\sn = W\sn/2$. The scaled coupling strength is $Vt_R=0.2$ in the main plot and the upper inset, and $V t_R=2$ in the lower inset. The scaled TLSs' switching rates are  $W\sn t_R = \exp(-3n/4)$, $n=1,...,10$,  with $\alpha=0.75$.  The  green lines are the full theory, Eq.~\eqref{eq:pair_full_1}. The orange lines show $\tilde r_2$ and $\tilde r_3$ for TLSs approximated by Gaussian noise,  Eq.~\eqref{eq:class_noise_results_1}, with the phase correlators $f_k$ obtained from the TLS spectrum \eqref{eq:TLS_power_spectrum}. The dashed black line is the approximation of small $Vt_R$ and $W\sn t_R$, Eq.~\eqref{eq:short_tR_r2}.
(b) The correlators $\tilde r_2(k)$ and $\tilde r_3(k, k+3)$ for the coupling to a single symmetric TLS,  $V t_R=0.75$ and $Wt_R=0.001$. The orange line refers to approximating the TLS by a Gaussian noise with the spectrum (\ref{eq:TLS_power_spectrum}). The purple dashed line is obtained assuming that the noise is Gaussian with the parameters $f_k$ found from the data on $r_1$ and  $\tilde{r}_2(k)$.
(c) Coupling to 5 symmetric TLSs, $Vt_R=2$ and $ W\sn t_R = \exp[-3(n+n_0)/4]$, $n=1,...,5$,  with $n_0=7$. The gray dashed lines show partial contributions of individual TLSs; their switching rates are $W\sn t_R =\exp(-3n/4)]$, where $n$ is indicated in the figure for each of them. 
(d) Comparison of the effects of the coupling to symmetric and asymmetric TLSs, for 5 TLSs, $Vt_R=0.2$. For the symmetric TLSs $W\sn t_R= \exp(-3 n/4)$ with $n=1,...,5$. For the asymmetric TLSs $W_{01}\sn t_R=\exp[-3 (n+1)]/4]/2$ and $W_{10}\sn t_R=\exp(-3n/4)/2$. In the upper inset the orange lines are the Gaussian approximation for symmetric (solid) and asymmetric (dashed) TLSs.}
    \label{fig:r2_r3_TLS}
\end{figure}
\end{widetext}

Coupling to the TLSs leads to Gaussian noise in the limit where the number of the TLSs is $\NTLS \gg 1$, whereas the qubit coupling to an individual TLS is small, $V\sn\propto \NTLS^{-1/2}$. However, generally noise from TLSs is non-Gaussian. 

In Fig.~\ref{fig:r2_r3_TLS} we compare the simulated values of the centered correlator $\tilde r_2(k)$  for the TLSs-induced fluctuations with the theory of Sec.~\ref{sec:summary_TLS} and present the results on $\tilde r_3(k,l)$. To reveal the non-Gaussianity of noise we compare the results for $\tilde r_2(k)$ with the analytical results for Gaussian noise with the power spectrum (\ref{eq:TLS_power_spectrum}). Importantly, we also compare the results for $\tilde r_3(k,l)$ with the expression (\ref{eq:class_noise_results_2}) in which the parameters $f_k$ are determined from the data on $\tilde r_2(k)$ assuming that noise is Gaussian. The difference between $\tilde r_3$ and the results of such construction is a direct indication of non-Gaussianity of the fluctuations.

It is seen from Fig.~\ref{fig:r2_r3_TLS} that the theory and the simulations of $\tilde r_2(k)$ are in excellent agreement in a broad range of parameters, both for symmetric and asymmetric TLSs, and both for a weak and strong coupling to the qubit. The values of $r_1$ are also in excellent agreement; the relative error was $\lesssim 10^{-4}$.

The main part of Fig.~\ref{fig:r2_r3_TLS}~(a) and the top inset show that, for the considered moderately weak coupling and symmetric TLSs, the results  are close to what is expected from  the Gaussian approximation. This is the case not only for $\tilde r_2(k)$, but also for the three-time correlator $\tilde r_3(k,l)$; here there is some  deviation from the Gaussian approximation, but it is small. This shows that a moderately weak noise from the TLSs reasonably well mimics Gaussian noise already for 10 TLS, if the TLSs are symmetric. As seen from the inset in  Fig.~\ref{fig:r2_r3_TLS}~(d), this holds even for a noise from 5 symmetric TLSs. However, the lower inset in Fig.~\ref{fig:r2_r3_TLS}~(a) shows that this is not the case for a strong coupling. 

It is also seen in Fig.~\ref{fig:r2_r3_TLS}~(a) that $\tilde{r}_2(k)$ for $Vt_R=0.2$ and symmetric TLSs is reasonably well described by Eq.~(\ref{eq:short_tR_r2}), which refers to the limit of small $V\sn t_R, W\sn t_R$. As explained in Sec.~\ref{subsec: short_Ramsey}, this limit is important for the analysis of weak low-frequency noise from TLSs. 

Figure~\ref{fig:r2_r3_TLS}~(b) shows that the Gaussian approximation does not apply for the noise coming from a single symmetric TLS, even though the effective coupling strength $V^2t_R^2$ is not much higher than in the main plot of panel (a). This is the case both for the  Gaussian noise with the correlator $f_k$ calculated using the TLS power spectrum and the Gaussian noise with $f_k$ found from the simulation data on $r_1$ and $\tilde r_2(k)$. 

Figure~\ref{fig:r2_r3_TLS}~(c) shows  the effect of the coupling to TLSs with significantly different decay rates. It is seen that the effect of 5 slowly varying TLSs is not a sum of the effects of the individual TLSs. This is in agreement with Eq.~\eqref{eq:pair_full_1}, which shows that  all possible subsets of the TLSs contribute to the correlator  $\tilde r_2(k)$. Respectively, for a comparatively short time $k\ts$, $\tilde r_2(k)$ decays nonexponentially. For a longer time, i.e., for large $k$, the decay is controlled by the decay of the slowest TLS.  

Figure~\ref{fig:r2_r3_TLS}~(d) demonstrates that the centered three-time correlator $\tilde r_3(k,l)$ behaves very differently for symmetric and asymmetric TLSs even for a comparatively weak coupling. This is an important feature of asymmetric TLSs which, if observed, enables identifying them. In particular, asymmetric TLSs lead to  a positive $\tilde{r}_3(k,l)$ for positive $\tilde r_2(k)$ and $\cos\phi_R>0$, a signature of non-Gaussianity of the corresponding noise that can be directly revealed in the experiment.


\subsection{Effect of a Gaussian noise}
\label{subsec:simulations_Gaussian}

For a Gaussian noise the values of the correlators of the measurement outcomes $r_2(k)$ and $r_3(k,l)$, as well as the higher-order correlators are fully determined by the power spectrum of noise. We expect therefore that simulations should match the expressions for the correlators given in Sec.~\ref{sec:class_noise_summary}. Here we present the results of the simulations for exponentially correlated noise with the power spectrum $S_q(\omega) =D_\mathrm{corr}/(1+\omega^2\tau_\mathrm{corr}^2)$, cf. Eq.~(\ref{eq:Lorentzian_spectrum}), and with the $1/f$ type power spectrum given by Eq.~(\ref{eq:1_f_spectrum}); $1/f$ noise is characterized by the soft-cutoff minimal frequency $\omega_{\min}$. For  both types of noise  the values of the probability $r_1$ to have ``1'' as an outcome of the measurements coincided with the theoretical values to an accuracy $\lesssim 10^{-4}$.  

\begin{figure}[h]
     \centering
 \includegraphics[width = 0.47\textwidth]{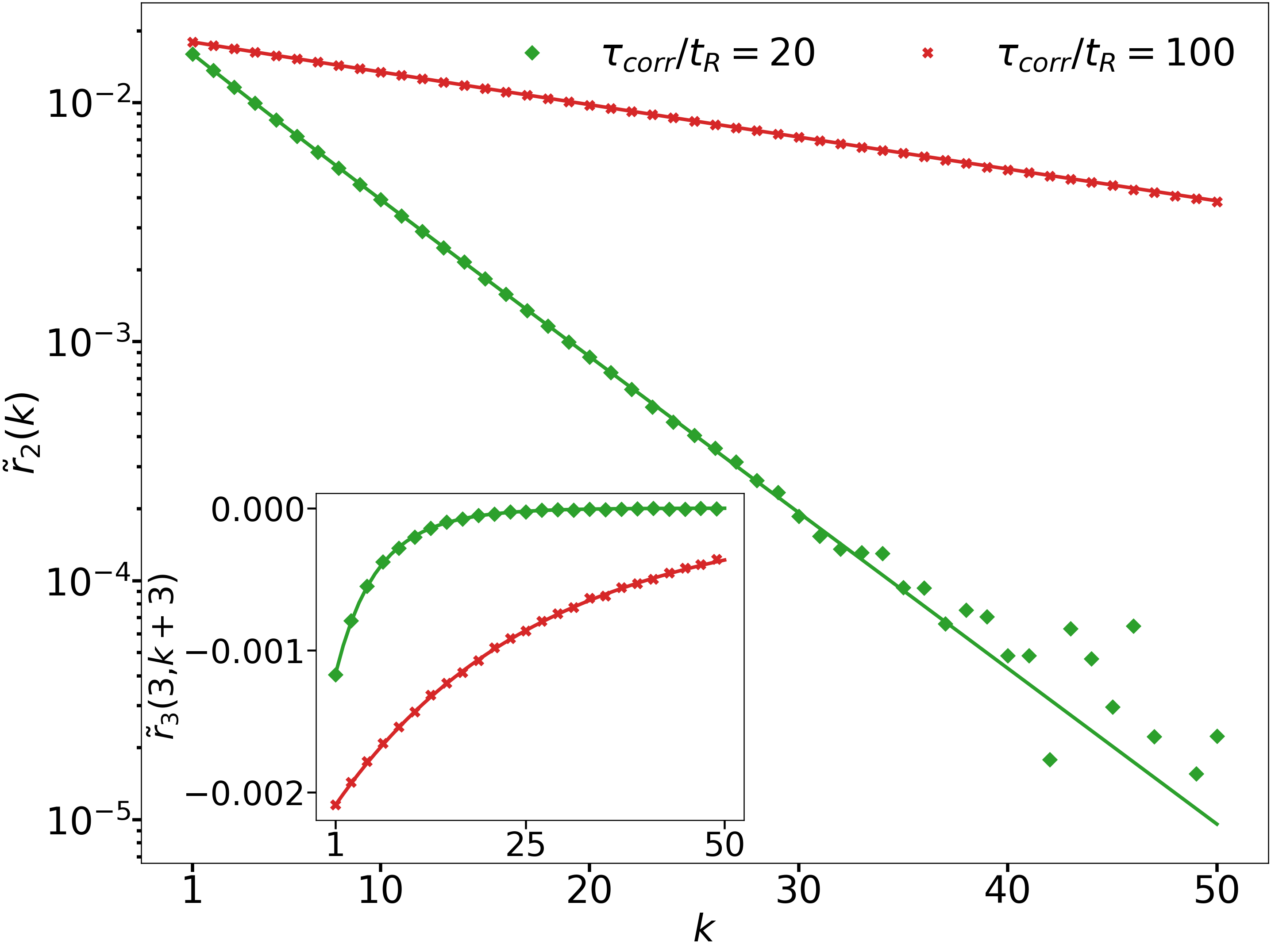}
     \caption{Two- and three-time centered correlators $\tilde r_2(k)$ and $\tilde r_3(k,k+3)$ for exponentially correlated noise. The solid lines show the theory and the diamonds show the simulations. 
     Noise intensity and correlation time for the green and red data are $D_\corr=6.51$, $\tau_\corr/t_R=20$ and $D_\corr=32.11$, $\tau_\corr/t_R=100$, respectively. They were adjusted so that the parameter $f_0$, which characterizes the coupling strength and gives the probability $r_1$, Eq.~(\ref{eq:class_noise_results_1}), be the same;   $f_0=0.16$.}
     \label{fig:r2_r3_expo}
 \end{figure}

In Fig.~\ref{fig:r2_r3_expo} we show the results for the centered correlators $\tilde r_2(k)$ and $\tilde r_3(k,k+3)$ for exponentially correlated noise. The plots refer to a comparatively weak coupling. For such coupling, as expected from the theoretical arguments, $\tilde r_2(k)$ falls off exponentially with $k$, i.e., the two-time correlator decays exponentially with time $k\ts$. The decay rate is determined by the decay rate of noise correlation $\tau_\mathrm{corr}^{-1}$. However, the three-time correlator decays non-exponentially. This  feature demonstrates the importance of studying a three-time correlator in order to identify and characterize  noise.      

\begin{figure}[h]
     \centering
      \includegraphics[width = 0.47\textwidth]{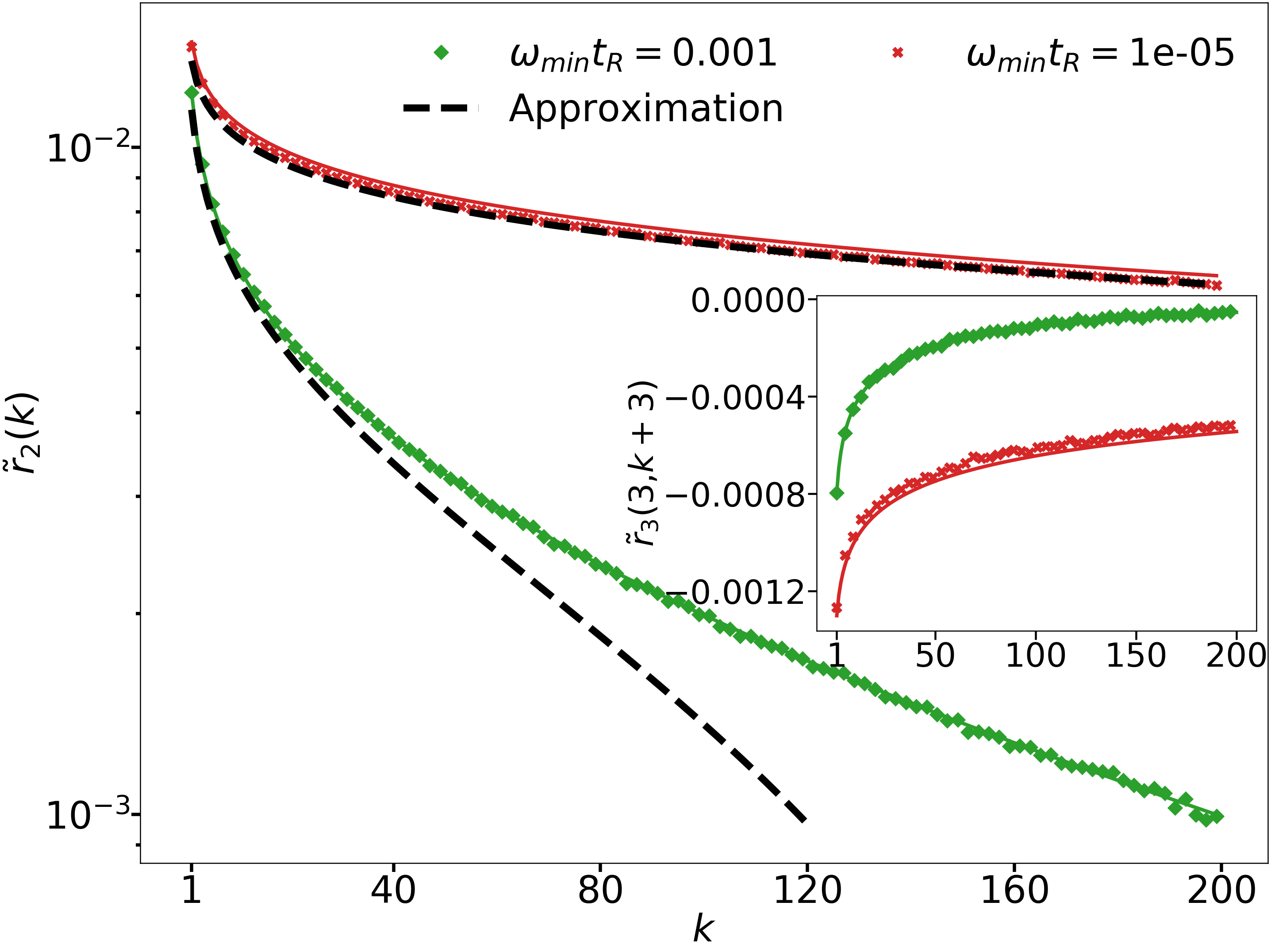}
     \caption{Two- and three-time centered correlators $\tilde r_2(k)$ and $\tilde r_3(k,k+3)$ for $1/f$-type noise with the spectrum (\ref{eq:1_f_spectrum}) that has  a smooth low-frequency cutoff at $\omega_{\min}$. The solid lines show the theory and the diamonds show the simulations. The dashed lines show the approximation (\ref{eq:1_f_intermediate}). The noise intensity and the cutoff frequency for the green and red data are $D_{\mathrm{fl}}=0.04087$, $\omegaMin t_R=0.001$ and $D_{\mathrm{fl}} = 0.02574$, $\omegaMin t_R=0.00001$, respectively. They were adjusted so that the parameter $f_0$, which characterizes the coupling strength and gives the probability $r_1$, Eq.~(\ref{eq:class_noise_results_1}), be the same;   $f_0=0.16$. 
     }
     \label{fig:r2_r3_flicker}
 \end{figure}

In Fig.~\ref{fig:r2_r3_flicker} we show the centered correlators for the model (\ref{eq:1_f_spectrum}) of $1/f$-type noise. In contrast to exponentially correlated noise in Fig.~\ref{fig:r2_r3_expo}, $\tilde r_2(k)$ does not fall off exponentially with the increasing $k$ for comparatively small $k$ even for studied weak noise. However, its decay approaches exponential for large $k$, where $k\omega_{\min}\ts \gg 1$, with the exponent determined by the low-frequency cutoff $\omega_{\min}$. For small $\omega_{\min}\ts$ this range is practically inaccessible, as $\tilde r_2(k)$ becomes extremely small. The dependence of $\tilde r_2(k)$ on $k$ is close to logarithmic for $t_R\ll k\ts \ll 1/\omega_{\min}$. The corresponding  expression~(\ref{eq:1_f_intermediate}) is shown by the dashed lines. As seen from the figure, the approximation (\ref{eq:1_f_intermediate}) actually  requires a more stringent condition, $\omega_{\min}k\ts \lesssim 0.1$.

The centered three-time correlator $\tilde r_3(k,l)$ displays a characteristic dependence on $k$ and $l$ for $1/f$ noise. As seen from the comparison of Figs.~\ref{fig:r2_r3_expo} and \ref{fig:r2_r3_flicker}, this dependence is very different for Gaussian exponentially correlated and $1/f$ noises.  

\section{Probability distribution for an intermediate acquisition time}
\label{sec:short_acquisition}

The above results on the measurement outcomes refer to the case where the data has been accumulated over  $M\gg 1$ Ramsey measurements. The duration of the data acquisition $M\ts$ was assumed to be sufficiently long, so that noise correlations decay, in which case the measurements are ``ergodic'': time-average coincides with the ensemble average. However, if noise has a slowly varying component, of interest is a distribution of the measurement outcomes for shorter times. It is obtained by $M\gg 1$ measurements for $M\ts$ smaller than the decay time of the noise correlations.

There is a similarity between the qubit measurements where noise remains constant in time and  transport measurements in condensed-matter systems in the presence of static disorder where electrons are elastically scattered by the disorder, but on the time scale of the measurement their energy is not changed. In the qubit case, the outcome of $M$ measurements is a snapshot of the static qubit frequency, which remains constant but varies from one series of $M$ measurements to another.

\subsection{Slow two-level systems}
\label{subsec:slow_TLSs_1}

In the case of noise from TLSs, the ergodic limit corresponds to such acquisition duration that all TLSs  have a chance to switch multiple times, 
$M\ts\gg 1/W\sn$  for all $n$.  However, in the presence of very slow TLSs the inequality $M\ts\gg 1/W\sn$ does not necessarily hold even for large, but not too large $M$.  For slowly switching TLSs the data will present a snapshot of their initial distribution. 

We will assume that there is a set $\{N_\mathrm{sl}\}$ of slowly switching TLSs, i.e., $M W\sn\ts\ll 1$ for $n\in \{N_\mathrm{sl}\}$. During  data accumulation they remain in the initially occupied states $\ket{0}\sn$ or $\ket{1}\sn$. Since the shift of the qubit frequency is determined by the operator   $\sum_n V\sn{} \tT_z\sn$ [cf. Eq.~(\ref{eq:q_TLS_dispersive})], the qubit will accumulate the same phase $\delta\omega_\mathrm{q} t_R$ during each of the $M$ Ramsey measurements (in the absence of other sources of noise). This phase  is 
\begin{align}
\label{eq:phase_initial}
\theta(\{j_n\}) =  \sum_{n\in\{N_\mathrm{sl}\}}V\sn{} (-1)^{j_n} t_R, 
\end{align}
where the parameters $j_n$ take on values 0 or 1; we choose $j_n=0$ if the occupied state is $\ket{0}\sn$ and $j_n=1$ if the occupied state is $\ket{1}\sn$. The probability to have a given $\theta(\{j_n\})$ is determined by the probabilities $w_{j_n}\sn$ of the occupation of the corresponding initial states of the TLSs, which are given in Eq.~(\ref{eq:stationary_populations}).

The value of $\theta \equiv \theta(\{j_n\})$ determines the probability $r_1$ of obtaining ``1'' in a Ramsey measurement. If there are no fluctuations of the qubit frequency $r_1= p(\theta)$, where $p(\theta)$ is given by Eq.~(\ref{eq:standard_probability}). Because $\theta\equiv \theta(\{j_n\})$ is random, $p(\theta)$ also becomes random. This means that the outcomes of $M\gg 1$ measurements will have a distribution, which is determined by the distribution of the values of $\theta(\{j_n\})$. In the simplest case where the slow TLSs are symmetric, $w_0\sn = w_1\sn=1/2$, and all $V\sn{} =V$ are the same [cf.~\cite{Ithier2005,Yoshihara2014}], $\theta(\{j_n\})$ takes on values $\theta_\mathrm{sl}(n)=Vt_R(2n-N_\mathrm{sl})$ with probabilities 
\[ P[\theta_\mathrm{sl}(n)] = 
2^{-N_\mathrm{sl}}  \binom{N_\mathrm{sl}}{n}
\]
($n$ is the number of the TLSs in the state $\ket{0}$.
We note a ``cumulative'' effect of the slow TLSs. The values of $\theta_\mathrm{sl}(n)$ are determined by the coupling constant $V$ multiplied by the difference of the number of TLSs in the states $\ket{0}$ and $\ket{1}$. Therefore they may be significantly larger than for a single TLS.

The resulting  probability distribution $\rho(m|M)$ to have ``1'' $m$ times in $M$ measurements is a supersposition of binomial distributions for different $\theta_\mathrm{sl}(n)$ weighted with the probability of having a given $\theta_\mathrm{sl}(n)$,
\begin{align}
\label{eq:m_out_of_M_TLS}
&\rho(m|M)=\binom{M}{m}\sum_{n=0}^{N_\mathrm{sl}} P[\theta_\mathrm{sl}(n)]p^m[\theta_\mathrm{sl}(n)]\nonumber\\
&\times \{1-p[\theta_\mathrm{sl}(n)]\}^{M-m}
\end{align}
For brevity,  we call ``static'' the limit in  which switching between the TLSs' states is disregarded and  Eq.~(\ref{eq:m_out_of_M_TLS}) applies.
\begin{figure*}
    \centering
    \includegraphics[width=0.47\textwidth]{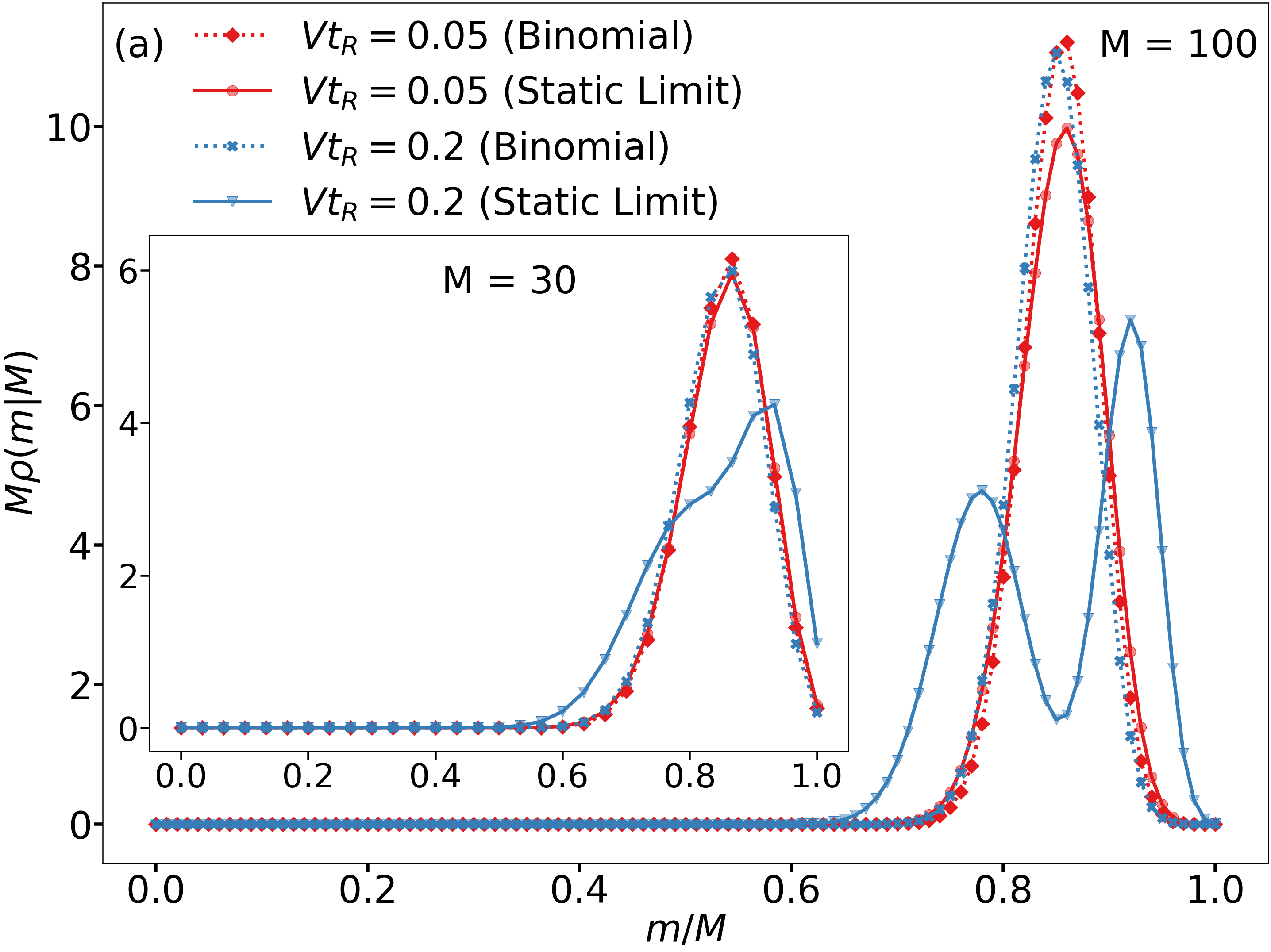}
    \includegraphics[width=0.47\textwidth]{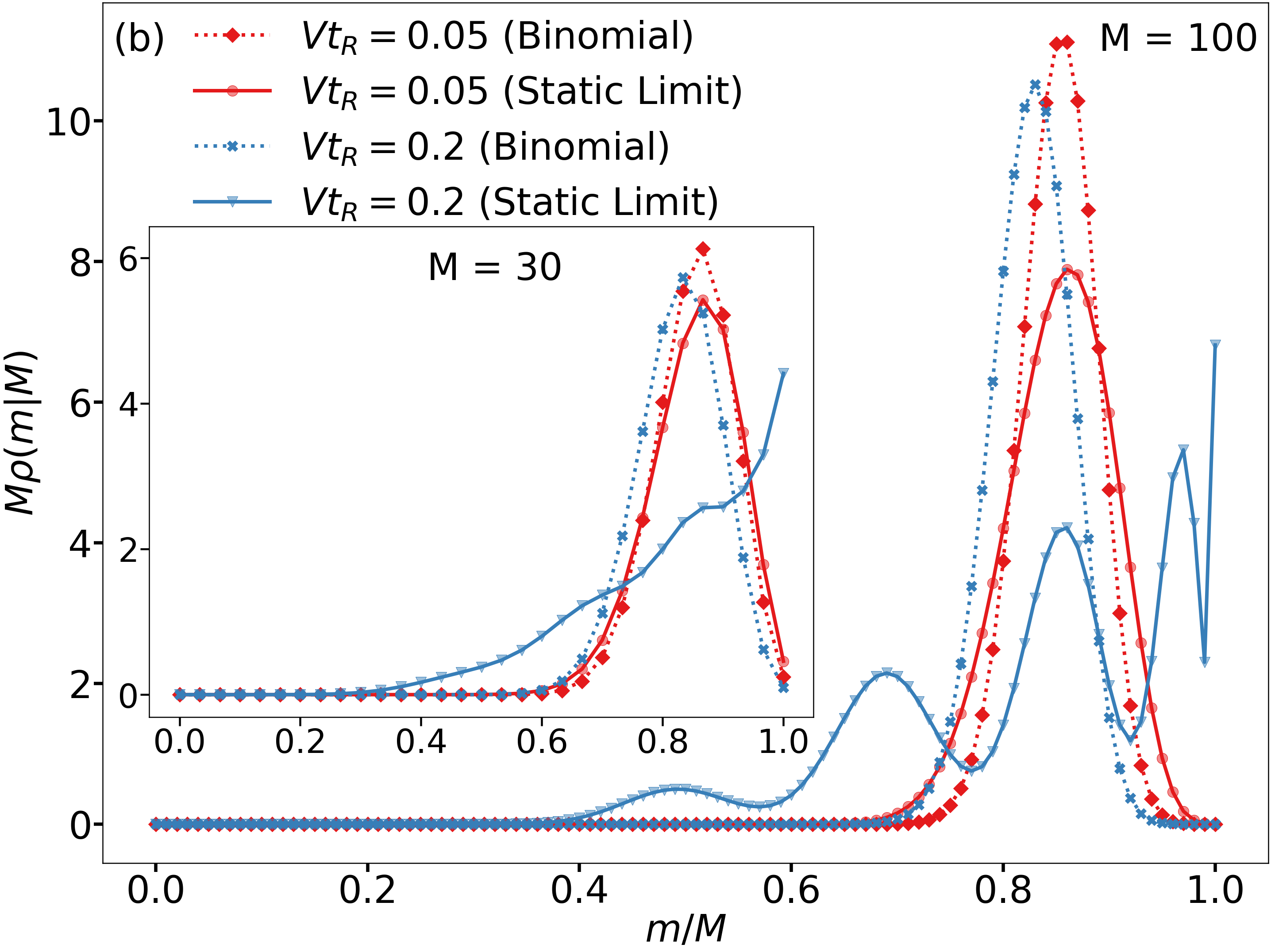}
    \includegraphics[width=0.47\textwidth]{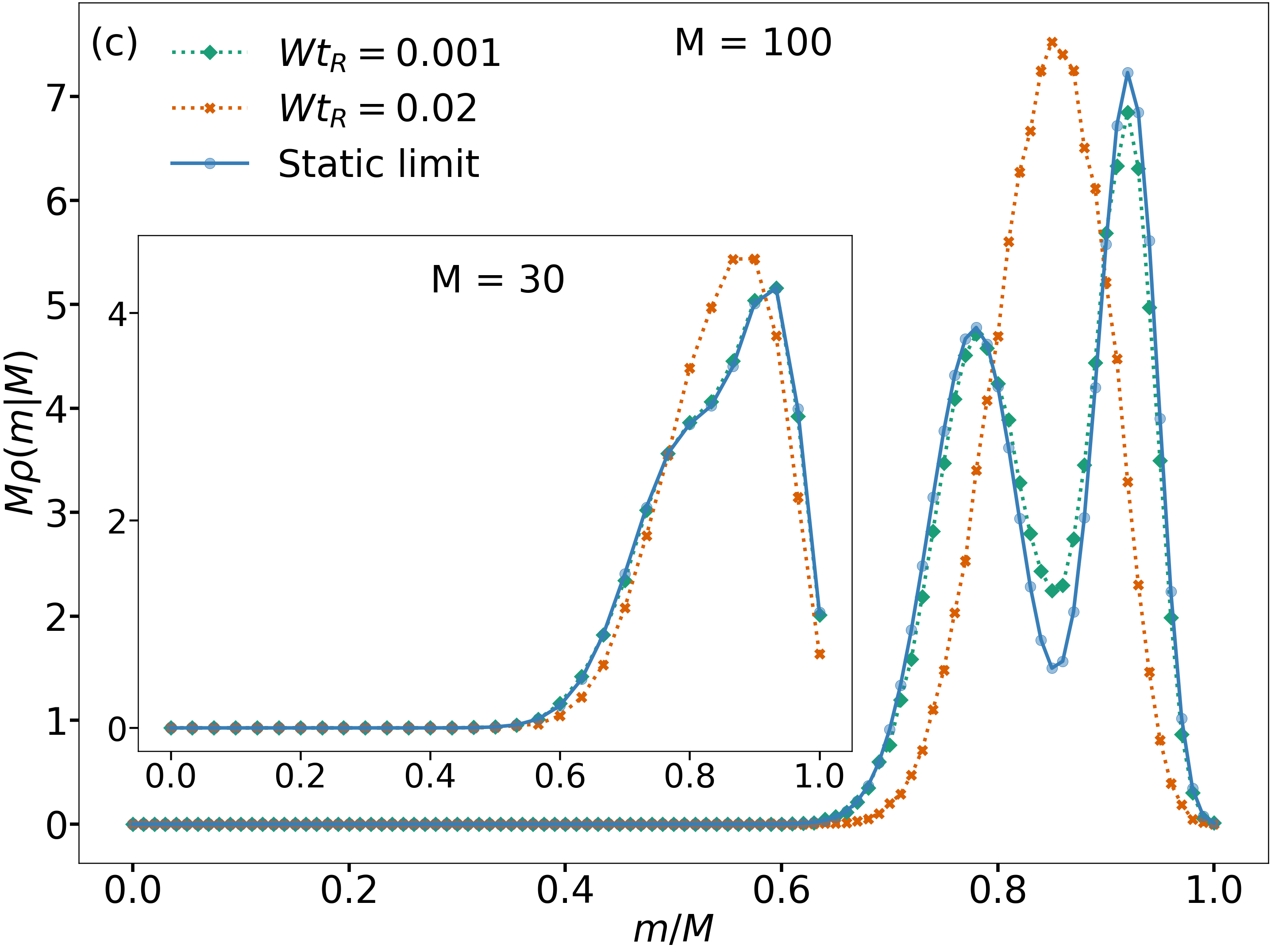}
    \includegraphics[width=0.47\textwidth]{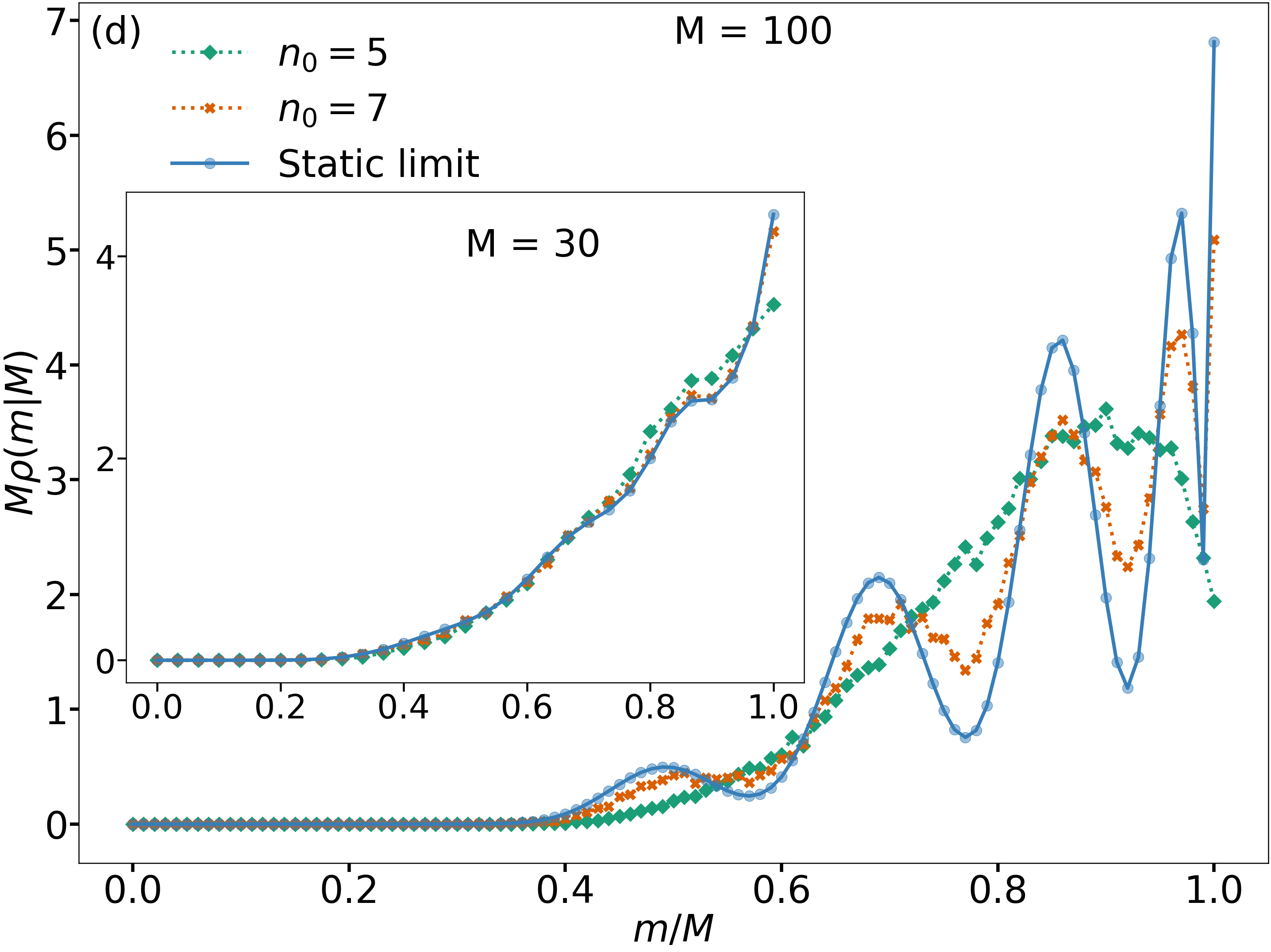}
      \caption{Probability distributions of the number of occurrences of ``1'' in $M$ Ramsey measurement for $t_R\ll T_2$. The main plots and the insets refer to $M=100$ and $M=30$, respectively. Panels (a) and (c) refer to the coupling to a single slowly switching TLS, whereas panels (b) and (d) refer to the coupling to four TLSs.
(a) Comparison of the static limit where the TLS switching over time $M\ts$ is disregarded, see Eq.~\eqref{eq:m_out_of_M_TLS}, and the binomial distribution \eqref{eq:binomial_trivial} for the probability $r_1$ calculated for the same TLSs and given by Eq.~\eqref{eq:TLS_r1_trivial}. 
(b) Same as in (a) for the coupling to four TLSs, $V\sn = V$.
(c) Comparison of the static limit, Eq.~\eqref{eq:m_out_of_M_TLS}, with the simulations for the scaled coupling $Vt_R=0.2$ and for the scaled switching rate $Wt_R=0.02$ (orange) and $Wt_R=0.001$ (light blue).  
(d) Comparison of the static limit, Eq.~\eqref{eq:m_out_of_M_TLS}, with the simulations for the scaled coupling $Vt_R=0.2$ and for the scaled switching rates $W\sn t_R = \exp[-3(n+n_0)/4]$ with $n=1,...,4$ and $n_0=5$ (light blue) and $n_0=7$ (orange). The simulation data for $n_0=7$ in the inset are indistinguishable from the  static limit result.
}
 \label{fig:acquisition_TLS}
\end{figure*}

In Fig.~\ref{fig:acquisition_TLS}~(a) and (b) we compare the distribution $\rho(m|M)$ in the static limit (\ref{eq:m_out_of_M_TLS})  with the binomial distribution $\rho_\mathrm{binom}(m|M)$ for the same $M$  calculated for the mean probability value $r_1$ of the measurement outcome. This value is obtained by averaging the measurement outcomes over a long time that largely exceeds all reciprocal switching rates of the TLSs, so that the TLSs have a chance to switch between their states multiple times.  The binomial distribution is given by the standard expression
\begin{align}
\label{eq:binomial_trivial}
 \rho_\mathrm{binom}(m|M)= \binom{M}{m}r_1^m (1-r_1)^{M-m}.
\end{align}
For $W\sn t_R\ll 1$ and $V\sn t_R \ll 1$ the value of $r_1$ in this expression  is given by Eq.~(\ref{eq:Pi_1_multi_TLS_explicit}) with $\Xi\sn(t_R) \approx \exp[-(V\sn t_R)^2/2]$, so that for $V\sn = V$
\begin{align}
\label{eq:TLS_r1_trivial}
r_1 \approx \frac{1}{2} + \frac{1}{2}\exp[-N_\mathrm{sl} (Vt_R)^2/2]\cos \phi_R.
\end{align}
We note that the binomial distribution disregards the effect of noise correlations. This effect leads to a broadening of the distribution, as discussed in Appendix~\ref{sec:distribution_broadening}. However, this broadening is much smaller than the effects discussed in this section.

As seen from Fig.~\ref{fig:acquisition_TLS} (a) and (b), for a very weak coupling, $Vt_R=0.05$, the distribution $\rho(m|M)$ is close to the binomial distribution $\rho_\mathrm{binom}(m|M)$ in the case of a single TLS even for $M$ as large as $M=100$. However, for four slowly switching TLSs, for such coupling the distributions are already somewhat different.  This is a consequence of the cumulative effect of the addition of the frequency shifts induced by different TLSs.

The difference of the distributions is much more pronounced for a stronger, but still weak, coupling. This is seen from the plots for $V\sn t_R=0.2$. Even for a single TLS and a moderately large $M$, where $M=30$, the maximum of $\rho(m|M)$ in the absence of switching is shifted from the maximum of the binomial distribution, which is located at  $m/M=r_1$. The distribution $\rho(m|M)$ is profoundly asymmetric and is much broader than $\rho_\mathrm{binom}(m|M)$. The difference with the binomial distribution is much stronger for four TLSs, where the shape of the distribution is qualitatively different from the shape of the binomial distribution.

For larger $M$ the distribution (\ref{eq:m_out_of_M_TLS}) shows a fine structure. The peaks of $\rho(m|M)$ correspond to the maxima of the binomial distributions calculated for different numbers $n$ of the TLSs in the same state. Respectively, they correspond to the different values of the accumulated qubit  phase $\theta_\mathrm{sl}(n) =Vt_R(2n-N_\mathrm{sl})$. The fine structure becomes more pronounced with the increasing $M$, because the variances of the ``partial'' binomial distributions $p[\theta_\mathrm{sl}(n)]\{1-p[\theta_\mathrm{sl}(n)]\}/M$  decrease with the increasing $M$.  

In Fig.~\ref{fig:acquisition_TLS} (c) and (d) we compare the static-limit result, Eq.~(\ref{eq:m_out_of_M_TLS}), with the simulation data. The data refer to  two sets of the switching rates. For all the data $\ts/t_R=3$. Therefore the actual parameter of the applicability of the static limit (\ref{eq:m_out_of_M_TLS}), , i.e., of the assumption that the TLSs do not switch over the data acquisition time, is the condition $3MWt_R\ll 1$.  For the single TLS, Fig.~\ref{fig:acquisition_TLS}~(c), $3MWt_R$ is small for $Wt_R=0.001$. The results of the simulations are then very close to Eq.~(\ref{eq:m_out_of_M_TLS}). However, for $Wt_R=0.02$ we have $3MWt_R=0.9$ for $M=30$. In this case the simulation data differ from the static limit, yet they are still different from the binomial distribution. For $M=100$ we have $3MWt_R=6$ and the result of the simulations is very close to the binomial distribution.

In the case of four TLSs, for $n_0=5$ and $M=30$ we have  $3MW\sn t_R\approx 1$ for the fastest-switching TLS, $n=1$. The results of the simulations are comparatively close to the static limit for such $M$. However, they become significantly different for $M=100$, even though $\rho(m|M)$ is still qualitatively different from the binomial distribution. For smaller switching rates, $n_0=7$, the simulation data are close to the static limit. Importantly, they show a fine structure, which is a signature of the  TLS-induced noise.


\subsection{Slow Gaussian fluctuations}
\label{subsec:slow_Gaussian}

The probability distribution $\rho(m|M)$ to have ``1'' $m$ times in $M$ measurements have a particular form also in the case of Gaussian frequency fluctuations if the fluctuations are slow, so that $\omq(t)$ does not change over $M$ Ramsey measurements.  It is seen from Eqs.~(\ref{eq:phase_k_classical}) and (\ref{eq:phase_correlation}) that the distribution of the qubit phase $\theta_\mathrm{sl}$ accumulated over a single Ramsey measurement  in this case is
\[ P_\mathrm{Gauss}(\theta_\mathrm{sl}) = (2\pi f_0)^{-1/2}\exp(- \theta_\mathrm{sl}^2/2 f_0).
\]
The value of $\theta_\mathrm{sl}$, even though it is random, remains the same for all $M$ measurements. Then the distribution $\rho(m|M)$ has the form of the binomial distribution integrated over $\theta_\mathrm{sl}$ with  weight $ P_\mathrm{Gauss}(\theta_\mathrm{sl})$, 
\begin{align}
\label{eq:m_out_of_M_Gauss}
&\rho(m|M)=\binom{M}{m}\int_{-\infty}^\infty d\theta_\mathrm{sl} P_\mathrm{Gauss}(\theta_\mathrm{sl})p^m(\theta_\mathrm{sl})\nonumber\\
&\times [1-p(\theta_\mathrm{sl})]^{M-m}\ .
\end{align}
As in the case of slowly switching TLSs, we refer to the limit where this expression applies as the static limit.

\begin{figure}
    \centering
    \includegraphics[width=0.45\textwidth]{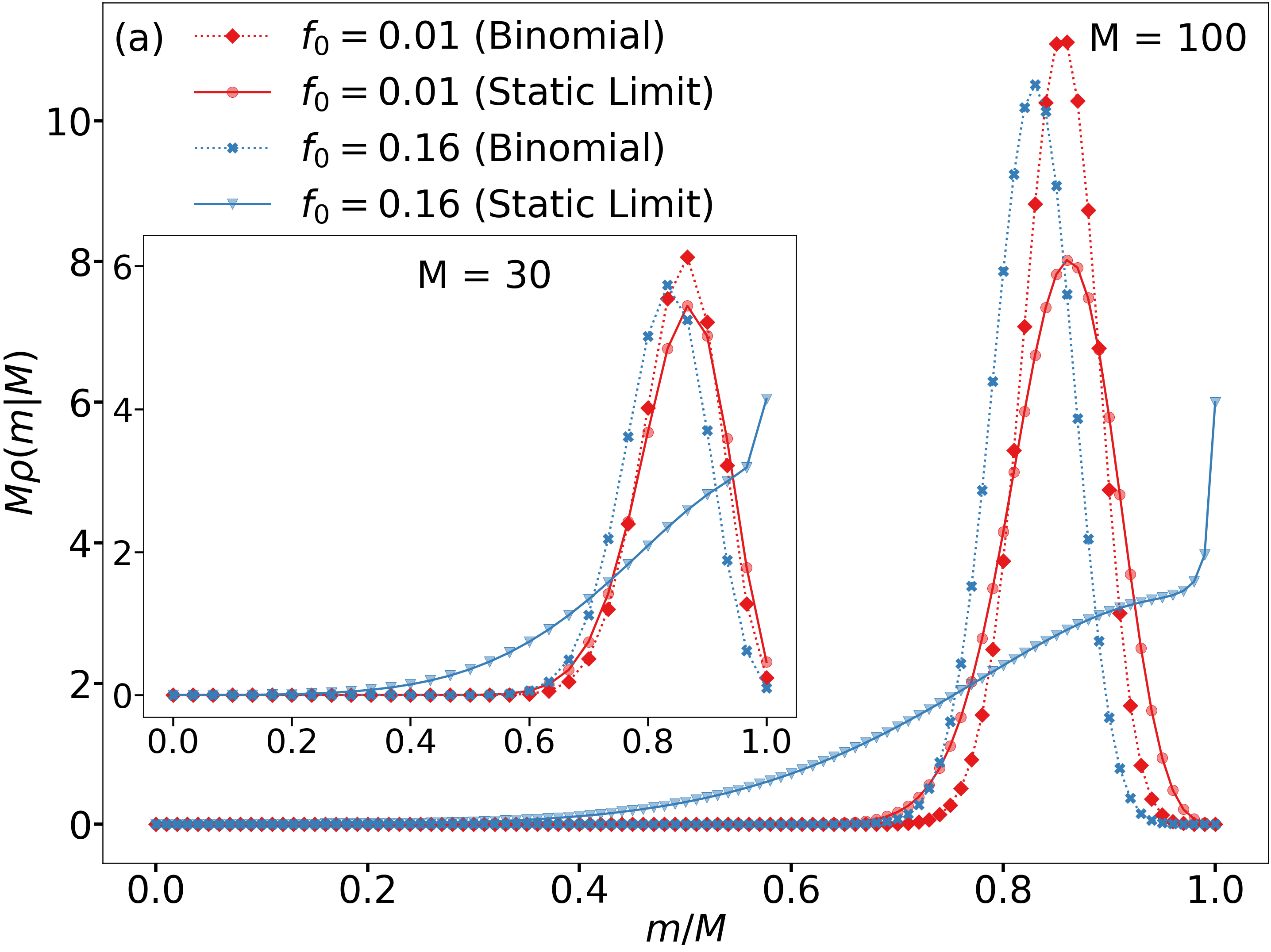}
    \includegraphics[width=0.45\textwidth]{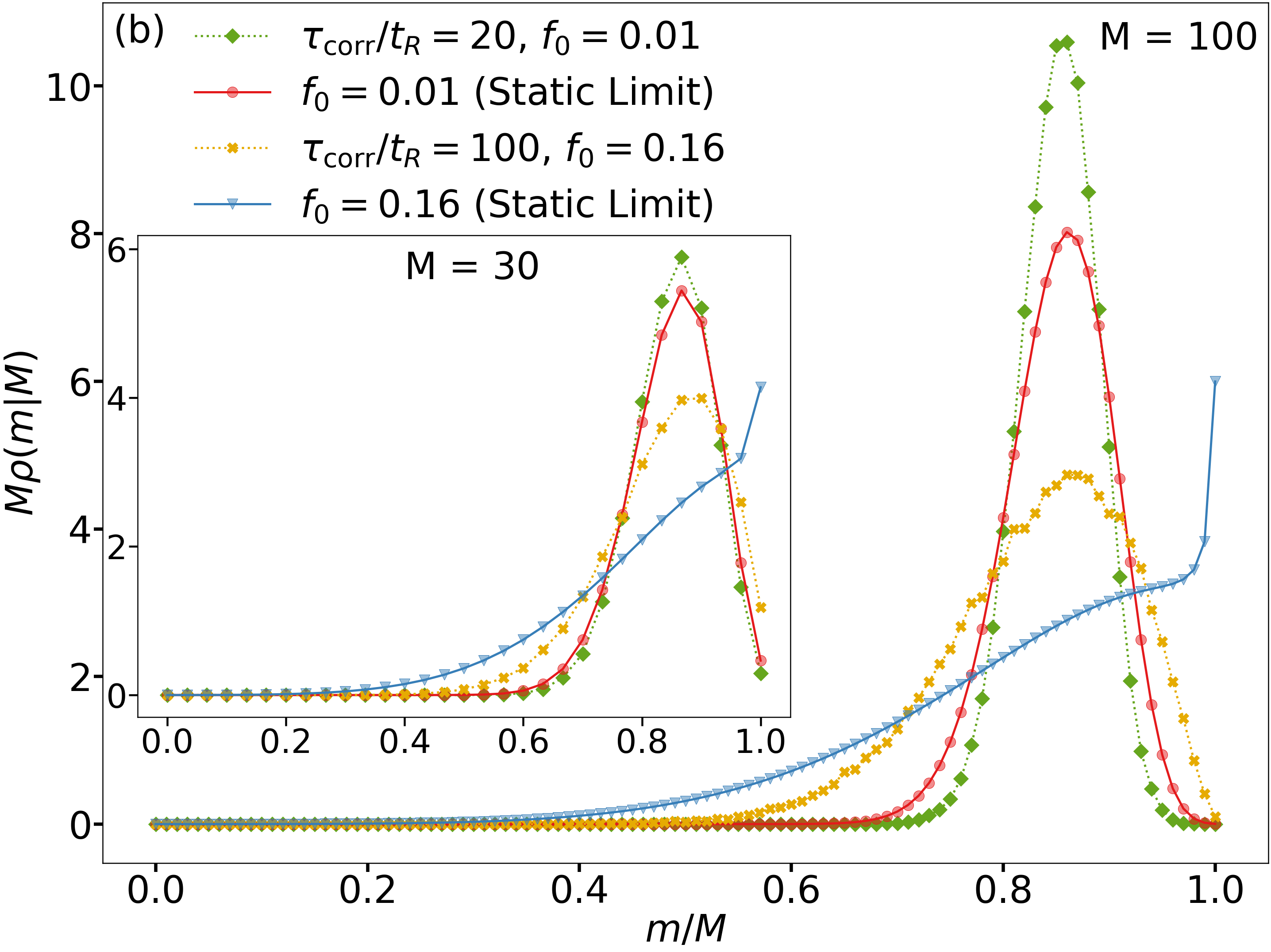}
    \includegraphics[width=0.45\textwidth]{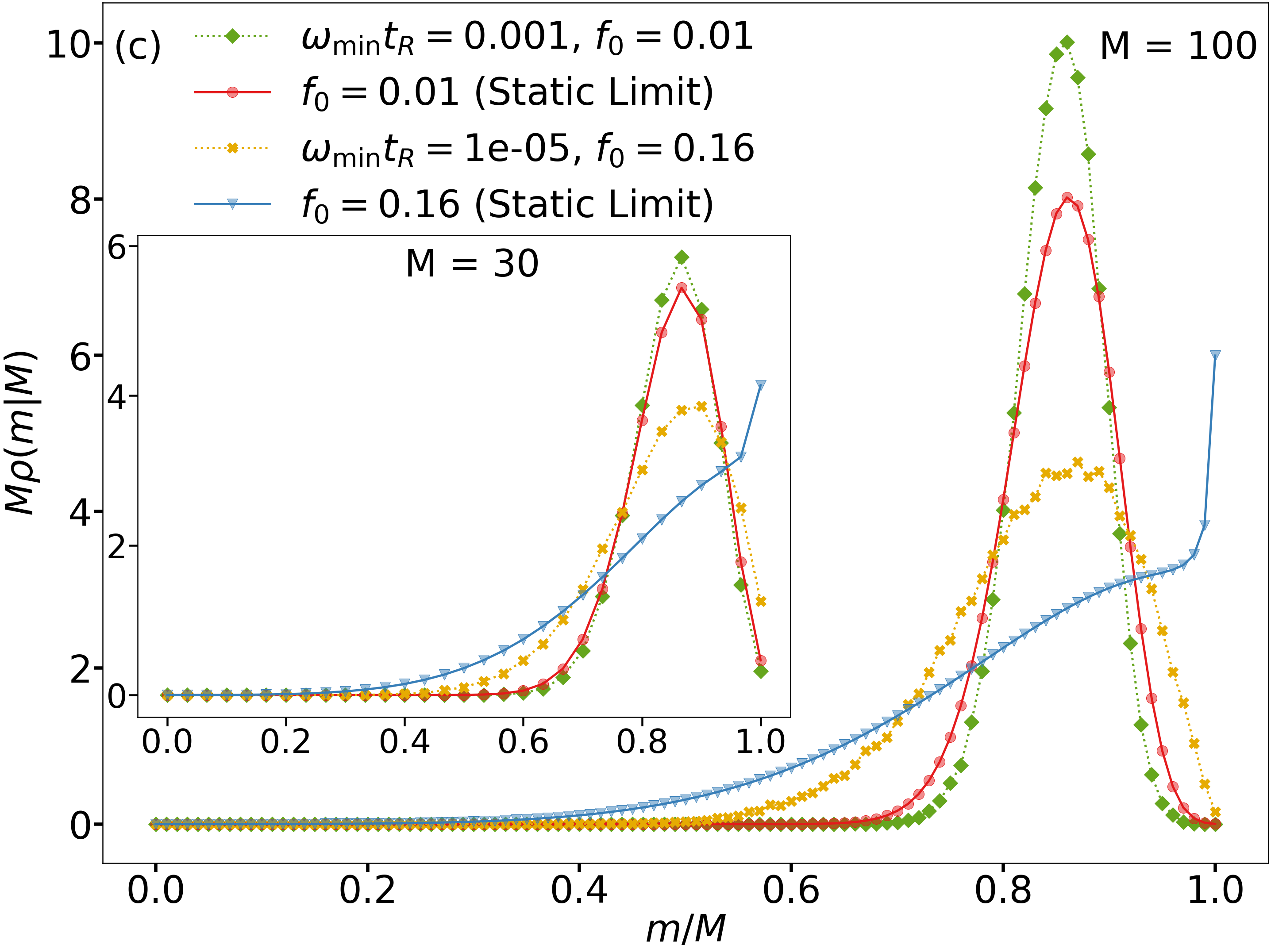}
    \caption{Probability distributions of the number $m$ of occurrences of ``1'' in $M$ Ramsey measurements for $t_R\ll T_2$, $\ts/t_R=3$. The main plots and the insets refer to $M=100$ and $M=30$, respectively.
(a) Comparison of the distribution $\rho(m|M)$ in the static limit \eqref{eq:m_out_of_M_Gauss} and the binomial distribution \eqref{eq:binomial_trivial} with the probability $r_1$ given by Eq.~\eqref{eq:class_noise_results_1}.
(b) Comparison of the static limit \eqref{eq:m_out_of_M_Gauss} with simulations of the exponentially correlated noise discussed in Sec.~\ref{subsec:exponentially_correlated_theory}. The green diamonds refer to $\tau_{\corr}/t_R = 20$ and $D_\corr=0.41$, whereas yellow crosses refer to $\tau_\corr/t_R = 100$ and $D_\corr=32.11$.
(c) Comparison of the static limit \eqref{eq:m_out_of_M_Gauss} with simulation of the $1/f$-type noise discussed in Sec.~\ref{subsec:1/f_theory} with $\omegaMin t_R=0.001$ and $D_{\mathrm{fl}}=0.002575$, so that $f_0=0.01$ (green diamonds), and $\omegaMin t_R=0.00001$ and $D_{\mathrm{fl}}=0.02574$, so that $f_0=0.16$ (yellow crosses). 
    }
      \label{fig:acquisition_gaussian}
\end{figure}

The characteristic shape of $\rho(m|M)$ for two types of slow Gaussian noises is illustrated in Fig.~\ref{fig:acquisition_gaussian}. We choose the noise intensities that give the same  $f_0= \langle\theta_\mathrm{sl}^2\rangle$ as the values of $f_0 = N_\mathrm{sl}V^2 t_R^2$ that characterize the noise from four slowly switching TLSs  in Fig.~\ref{fig:acquisition_TLS},  i.e.,  $f_0=0.01$ and $f_0=0.16$.  To further facilitate a comparison of the effects of a Gaussian noise and the noise from the TLSs, we present the results for $M=30$ and $M=100$, the same values of $M$ as in Fig.~\ref{fig:acquisition_TLS}. In (b) and (c) the data are collected from 100,000 repetitions of 100 simulated Ramsey measurements.

As it was done in Fig.~\ref{fig:acquisition_TLS}~(a) and (b) for the TLSs, in Fig.~\ref{fig:acquisition_gaussian} (a) we compare $\rho(m|M)$ for Gaussian fluctuations with the binomial distribution $\rho_\mathrm{binom}(m|M)$ (\ref{eq:binomial_trivial}) for the same $M$ calculated for the ``ergodic'' value of the probability $r_1$ of the measurement outcome. The binomial distribution is symmetric for $M\gg 1$, has a typical width $[r_1(1-r_1)/M]^{1/2}$, and a maximum at $m/M=r_1$. 

For very weak fluctuation intensity, where the phase variance $f_0=0.01$, the distributions $\rho(m|M)$ and  $\rho_\mathrm{binom}(m|M)$ are close to each other. However, for a  larger, even though still small variance, $f_0=0.16$, the shapes of the distributions become very different and the maxima are located at different $m/M$. In a qualitative distinction from the case of slow TLSs, the distribution $\rho(m|M)$ for a quasistatic Gaussian noise does not have a fine structure.

In Fig.~\ref{fig:acquisition_gaussian}~(b) and (c) we compare the results of the static limit of the Gaussian noise with the simulated  distributions for exponentially correlated and $1/f$-type noises. In the simulations the noise parameters were selected so that they give the same values of the phase variance $f_0=0.01$ and $f_0=0.16$ as in the static limit. However, the simulated noise, along with $f_0$, is characterized by the correlators $f_k=\langle\theta_0\theta_k\rangle$ with $k\geq 1$. Only a ``portion'' of the power  spectrum of the fluctuations of  $\omq(t)$ meet the condition $M\omega\ts \ll 1$. Therefore there is a significant deviation of the simulated distributions $\rho(m|M)$ from the static limit for $f_0=0.16$.  Somewhat surprisingly, the distributions for the exponentially correlated and $1/f$-type noises look similar.  Still they re somewhat different from each other. Also, they are strongly asymmetric and differ very significantly from the binomial distribution in  Fig.~\ref{fig:acquisition_gaussian}~(a). They also differ significantly from the distribution for the TLS-induced noise, they do not have a fine structure.

For a very weak noise, $f_0=0.01$, the simulated distribution, the distribution in the static limit, and the binomial distribution become similar. Their shape is determined primarily by the uncertainty of the quantum measurements, which, for $\phi_R=\pi/4$ gives $p(\theta) \approx (1+1/\sqrt{2})/2\approx 0.85$.

 
 \section{Periodic modulation of the qubit frequency}
 \label{sec:periodic}

A potentially important cause of the time dependence of the qubit frequency is a low-frequency periodic signal. It can come, for example, from an AC power supply or from other low-frequency sources. For a sinusoidal modulation at frequency $\omega_p$, so that $\omq(t) = a_p\cos(\omega_pt+\phi_p)$, the phase accumulation during the $k$th Ramsey measurement is
\begin{align}
\label{eq:frequency_sinusoidal}
&\theta_k= \int_{k\ts}^{k\ts +t_R}\omq(t) dt=A_p\cos(k \omega_p\ts +\tilde \phi_p)\ ,
\nonumber\\
&A_p= \frac{2a_p}{\omega_p}\sin\frac{\omega_p t_R}{2},\quad \tilde\phi_p=\phi_p+\frac{\omega_pt_R}{2}\ ,
%
%
\end{align}
where $a_p$ and $\phi_p$ are the amplitude and phase of the qubit frequency modulation.

A periodic frequency modulation can be revealed by studying the power spectrum of the measurement outcomes $x_n$, with  $x_n$  taking the values $0$ or $1$.  The spectrum obtained in a series of $N$ measurements is related to the discrete Fourier transform $X(m)$ of $x_n$,
\begin{align}
\label{eq:fast_Fourier_general}
&X(m) = \sum_{n=0}^{N-1}x_n\exp(2\pi i mn/N)\ . 
\end{align}
To obtain the power spectrum, one has to calculate $|X(m)|^2$ and average the result over a repeated series of measurements. The outcome sensitively depends on  how the averaging is done. If the measurements are synchronized with the modulation $\omq(t)$, i.e., all of them refer to the same phase $\phi_p$, the result depends on this phase. However, generally the modulation is unknown. In fact, the measurements are done to reveal its presence. Then $|X(m)|^2$ is calculated each time for a different phase and there occurs averaging over $\phi_p$. It is this case that we consider, and the phase averaging is implied when we use the notation $\mathbb{E}[\cdot]$. The power spectrum $R(m)$ is then defined as 
\begin{align}
\label{eq:qubit_spectrum_periodic_general}
R(m) = &\mathbb{E}[|X(m)|^2] = \sum_{n_1,n_2=0}^{N-1}e^{2\pi im(n_1 - n_2)/N}\nonumber\\
&\times\langle p(\theta_{n_1})
p(\theta_{n_2})\rangle_\phi\ ,
\end{align}
where $p(\theta_n)$ is the probability for $x_n$ to be equal to 1, cf. Eq.~(\ref{eq:standard_probability}), whereas $\langle\cdot\rangle_\phi$ means averaging over $\phi_p$. The averaging may also include averaging over noise in $\theta$; in this section we disregard this noise, for simplicity.

Since $p(\theta$) contains $\cos\theta$, the product $p(\theta_{n_1}) p(\theta_{n_2})$ in Eq.~(\ref{eq:qubit_spectrum_periodic_general})  contains terms $\propto \exp[\pm i k(n_1 -n_2)\omega_p\ts]$ with amplitudes $\propto J_k^2(A_p)$, where $J_k$ is the Bessel function. They lead to peaks in $R(m)$ at the values of $m$ that correspond to multiples of $N\omega_p\ts/2\pi$. 

We will consider weak and slow frequency modulation, which is of utmost interest for the experiment,
\[\omega_p t_R\ll 1, \qquad A_p\approx a_pt_R\ll 1.\]
In this case, to the leading order in $A_p$,  we find that $R(m)$ has two sharp mirror-symmetric peaks at $m>0$, one near $N\omega_p\ts/2\pi$ and the other near $N- (N\omega_p\ts/2\pi)$. For $|2\pi m/N)-\omega_p\ts|\ll 1 $, 
\begin{align}
\label{eq:spectral_peaks}
&R(m) \approx \frac{1}{8}A_p^2\sin^2 \phi_R e^{-2t_R/T_2} 
\nonumber\\
&\times\frac{1-\cos(N\omega_p\ts)}{[(2\pi m/N)-\omega_p\ts]^2}.
\end{align}
The height of the peak is $\propto N^2$ for a generically noninteger $N\omega_p\ts/2\pi$. It is well beyond the noise floor. Observing the peak should allow identifying the presence of a slow periodic modulation of the qubit frequency. 

The spectral peak (\ref{eq:spectral_peaks}) can be broadened and its height can be reduced by fluctuations of the modulation frequency $\omega_p$ or by fluctuations of the time intervals between the measurements $\ts$. For an illustration, we consider the simplest model where $\ts$ fluctuates from cycle to cycle with variance $\langle (\ts - \langle \ts\rangle) ^2\rangle = \sigma_\mathrm{cyc}^2$ and where these fluctuations are Gaussian and there is no correlation between different cycles. A straightforward calculation shows that, in this case, Eq.~(\ref{eq:spectral_peaks}) for the shape of the spectral peak near $m \approx  N\omega_p\ts/2\pi$ is modified to 
\begin{widetext}
\begin{align}
\label{eq:spectral_peaks_decay}
&R(m) \approx \frac{1}{8}A_p^2\sin^2 \phi_R e^{-2t_R/T_2}(\Delta_m^2 + \Delta_\mathrm{cyc}^2)^{-2} 
\left\{
(\Delta_m^2 - \Delta_\mathrm{cyc}^2)\left[1-e^{-N\Delta_\mathrm{cyc}}\cos(N\omega_p\ts)\right]
\right.
\nonumber\\
&\left.
 -2\Delta_m\Delta_\mathrm{sec}e^{-N\Delta_\mathrm{cyc}}\sin(N\omega_p\ts)
+N \Delta_\mathrm{cyc}(\Delta_m^2 + \Delta_\mathrm{cyc}^2)
\right\}
\end{align}
\end{widetext}
where
\[\Delta_m = (2\pi m/N)-\omega_p\ts, \quad \Delta_\mathrm{cyc} = \omega_p^2\sigma_\mathrm{cyc}^2/2
\]
Even for a small variance of $\ts$, where $\Delta_\mathrm{cyc}\ll 1$, the spectrum (\ref{eq:spectral_peaks_decay}) can be dramatically different from the spectrum (\ref{eq:spectral_peaks}). Not only is the peak broadened but, for $N\Delta_\mathrm{cyc}\gg 1$, the maximum of the  peak scales with $N$ as $N$ rather than as $N^2$ . This suggests a sensitive way of revealing a weak aperiodicity of the pulse sequence and thus characterizing the relevant gate operations.


\section{The master equation}
\label{sec:master_equation}

We now proceed with the derivation of the results for the correlators of the periodically repeated Ramsey measurements. This and the next two  sections can be read prior to Secs.~\ref{sec:summary_TLS} and \ref{sec:class_noise_summary} where there are presented the results of the theory described below.

\subsection{The Hamiltonian}

\label{subsec:Hamiltonian}

We consider a qubit, which is coupled to TLSs, that has a fluctuating frequency, and is subjected to the control pulses in Fig.~\ref{fig:pulse_sequence}. Its Hamiltonian is  $H=H_\mathrm{q}+H_\mathrm{q-TLS}$. Here $H_\mathrm{q-TLS}$ is given by Eq.~(\ref{eq:q_TLS_dispersive}) and describes the dispersive coupling to the TLSs,  whereas $H_\mathrm{q}$  has the form
\begin{align}
\label{eq:Hamiltonian}
&H_\mathrm{q} = 
 -\frac{1}{2}\omq(t) \sigma_z\nonumber\\
&+ \frac{\pi}{4}\sigma_y \sum_k[\delta(t-k\ts) + \delta(t-k\ts-t_R)].
\end{align}
The term $\propto \sigma_y$ in Eq.~(\ref{eq:Hamiltonian}) describes the periodically repeated pairs of Ramsey pulses of rotation about the $y$-axis, whereas $\omq(t)$ describes the qubit frequency fluctuations  due to external classical noise. Here we somewhat conditionally distinguish such classical noise from noise stemming from the coupling to TLSs; see however Appendix~\ref{sec:Yaxing_method} where the effect of the TLSs is described by modeling $\omq(t)$ by a  telegraph noise

It is convenient to assume that $\langle \omq(t)\rangle =0$. If $\langle \omq(t)\rangle$ were nonzero, it could be interpreted as the detuning of the mean qubit frequency from the frequency $\omega_\mathrm{ref}$ of the reference signal used in the Ramsey measurements.  Such detuning  leads to a phase accumulation $\phi_R$ during a measurement given by Eq.~(\ref{eq:phi_R_as_detuning}). It can be controlled in the experiment and can be implemented by incorporating into $H_\mathrm{q}$ the term
\begin{align}
\label{eq:H_R}
H_R = -\frac{1}{2}\phi_R\sigma_z \sum_k \delta(t-k\ts-t_R^-).
\end{align}
The Hamiltonian $H_R$ describes rotations of the qubit around  the $z$-axis prior to the second Ramsey pulse within a cycle, i.e., the pulse applied  at time $k\ts + t_R$, see Eq.~(\ref{eq:Hamiltonian}) and also Fig.~\ref{fig:pulse_sequence}. It mimics the effect of the detuning of the qubit frequency $\omega_\mathrm{q}$ from the reference frequency $\omega_\mathrm{ref}$ in the Ramsey measurement, with $\phi_R= (\omega_\mathrm{q} -\omega_\mathrm{ref})t_R$, see Eq.~(\ref{eq:phi_R_as_detuning}).

We note that, if $\langle \omq(t)\rangle \neq 0$, the qubit frequency measured in the experiment incorporates $\langle\omq\rangle$. It is incremented by $\langle\omq\rangle$ compared to the value in the absence of the coupling to the noise source, in particular, in the absence of the coupling to the TLSs. In a way, the nonzero $\langle\omq\rangle$ is an artifact of the model. To relate the Hamiltonian $H_R$ to the frequency detuning in the Ramsey measurement, one has to subtract the increment $\langle\omq\rangle$ from the experimental value of $\omega_\mathrm{q}$ when calculating $\phi_R$. This means that  the phase $\phi_R$ in Eq.~(\ref{eq:H_R}) has to be replaced by $\tilde\phi_R = \phi_R - \langle \omq(t)\rangle t_R$. 

We consider a classical noise where $\langle \omq(t)\rangle =0$. However, the above argument applies also to the noise from the coupling to the TLS, in which case the qubit frequency shift is $\sum_n V\sn \langle \tT_z\sn \rangle$, cf. Eq.~(\ref{eq:phi_R_TLS}). On the practical side, one can consider the qubit-TLS coupling of the form
\begin{align}
    \label{eq:q_TLS_coupling_prime}
    H'_\mathrm{q-TLS} = -\frac{1}{2}\sigma_z \sum_nV\sn\left(\tT_z\sn - \langle \tT_z\sn\rangle\right),
\end{align}
which does not lead to a renormalization of the mean qubit frequency. In this sense it is more relevant from the viewpoint of the experiment. Then in Eq.~(\ref{eq:H_R}) one should use $\phi_R$ rather than $\tilde\phi_R$. In particular, in the experiment one should use $\phi_R$. The result of the calculation will not change.

To analyze the dephasing due to the dispersive coupling to the TLSs, Eq.~(\ref{eq:q_TLS_dispersive}), we write  the states of an $n$th TLS as 
\[\ket{0}\sn\equiv \left(\begin{array}{c}
1\\
0
\end{array}\right)\sn, \qquad \ket{1}\sn\equiv\left(\begin{array}{c}
0\\
1
\end{array}\right)\sn,\] 
and we use the Pauli operators $\tT_\varkappa\sn$ to describe the TLS dynamics. Here
$\varkappa =0, x,y,z$, with $\tT_0\sn\equiv\hI_\tau\sn$ being a unit operator. 


\subsection{Dynamics during the Ramsey measurement}
\label{subsec:QKE}

We first consider the qubit dynamics during a Ramsey measurement, i.e., in the time interval $n\ts < t < n\ts+t_R$, cf. Fig.~\ref{fig:pulse_sequence}. We assume that, in slow time compared to $1/\omega_\mathrm{ref}$, relaxation of the qubit and the TLSs is Markovian. The kinetic equation for the density matrix then has the form
\begin{align}
\label{eq:q_TLS_full}
&\partial_t\rho= i[\rho, H_\mathrm{q} + H_\mathrm{q-TLS}] +2\Gamma\DD[\sigma_+]\rho \nonumber\\
&+\frac{1}{2}\Gamma_\phi\DD[\sigma_z]\rho
+ \sum_n\mathcal{L}_{TLS}\sn\rho\ , 
\end{align}
where the last term describes relaxation of the TLSs,
\begin{align}
\label{eq:master_TLS}
\mathcal{L}_{\rm TLS}\sn\rho =& W_{10}\sn{}\DD[\tT_+ \sn]\rho + W_{01}\sn{}\DD[\tT_-\sn]\rho\nonumber\\
& + \frac{1}{2}\Gamma_{\phi\mathrm{TLS}}\sn{}\DD[\tT_z\sn]\rho\ .
\end{align}
We use the conventional notation $\DD[F]\rho = F\rho F^\dagger- (F^\dagger F\rho + \rho F^\dagger F)/2$ for the relaxation operator; $\sigma_\pm=(\sigma_x \pm i\sigma_y)/2, \tT_\pm\sn = (\tT_x\sn \pm i\tT_y\sn)/2$.

Parameters $\Gamma$ and $\Gamma_\phi$ describe the qubit decay rate and the rate of dephasing due to fast dephasing processes. They give the familiar parameters of the Bloch equation  for the qubit, 
\[T_1 = 1/2\Gamma, \quad T_2 = 1/(\Gamma+ \Gamma_\phi).\]

Parameters $W_{ij}\sn$ describe the rates of transitions $\ket{i}\sn \to \ket{j}\sn$ between the states of an $n$th TLS ($i,j$ take on the values $0,1$), whereas $\Gamma_{\phi\mathrm{TLS}}\sn$ is the TLS dephasing rate. 
To make the TLSs fully incoherent this rate has to be much larger than other relaxation rates. In this case  the off-diagonal matrix elements of $\rho$ with respect to the TLS states,  $\sn\!\!\bra{i}\rho\ket{j}\sn$ with $i \neq j$, will decay fast and can be disregarded. However, as seen from the analysis below, for the considered dispersive qubit-to-TLS coupling these matrix elements do not affect the outcomes of qubit measurements. Therefore they will not be discussed, i.e., we will consider only the matris elements  $\sn\!\!\bra{i}\rho\ket{j}\sn$ with $i=j$.

In deriving Eq.~(\ref{eq:master_TLS}) we assumed that each TLS is coupled to its individual thermal reservoir, that is, not only there is no direct interaction between the TLSs, but there is also no interaction mediated by a common thermal reservoir. In the case of phononic thermal reservoirs, this model goes back to the original papers \cite{Anderson1972,Phillips1972}.

We assume that at $t=0^-$ the qubit is in the ground state $\ket{0}$ and the TLSs are in their stationary states.  From Eq.~(\ref{eq:master_TLS}) the stationary density matrix of an $n$th TLS is
\begin{align}
\label{eq:TLS_populations}
&\rho_\mathrm{st}\sn = \frac{1}{2}\Bigl[\hI\sn +\frac{\Delta W\sn}{W\sn}\tT_z\sn)\Bigr]\ , \\ 
&W\sn = W_{01}\sn + W_{10}\sn, \quad \Delta W\sn = W_{10}\sn - W_{01}\sn\ . \nonumber
\end{align} 
The density matrix of the whole system, the qubit and the TLSs, before the rotation around the $y$-axis at $t=0$ is 
\[
\rho(t=0^-) = \frac{1}{2}(\hI_q + \sigma_z)\prod_n \rho_\mathrm{st}\sn.\]

At $t=0$  the qubit is rotated around the $y$-axis, as described by the term $\propto \sigma_y$ with $n=0$ in the Hamiltonian (\ref{eq:Hamiltonian}). The TLSs are not affected by unitary transformations on the qubit. The density matrix after the transformation becomes
\begin{align}
\label{eq:initial_rho}
\rho(t=0^+) = \frac{1}{2}(\hI_q + \sigma_x)\prod_n \rho_\mathrm{st}\sn.
 \end{align}
This equation provides the initial condition for the evolution of the density matrix during the first Ramsey measurement, i.e., in the time interval $0<t<t_R$. The solution of Eq.~(\ref{eq:q_TLS_full}) in this time interval can be sought in the form
\begin{align}
\label{eq:many_TLS_general}
&\rho(t) = \frac{1}{2}(\hI_q +\sigma_z)\rho_I + \frac{1}{2}e^{-t/T_1}\sigma_z\rho_z 
\nonumber\\
&+ \frac{1}{2}e^{-t/T_2}\sum_{\alpha=\pm}\exp\left[i\alpha\int_0^t dt'\omq(t')\right]\sigma_\alpha\rho_\alpha
\end{align}
with the operators $\rho_\lambda$ defined as
\begin{align}
\label{eq:many_TLS_Schmidt}
\rho_\lambda =  \prod_n \sum_{\varkappa=0,z}C_{\lambda\varkappa}\sn{}\tT_{\varkappa}\sn. 
\end{align}
Here $\lambda=I, z, \pm$ enumerates the components of the qubit-dependent part of the density matrix, whereas $\varkappa =0, z$ enumerates the TLS operators  $\tT_0\sn\equiv\hI\sn$ and $\tT_z\sn$. The operators $\rho_\lambda$ depend only on the TLS variables. The coefficients $C_{\lambda\varkappa}\sn \equiv C_{\lambda\varkappa}\sn(t)$ describe the evolution of the density matrix in time.

The components of the density matrix $\rho$ that contain $\tT_\pm\sn$ are uncoupled from other components of $\rho$. They do not get coupled by the coupling to the qubit and by the gate operations on the qubit. They decay with the rates $W\sn + \Gamma\sn_{\phi\mathrm{TLS}}$  and are not discussed below. This is why $\varkappa$ in Eq.~(\ref{eq:many_TLS_Schmidt}) runs through $0$ and $z$ only.

The equations for $C_{\lambda\varkappa}\sn$ are obtained by substituting Eq.~(\ref{eq:many_TLS_general}) into the full master equation (\ref{eq:q_TLS_full}), multiplying the left- and right-hand sides in turn by $\hI_q, \hI_q-\sigma_z,  \sigma_\pm$, and taking trace over the qubit states. Because the TLSs decay is independent of each other, the resulting equations for $\rho_\lambda$ separate into  equations for individual TLSs (see Appendix~\ref{sec:C_eqns}). Equations~(\ref{eq:master_TLS}) and (\ref{eq:many_TLS_general}) then reduce to the equation
\begin{align}
\label{eq:single_TLS_decay}
&\sum_\varkappa \dot C_{\lambda\varkappa}\sn \tT_\varkappa\sn =
\mathcal{L}_\mathrm{TLS}\sn \sum_\varkappa C_{\lambda\varkappa}\sn \tT_\varkappa\sn\nonumber\\
&+i\sum_{\alpha=\pm}\delta_{\lambda\alpha}\alpha V\sn{} 
\left(C_{\alpha 0}\sn \tT_z\sn +C_{\alpha z}\sn \tT_0\sn\right)
\end{align}
with $\varkappa = 0,z$. Multiplying this equation in turn by the TLS operators $\hI\sn$ and $\tT_z\sn$ and taking trace over the TLS states, we obtain equations for each of the coefficients $C_{\lambda\varkappa}\sn$.  

The last term in Eq.~(\ref{eq:single_TLS_decay}) describes the effect of the coupling to the qubit on the TLS dynamics. This term comes from the components of the density matrix $\rho$, which are proportional to $\sigma_\pm$. It determines the accumulation of the phase of the qubit between the Ramsey pulses at $t=0$ and $t=t_R$. 

The initial conditions for $C_{\lambda\varkappa}\sn$ follow from Eq.~(\ref{eq:initial_rho}) and are given  by Eq.~(\ref{eq:after_pulse_multi_TLS}) in Appendix \ref{sec:C_eqns}. The  coefficients $C_{\lambda\varkappa}\sn(t)$ in terms of $C_{\lambda\varkappa}\sn(0)$ are given in  Eqs.~(\ref{eq:solution_many_TLS_0}) and (\ref{eq:sigma_plus_many}).  Using these expressions we find that, by the end of the interval between the Ramsey pulses, i.e., for $t\to t_R^-$, we have, in particular,
\begin{align}
\label{eq:C0_coeff}
C_{I0}\sn{}(t_R)=\frac{1}{2},\quad C_{+0}\sn{}(t_R)=\frac{1}{2}\Xi\sn(t_R), 
\end{align}
where function $\Xi_n(t_R)$ is given in Eq.~(\ref{eq:Xi_n}). This function describes the effect of the coupling to an $n$th TLS on the probability of the Ramsey measurement outcome.
It depends on the interrelation between the strength of the TLS-to-qubit coupling $V\sn{} $ and the TLS relaxation rate $W\sn$, that is, for a TLS to be effectively strongly coupled to the qubit it suffices to have $|V\sn{} |$ large compared to the TLS relaxation rate $W\sn$.

The explicit expressions for the coefficients $C_{\lambda\varkappa}\sn(t_R)$  determine the operators $\rho_\lambda(t_R)$, as seen from Eq.~(\ref{eq:many_TLS_Schmidt}). They thus describe the change of the density matrix $\rho(t)$ over the time after the qubit was prepared in the state $(\ket{0} + \ket{1})/\sqrt{2}$ and before it is going to be measured.


\subsection{The probability of a Ramsey measurement outcome}
\label{sec:Ramsey_probability}

The above results allow us to find the probability  $ r_1$ of obtaining ``1'' as an outcome of the  Ramsey measurement. In the Bloch sphere representation, the involved steps include the rotation  of the density matrix $\rho(t_R)$ about the $z$-axis by the angle $\tilde\phi_R$. The corresponding unitary transformation is determined by Eq.~(\ref{eq:H_R}) in which we replace $\phi_R$ with $\tilde\phi_R$ to allow for the shift of the average qubit frequency due to the coupling to the TLSs. 

The rotation about the $z$-axis is  followed by the rotation about the $y$-axis by $\pi/2$, as prescribed by the term $(\pi/4)\sigma_y\delta(t-t_R)$ in Eq.~(\ref{eq:Hamiltonian}). Finally, the transformed density matrix $\rho(t_R^+)$ has to be multiplied by the operator  $\hat\pi =( \hI_q-\sigma_z)/2$ and the trace over the states of the qubit and the TLSs has to be taken along with the averaging over classical noise of the qubit frequency $\omq$.  

The aforementioned unitary transformations refer to the operators  $\sigma_{x,y,z}$ in the density matrix $\rho(t)$ in Eq.~(\ref{eq:many_TLS_general}). The operators $\rho_\lambda(t_R)$ are operators in the space of the TLSs, they are not affected by the gate operations on the qubit at time $t_R$, i.e., $\rho_\lambda(t_R^-) = \rho_\lambda(t_R^+) = \rho_\lambda(t_R)$  (we remind that $\lambda$ takes on the values $I,z,\pm$) . In terms of these operators 
\begin{align}
\label{eq:post_Ramsey_multi_TLS}
&\hat\pi\rho(t_R^+)= \RR(t_R^+)+ \hat m, \quad \RR(t) = \frac{1}{4}(\hI_q+\sigma_z) \nonumber\\
&
\times \left[\rho_I(t) + \frac{1}{2}e^{-t/T_2}\sum_\alpha e^{i\alpha(\theta+\tilde\phi_R)}\rho_\alpha(t)\right]\ .
\end{align}
Here, $\hat m$ is a sum of the terms proportional to $\sigma_x, \sigma_y$, and $\sigma_z$; therefore $\tr\,\hat m =0$. The term $
\theta = \int_0^{t_R}\omq(t)dt
$
is the phase accumulated because of slow classical qubit frequency noise. It {\it does not} include the contribution from the TLSs.

From Eqs.~(\ref{eq:C0_coeff}) and (\ref{eq:post_Ramsey_multi_TLS}) we find that the probability of obtaining ``1'' in a Ramsey measurement is given by Eq.~(\ref{eq:Pi_1_multi_TLS_explicit}). To allow for a classical qubit frequency noise, one has to replace the factor $\exp(i\tilde\phi_R)$ in  equation Eq.~(\ref{eq:Pi_1_multi_TLS_explicit}), 
\begin{align}
\label{eq:r_1_random_phase}
\exp(i\tilde\phi_R) \to \exp(i\tilde\phi_R)\langle e^{i\theta}\rangle \ ,
\end{align}
where $\langle\cdot\rangle$ implies averaging over the  classical frequency noise. This noise does not affect the dynamics of the TLSs and therefore its effect is described just by an extra factor.


\section{The pair correlation function for the coupling to two-level systems}
\label{sec:r_2_TLS}

In this section we discuss the effect of the TLSs on the pair correlation function of the qubit measurements $r_2(k)$.  To simplify the notations we will set $\omq = 0$. We start  with the correlator for neighboring cycles, i.e., $r_2(1)$, and as we move on we will extend the analysis to $r_2(k)$ for an arbitrary $k$. 

It is clear from the definition (\ref{eq:correlators1_defined}) that  finding $r_2(1)$  involves the following steps. After we have found  $\hat\pi\rho(t_R^+)$, we have to find how this operator   evolves in the time interval from $t_R^+$ to $\ts$ during which the qubit is reset to the ground state $\ket{0}$. At $\ts$ the qubit is rotated to $(\ket{0}+\ket{1})/\sqrt{2}$. We then have to consider the dynamics in the interval from $\ts^+$ to $\ts^+ +t_R$.  At $\ts +t_R$ the qubit is again rotated and the evolved operator $\hat\pi\rho(\ts+t_R^+)$ is multiplied by $\hat\pi$.  The value of $r_2(1)$ is given by the trace of the result. We will discuss each of these steps separately.

\subsection{Evolution during the reset, $t_R<t<\ts$}
\label{subsec:reset}

The dynamics of the system during the reset of the qubit can be formally described by the master equation (\ref{eq:q_TLS_full}) written for $\hat\pi\rho(t)$. In this equation one can assume that the qubit decay rate $\Gamma$ is large, $\Gamma (\ts - t_R)\gg 1$. In this limit the part of $\hat\pi\rho(t)$ described by the operator $\hat m$ in Eq.~(\ref{eq:post_Ramsey_multi_TLS}) and thus proportional to $\sigma_{x,y,z}$ will decay to zero. Therefore of interest is only the evolution of the operator $\RR(t)$ in Eq.~(\ref{eq:post_Ramsey_multi_TLS}). 

In the operator $\RR(t)$ the qubit-dependent factor $\hI+\sigma_z$ does not change. However, the TLSs are not in their stationary states at $t_R$, and they keep evolving for $t>t_R$, each with its own rate. To describe this evolution it is convenient to first separate out the part $\Rst$ of $\RR(t)$ that would describe the system if the TLSs were in the stationary states, i.e., they were described by the density matrices  $\rho_\mathrm{st}\sn$. Using the explicit expressions  (\ref{eq:C0_coeff}), (\ref{eq:C_coeff_explicit}),  and (\ref{eq:C_+z}) for the operators $\rho_I, \rho_\pm$ we obtain
\begin{align}
\label{eq:G_t_general}
&\RR(t) = r_1 \Rst(t) + \Rcorr(t), \quad t_R^+ \leq t < \ts\ ,\nonumber\\
&\Rst(t_R) = \frac{1}{2}(\hI_q+\sigma_z)\prod_n \rho_\mathrm{st}\sn\ , 
\end{align}
and
\begin{align}
\label{eq:Rcorr}
&\Rcorr(t_R)=\frac{1}{2}(\hI_q+\sigma_z)\sum_{s\geq 1}^{N_\mathrm{TLS}}\sum_{\{m\}_s}\mK\nonumber\\
&\times \tT_z^{(m_1)}...\tT_z^{(m_s)}\prod_{n\notin \{m\}_s}\rho_\mathrm{st}\sn\ .
\end{align}
Here we have introduced sets $\{m\}_s$. Their components   $m_1,m_2,...,m_s$ enumerate different TLSs. The values of $m_i$ ($i=1,...,s$) within each set run from 1 to $N_\mathrm{TLS}$. The sum
 $\sum_{\{m\}_s} $ is taken over all $m_i$, for example, one can think of it as 
\[\sum_{\{m\}_s} \equiv \sum_{N_\mathrm{TLS}\geq m_1>m_2...>m_s\geq 1}.\] 
The  parameters $\mathbb{K}_{{m}_s}$ are given in Appendix~\ref{sec:K_term_evolution}.

The form of $\Rcorr$ can be understood by noting that, to describe the evolution of $\RR(t)$, we have to take into account the decay of all possible combinations of the TLSs. The parameter $s$ in Eq.~(\ref{eq:Rcorr}) gives the number of the TLSs included into a combination, $1\leq s \leq N_\mathrm{TLS}$. By construction, the trace of any term in the sum over $s$ is equal to zero.  

The operator $\Rst(t)$  describes the qubit in its ground state and the TLSs in their stationary states. It is not changed during reset, i.e. 
\[\Rst(\ts^-) = \Rst(t_R^+)\ .\]  

In contrast, the TLS operators $\tT_z^{(m_i)}$ in $\Rcorr$ exponentially decay with the rates $W^{(m_i)}$ because of the transitions between the states of the TLSs, as seen from Eq.~(\ref{eq:master_TLS}). Since the TLSs are independent,  over the duration of the reset $\ts - t_R$ each $\tT_z^{(m_i)}$ in Eq.~(\ref{eq:Rcorr}) acquires a factor $\exp[-W^{(m_i)}(\ts-t_R)]$, so that at the end of the reset period, i.e., at the end of the cycle the expression for $\Rcorr(\ts^-)$ is given by Eq.~(\ref{eq:Rcorr}) in which one replaces
\begin{align}
\label{eq:Rcorr_ts}
\tT_z^{(m_i)} \to \tT_z^{(m_i)}\exp[-W^{(m_i)}(\ts-t_R)]\ ,
\end{align}
for all $m_i\in\{m\}_s$. Note that the terms $\rho_\mathrm{st}\sn$ with $n\notin \{m\}_s$ do not change.


\subsection{The second Ramsey measurement}
\label{subsec:second_Ramsey}

To find  the dynamics of the operator $\RR(t)$ in the time interval $\ts <t \leq \ts+t_R^+$ we should take into account that at time $\ts$ the qubit undergoes a unitary transformation of rotation around the $y$-axis, as seen from the Hamiltonian (\ref{eq:Hamiltonian}). Respectively, in the expression for $\RR(t)$ the operator $\hI_q + \sigma_z$ is transformed into $\hI_q +\sigma_x$.
%

\subsubsection{The contribution of the term $\Rst$}
\label{subsubsec:Rst}

The evolution of the operator $\Rst(t)$ after the qubit rotation at $\ts$ is described in the same exact way as it was done in Sec.~\ref{subsec:QKE} for $\rho(t)$. Indeed, $\Rst(\ts^+)$ has the same form as the density matrix $\rho(0^+)$,  Eq.~(\ref{eq:initial_rho}), except that $\Rst$ has an extra factor $r_1$.  Thus  the evolution of $\Rst(t)$ in the time interval $\ts<t<\ts +t_R$  is given by Eqs.~(\ref{eq:many_TLS_general}) and (\ref{eq:many_TLS_Schmidt}) with the coefficients $C_{\lambda\varkappa}\sn$  multiplied by $r_1$.

It follows from the above argument that if, after the next Ramsey  rotation at  $\ts+t_R$,  the transformed $\Rst$ [i.e., $\Rst(\ts+t_R^+)$] is multiplied by $\hat\pi$ and the trace is taken over the qubit and the TLSs, the result will be  $r_1^2$. This is the contribution of $\Rst$ to $r_2(1)$.

Further, to find the contribution of $\Rst$ to $r_2(k)$  with $k>1$ we note that, by applying the decomposition of the density matrix (\ref{eq:many_TLS_general}) to $\Rst$, one can write $\Rst(\ts+t_R^+)$ as 
\[\Rst(\ts+t_R^+) = \Rst(t_R^+) + \hat m'\ ,\]
where $\hat m'$ is a sum of the terms proportional to $\sigma_x,\sigma_y$, and $\sigma_z$. 
Evaluating $r_2(k)$ involves resetting the qubit after each $n\ts+ t_R^+$ with $n< k$. After the reset, $\hat m'$ will go to zero.  Therefore by the end of the second cycle, $t\to 2\ts$, we will have $\Rst(2\ts^-) = \Rst(\ts^-) =\Rst(t_R^+)$. The operator $\Rst$ will evolve in the same way during the following cycles. The cycling does not change $\Rst$, that is $\Rst(n\ts + t_R^+) = \Rst(m\ts^-)$ for any $m$ and $n$.  Therefore the contribution of $\Rst$ to the correlator $r_2(k)$ with $k>1$ is the same as for $k=1$. It is equal to $r_1^2$ independent of $k$. 

\subsubsection{The contribution of the term $\Rcorr$}
\label{subsubsec:Rst}

The term $\Rcorr$ describes the effect of correlations in the TLS dynamics on the outcome of the qubit measurements. To analyze this effect we notice first that the outcome of the qubit rotation at $t=\ts$ can be written as 
\begin{align}
\label{eq:transform_qubit}
\hI_q+\sigma_z \to \hI_q + \sigma _z + \sum_{\alpha=\pm}\sigma_\alpha - \sigma_z\ .
\end{align}
We saw above that, after the rotation at $\ts+t_R$, the last term, $\sigma_z$, is transformed into the terms that decay on reset; these terms also do not contribute to the trace over the qubit states if multiplied by $\hat\pi$. Therefore we will not consider the contribution from the term $\propto\sigma_z$ in $\Rcorr(\ts^+)$.   

As seen from the master equation (\ref{eq:q_TLS_full}), the terms  $\propto (\hI_q + \sigma_z) \tT^{m_1}_z...\tT^{m_s}_z$ in $\Rcorr(\ts^+)$ commute with the Hamiltonian and therefore do not lead to mixing of the qubit and TLS states. We denote this part of $\Rcorr(\ts^+)$ as $\Rcorr'(\ts^+)$. Over time $t_R$ the operator $\Rcorr'(\ts^+)$ will decay as $\exp(-\sum_i W^{(m_i)}t_R)$. With the account taken of Eq.~(\ref{eq:Rcorr_ts}), after the Ramsey pulse at $\ts + t_R$, $\Rcorr'(\ts+t_R^+)$ will have the form of Eq.~(\ref{eq:Rcorr}) in which $\hI_q+\sigma_z$ is replaced by $\hI_q + \sigma_x$ and also there is made a replacement
\begin{align*}
&\tT_z^{(m_i)}
\to \tT_z^{(m_i)}\exp(-W^{(m_i)}\ts), \quad m_i\in \{m\}_s \ .
\end{align*}
The trace of $\hat\pi\Rcorr'(\ts+t_R^+)$ over the TLSs is zero, therefore this term will not contribute to $r_2(1)$.

However, this term determines the values of $r_2(k)$ with $k>1$. 
To see this we first note that, after the qubit reset at $\ts+t_R^+$, by the time $2\ts$ the operator $\hI_q +\sigma_x$ in $\Rcorr'$ will transform into $\hI_q + \sigma_z$ and there will emerge an extra factor $\exp \Bigl[-\sum_{m_i}W^{(m_i)}(\ts-t_R)\Bigr]$  in the sum over $\{m\}_s$. Thus $\Rcorr'(2\ts^-)$ will have the same structure as $\Rcorr(\ts^-)$. It is seen from Eq.~(\ref{eq:transform_qubit}) that this structure will be reproduced from cycle to cycle, so that
\begin{align}
\label{eq:Rcorr_prime}
&\Rcorr'(k\ts^-)=\frac{1}{2}(\hI_q+\sigma_z)\sum_{s\geq 1}\sum_{\{m\}_s}\mK \tT_z^{(m_1)}...\tT_z^{(m_s)}\nonumber\\
&\times \exp\left[-\sum_{m_i \in\{m\}_s}W^{(m_i)}(k\ts-t_R)\right]\prod_{n\notin \{m\}_s}\rho_\mathrm{st}\sn\ ,
\end{align}
 Moreover, it is easy to see that $\Rcorr(k\ts^-) = \Rcorr'(k\ts^-)$ provided no measurements are done at $n\ts+t_R^+$ with $n<k$. This is because the terms $\propto \sigma_\pm$ in $\Rcorr$ vanish on reset. Indeed, the rotation around the $y$-axis at $n\ts+t_R$ transforms $\sigma_\pm$ into $\sigma_x,\sigma_y, \sigma_z$ with different coefficients, which all decay on reset.

The accumulation of the decay of different TLSs described by Eq.~(\ref{eq:Rcorr_prime}) ultimately determines the correlator $r_2(k)$.
However, the very values of $r_2(k)$ are determined by the terms in $\Rcorr(k\ts^+)$, which are  $\propto\sigma^\pm$ and emerge after  the transformation Eq.~(\ref{eq:transform_qubit}).  They have to be studied separately for each product of the TLS operators $\tT_z^{(m_1)}...\tT_z^{(m_s)}$ in $\Rcorr$. The analysis is similar to that in Sec.~\ref{subsec:QKE} and is described in Appendix~\ref{sec:K_term_evolution}. The result is Eq.~(\ref{eq:pair_full_1}) for the centered correlator $\tilde r_2(k)$.

The described method allows one to calculate higher-order correlators as well. However, the expressions are combersome and will not be provided here. In Appendix~\ref{sec:Yaxing_method} we describe an alternative approach to calculating the probability $r_1$ and the correlator $r_2(k)$, which is based on the properties of telegraph noise that drives a qubit and mimics the coupling to the TLSs.



\section{Correlators of measurement outcomes for a Gaussian frequency noise}
\label{sec:Gaussian_general}

We now consider the effect of Gaussian fluctuations of the qubit frequency $\omq(t)$ on the probability $r_1$ of the Ramsey measurement outcomes and their 2- and 3-time correlation functions $r_2(k)$, $r_3(k,l)$. We will express these probabilities in terms of the correlation functions $f_k$ of the phases $\theta_n$ accumulated by the qubit between the Ramsey pulses applied at times $n\ts$ and $n\ts+t_R$ with $n=0,1,...$,
\[\theta_n = \int_{n\ts}^{n\ts+t_R}\omq (t)dt, \quad
f_{k}\equiv \langle\theta_n\theta_{n+k}\rangle\ ,
\]
cf. Eqs.~(\ref{eq:phase_k_classical}). Equation (\ref{eq:phase_correlation}) relates the correlators $f_k$  to the power spectrum $S_q(\omega)$ of noise $\omq(t)$. The probability distribution of the phases has the form 
\begin{align}
\label{eq:phase_distribution}
&P(\{\theta\}) = Z^{-1}\exp\left[-\frac{1}{2}\sum_{m,n}(\hat f{}^{-1})_{n-m}\theta_n\theta_m\right],\nonumber\\
&\sum_m(\hat f{}^{-1})_{n-m} f_{m-k} = \delta_{m,k}\ ,
\end{align}
where $\{\theta\}$ is the set of $\theta_n$ with different $n$, while  $Z$ is the normalization factor.

The effect of classical noise can be easily described using the master equation approach. One does not have to care about the evolution of the TLS-dependent part of the density matrix, which significantly simplifies the calculation. 

We begin with the first cycle  that starts at $t=0$. Before there is applied the  first Ramsey pulse (the rotation around the $y$-axis by $\pi/2$), the qubit is in the ground state. Its density matrix is 
\[\rho(0^-) = (\hI_q+\sigma_z)/2\ .\]
After the first Ramsey pulse at $t=0$ we have $\rho(0^+) = (\hI_q+\sigma_x)/2$. The evolution of the system at $0<t<t_R$   is described by Eq.~(\ref{eq:many_TLS_general}) in which one replaces the TLS operators by the numbers determined by the form of $\rho(0^+)$, i.e., $\rho_I \to 1, \rho_z \to -1, \rho_\pm \to 1$. 

After the Ramsey pulse at  $t=t_R$, we have, as seen from Eq.~(\ref{eq:post_Ramsey_multi_TLS}), 
\begin{align}
\label{eq:Ramsey_1_noise}
&\hat\pi \rho(t_R^+) = \RR_0(t_R^+) +\hat m_0, \nonumber\\
&\RR_0(t_R^+)=\frac{1}{2} (\hI_q+ \sigma_z)p(\theta_0),
\end{align}
where the probability $p(\theta)$ is given by the standard expression (\ref{eq:standard_probability}) and $\hat m_0$ is a sum of terms proportional to the operators $\sigma_{x,y,z}$. Taking a trace over the qubit states and averaging the result over the phases $\theta$ using Eq.~(\ref{eq:phase_distribution}) gives Eq.~(\ref{eq:class_noise_results_1}) for
\[r_1=\langle p(\theta_0)\rangle \equiv \int  p(\theta_0)\prod_n d\theta_n P(\{\theta\})\]
(the integral goes over all $\theta_n$, for the correlated phases $\theta_n$).

If the trace and the averaging are not done and instead the qubit is reset, the term $\hat m$ in Eq.~(\ref{eq:Ramsey_1_noise}) will decay whereas the operator $\hI_q +\sigma_z$, and thus $
\RR_0(t_r^+)$, does not change. Then by the end of the first cycle, $t\to \ts$, the operator $\hat\pi \rho(\ts^-)$ will become $\hat\pi \rho(\ts^-)=\RR_0(t_R^+)=\RR_0(\ts^-)=p(\theta_0)\rho(0^-)$. 

To describe the dynamics during the next cycle  we use again Eq.~(\ref{eq:many_TLS_general}). The analysis is identical to that for the previous cycle, except that $\rho(0^-)$ is replaced with $p(\theta_0)\rho(0^-)$.  After the pair of the Ramsey pulses applied during the cycle we have
\begin{align}
\label{eq:Ramsey_2_noise}
 &\RR_0(\ts+t_R^+) = \RR_1(\ts+t_R^+)  + \hat m_1,\nonumber\\
&\RR_1(\ts + t_R^+)=  \frac{1}{2} p(\theta_0)\Bigl[\hI_q \nonumber\\
& - 
 \sigma_z e^{-t_R/T_2} \cos(\phi_R+\theta_1)\Bigr],
 \end{align}
 where $\hat m_1$ has terms proportional only to $\sigma_x$ and $\sigma_y$. Note that the random phase $\theta_1$ has been accumulated over the time interval $(\ts,\ts+t_R)$, which is different from the time interval $(0,t_R)$ over which $\theta_0$ was accumulated.
 
As a result of the reset during the time interval $(\ts+t_R^+, 2\ts)$,  by the end of the second cycle we will have again $\RR_0(2\ts^-) =   p(\theta_0) \rho(0^-)$. The further evolution is just a repetition of the previous steps. After $k$ pairs of the Ramsey pulses the expression for $ \RR_0(k\ts+t_R^+)$ will have the same form as Eq.~(\ref{eq:Ramsey_2_noise}) except that $\theta_1$ will be replaced by $\theta_k$.
 
To find the pair correlator $r_2(k)$ one has to multiply $\RR_0(k\ts+t_R^+)$ by the projection operator $\hat\pi$, which gives, as seen by extending Eq.~(\ref{eq:Ramsey_2_noise}) from $\ts+t_R^+$ to $k\ts + t_R^+$, 
 \begin{align}
 \label{eq:Ramsey_3_noise}
  &\hat\pi \RR_0(k\ts+t_R^+)=\RR_2(k\ts + t_R^+) + \hat m_2, \nonumber\\
&\RR_2(k\ts + t_R^+)=  \frac{1}{2}\hI_q p(\theta_0) p(\theta_k)\ .
 \end{align}
 Here again $\hat m_2$ is a sum of the terms proportional to $\sigma_x,\sigma_y$, and $\sigma_z$. 
Taking a trace over the qubit variables and averaging the result over the correlated phases $\theta_0,\theta_k$ gives Eq.~(\ref{eq:class_noise_results_1}) for the correlator $r_2(k)=\langle p(\theta_0)p(\theta_k)\rangle$.

To find the three-time correlator $r_3(k,l)$ we have to follow the evolution of the operator $ \RR_2(k\ts+t_R^+)$ for the next $l-k$ cycles. There is no difference from the previous steps, as after reset we again express this operator in terms of the density matrix of the qubit in the ground state $\rho(0^-)$,
\[\RR_2[(k+1)\ts^-] = p(\theta_0)p(\theta_k)\rho(0^-)\ ,\]
(we note that, as a result of the reset, the operator $\hI_q$ in $\RR_2$ goes into $\hI_q +\sigma_z$). After $l-k$ cycles we will have, similar to Eq.~(\ref{eq:Ramsey_3_noise}),
 \begin{align}
 \label{eq:Ramsey_4_noise}
  &\hat\pi \RR_2(l\ts+t_R^+)=\RR_3(l\ts + t_R^+) + \hat m_3, \nonumber\\
&\RR_3(l\ts + t_R^+)=  \frac{1}{2}\hI_q p(\theta_0) p(\theta_k)p(\theta_l)\ .
 \end{align}
 This leads to the expression $r_3(k,l) = \langle  p(\theta_0) p(\theta_k)p(\theta_l)\rangle$. The explicit form of the centered correlator $\tilde r_3(k,l)$ in terms of the correlators $f_k$ for Gaussian noise is given in Eq.~(\ref{eq:class_noise_results_2}).
 


\section{Conclusions}
\label{sec:conclusions}

This paper describes several features of slow qubit frequency fluctuations that allow one to characterize the mechanism of the fluctuations. Of primary interest are fluctuations with the correlation time that exceeds the decoherence time of the qubit due to its decay and dephasing by fast processes. The approach is based on periodically repeated Ramsey measurements. In such measurements one can vary the duration of the single measurement $t_R$ and the period of the measurements $\ts$.  One can find the probability $r_1$ of observing ``one'' as a measurement outcome, and the two- and three-time correlation functions  $r_2(k)$ and $r_3(k,l)$ of repeatedly observing ``one'' over time $k\ts$, and $k\ts$ and $l\ts$. 

We have developed a fairly general analytical approach which allowed us to describe the qubit dynamics in the presence of an evolving noise with the account taken of the gate operations involved in the repeated measurements. This approach enabled calculating the probabilities $r_1$ and $r_2(k)$ for the dispersive coupling to the TLSs in the explicit form, and also finding $r_{1,2,3}$  for a Gaussian noise. 
The results cover a broad range of the noise  characteristics. For the TLSs, those are  the strength of the coupling to the qubit (in the units of frequency) compared to $t_R^{-1}$ and to the TLSs switching rates, the difference of the switching rates between the TLSs' states, as well as the actual number of the TLSs that contribute to the noise. For Gaussian noise these are the noise intensity and the noise power spectra. 

A distinguishing  feature of a Gaussian noise is the relation between the correlators. As we show, once $r_1$ and $r_2(k)$ have been measured, they determine the form of $r_3(k,l)$. We find the corresponding relation. If it does not hold, the noise is non-Gaussian. We also find analytically and confirm by simulations the form of $r_2(k)$ for several important types of noise, in particular for the noise with the power spectrum of the form of a Lorentzian peak at zero frequency and for a $1/f$ noise with a soft low-frequency cutoff. The cutoff we used is characteristic of the noise that results from the dispersive coupling to a large number of symmetric TLSs with the log-normal distribution of the switching rates.

The analytical results and the results of the simulations show explicitly the sensitivity of the correlators to various noise parameters, which we illustrate in the figures throughout the paper. We also show that the power spectrum of the measurement outcomes allows one to reveal a slow periodic modulation of the qubit frequency. The emerging narrow spectral peak is very sensitive to the periodicity of the measurements.  

An advantageous feature of the proposed method is that it accounts for the  evolution of the qubit step by step during and between the gate operations. Therefore it can be immediately extended to allow for gate errors and for measurement errors.  

Along with long measurement sequences, one can also study the distribution of the instances of observing ``one''  in a relatively small number of measurements $M$ (still $M\gg 1$), provided the total duration of the data acquisition $M\ts$ is comparable to the noise correlation time. As we show analytically in the limiting cases and by simulations in the general case, this distribution can be significantly different from the conventional binomial distribution. In the case of coupling to slow TLSs, the distribution can have a fine structure that corresponds to the TLSs mostly staying in their initially occupied states during the data acquisition. On the other hand, even where $M\ts$ largely exceeds the noise correlation time, the distribution of the instances of observing ``one'' is significantly broadened by the noise correlations compared to the binomial distribution. This is a simple test of the presence of noise correlations.

The noise from TLSs is often assumed to be the cause of slow fluctuations of the qubit frequency. The presented analysis provides a tool for testing this assumptions in various types of qubits. We believe that the developed methods can be extended to other types of noise of potential interest. In particular, this refers to a shot noise. In a way, this paper is the first step toward creating a ``map'' of the effects of the mechanisms of different types of slow qubit frequency fluctuations.


\section*{Acknowledgements}
We are grateful to Vadim Smelyanskiy for helpful and inspirational discussions, and to Juan Atalaya, who participated in the work at the early stage.  
FW and MID are thankful for the support from NASA Academic Mission Services, Contract No. NNA16BD14C and from Google under NASA-Google SAA2-403512. MID acknowledges  financial support from Google via PRO Unlimited.

\appendix

\section{Effect of a telegraph noise on periodically repeated Ramsey measurements}
\label{sec:Yaxing_method}  

In this section, we discuss an alternative method of deriving our major result in Eqs.~(\ref{eq:Pi_1_multi_TLS_explicit}) and (\ref{eq:pair_full_1}) for the effect of the coupling to the TLSs  on the probability of a Ramsey measurement outcome and the correlator of the outcomes. We describe  this effect as resulting from classical telegraph noise $\omq(t)$ that is added to the qubit frequency. Noise comes from  random uncorrelated switching of the TLSs between their two states $\ket{0}\sn$ and $\ket{1}\sn$.   

The method is based on the relation (\ref{eq:trivial_probability}) between the sought parameters $r_1$ and $r_2$  and  the random phases \[\theta_k =\int_{k\ts}^{k\ts + t_R} \delta\omega_\mathrm{q}(t)dt\] accumulated during the $k$th Ramsey measurement.  The idea is to relate $r_1,r_2$ to  the characteristic function of the phases $\{\theta_k$\}. We note that for the telegraph noise, generally, $\langle\theta_k\rangle\neq 0$. 

The characteristic function is defined as the average over random phases $\theta_k$,
\begin{align}
\label{eq:Phi_q}
    \Phi(\vec q) = \langle e^{i \vec q \cdot \vec \theta} \rangle, 
\end{align}
where we consider the values $\theta_0,\theta_1, ...$ and numbers $q_1, q_2,...$ as components of vectors $\vec \theta = (\theta_0,\theta_1,\theta_2,...)$ and  $\vec q = (q_0, q_1, q_2...)$. 

From Eq.~(\ref{eq:standard_probability}), $r_1$ and $r_2$ can be written in terms of the characteristic function as  
\begin{widetext}
\begin{align}
    r_1 &= \frac{1}{2} + \frac{1}{2}e^{-t_R/T_2}\mathrm{Re}[e^{i\tilde\phi_R}\Phi(q_0=1,q_{k\neq 0}=0)], \label{eq:r1_general} \\
     r_2(k) &= r_1^2 + \frac{1}{8} e^{-2t_R/T_2} \left\{ \Phi(q_0=-1,q_k = 1,q_{k'\neq 0,k}=0) - |\Phi(q_0=1,q_{k\neq 0}=0)|^2 \right.\nonumber\\
     & \left.+\mathrm{Re} \left[e^{2i\tilde\phi_R} \Phi(q_0 = 1,q_k=1,q_{k'\neq 0,k}=0)-e^{2i\tilde\phi_R}\Phi^2(q_0 =1,q_{k\neq 0}=0)\right] \right\}.\label{eq:r2_general}
\end{align}
\end{widetext}
In what follows we derive explicit expressions for the characteristic function when the qubit frequency is subject to telegraph noise. 


\subsection{Characteristic function for one TLS}
\label{subsec:characteristic_one_TLS}

We first consider the case where  the qubit is coupled to one classical TLS, i.e., $\delta\omega_\mathrm{q}(t) = V^{(n)} \tau_z^{(n)}(t)$. Here $\tau_z^{(n)}(t)$ is a classical random variable, telegraph noise that takes values  $\pm 1$ depending on whether the considered $n$th TLS is in the state $\ket{0}\sn$ or $\ket{1}\sn$, i.e., $\tau_z^{(n)}$ is the eigenvalue of $\tT_z\sn$ on the corresponding states. It follows from the definition of $\theta_n$ 
that the characteristic function in Eq.~(\ref{eq:Phi_q}) can be written in the form 
\begin{align}
\label{eq:Phi_q_oneTLS}
    \Phi(\vec q) = \langle \exp\left[i\int_0^\infty \alpha (t) \tau_z^{(n)}(t) dt \right] \rangle, 
\end{align}
where $\alpha(t)$ is a piecewise-constant function of time and is only non-zero in between the two Ramsey pulses within each cycle,
\begin{align}
\label{eq:alpha_t}
    \alpha(t) = q_k V^{(n)},\quad kt_{\rm seq} \leq t \leq kt_{\rm seq} + t_R. 
\end{align}
whereas $\alpha(t)=0$ for $k\ts + t_R\leq t \leq (k+1)\ts$.


\subsubsection{Auxiliary functions}

Telegraph noise $\tau_z\sn (t)$ is a Markov process. The Markovian property and the feature that noise takes the values $\pm 1$ allow one to derive an important relation \cite{Gurvitz2019}, which extends to asymmetric TLSs that was previously obtained in \cite{Shapiro1978} for symmetric TLSs,
\begin{align}
\label{eq:Gurvitz}
    &\frac{d}{dt}\langle \tau_z^{(n)}(t) F[\tau_z^{(n)}] \rangle = - W^{(n)} \langle \tau_z^{(n)}(t) F[\tau_z^{(n)}] \rangle  \nonumber \\ &+\langle \tau_z^{(n)}(t) \frac{d}{dt} F[\tau_z^{(n)}]\rangle+ \Delta W^{(n)} \langle F[\tau_z^{(n)}] \rangle,
\end{align}
where $F[\tau_z\sn]$ is an arbitrary function of $t$ and an arbitrary functional of $\tau_z\sn(t')$ for $ 0 \leq t' \leq t$, with $t = 0$ being the moment of imposing initial conditions. The last term in Eq.~(\ref{eq:Gurvitz}) can be understood by noting that $\langle\tau_z\sn\rangle = \Delta W\sn /W\sn$. 

We will use the relation (\ref{eq:Gurvitz}) to reduce the calculation of the characteristic function $\Phi(\vec q) $ to a set of ordinary differential equations. To this end  
we introduce the following functions: 
\begin{align}
\label{eq:chi_t}
    \chi\sn(t) &= \left\langle \exp\left[i\int_0^t \alpha(t')\tau_z^{(n)}(t')dt'\right] \right\rangle,\\
    \mathcal{X}\sn(t) &=  \left\langle \tau_z^{(n)}(t) \exp\left[i\int^t_0 \alpha(t') \tau_z^{(n)} (t') dt'\right]\right\rangle.
\end{align}

From Eq.~(\ref{eq:Gurvitz}), functions $\chi\sn(t)$ and $\mathcal{X}\sn(t)$ satisfy a system of coupled differential equations, 
\begin{align}
\label{eq:dot_chi_t}
        \frac{d}{dt}\chi\sn(t) &= i \alpha(t) \mathcal{X}\sn(t),  \\
\label{eq:dot_Y_t}        
        \frac{d}{dt}\mathcal{X}\sn(t) &= -W^{(n)} \mathcal{X}\sn(t) +i\alpha(t) \chi\sn(t)  \\ & + \Delta W^{(n)} \chi\sn(t).\nonumber
\end{align}
where we used that $\tau_z\sn{}^2 = 1$.


\subsubsection{One-time characteristic function}

With Eqs.~(\ref{eq:dot_chi_t}) and (\ref{eq:dot_Y_t}) in hand, we are ready to derive the expression for $\Phi(q_0=1,q_{k\neq 0}=0)$ in Eq.~(\ref{eq:r1_general}), which we refer to as the one-time characteristic function. 

It follows from Eqs.~(\ref{eq:Phi_q_oneTLS}) and  (\ref{eq:chi_t}) that, to compute $\Phi(q_0=1,q_{k\neq 0}=0)$, we simply need to compute $\chi\sn(t)$ assuming that $\alpha (t) =V\sn$ for $ 0 \leq t\leq t_R$ and $\alpha(t)=0$ for $t>t_R$. For such $\alpha(t)$, we have 
\begin{align}
\label{eq:Phi_one_TLS}
    \Phi(q_0=1,q_{k\neq 0}=0) = \chi\sn(t_R).
\end{align}
Solving  Eqs.~(\ref{eq:dot_chi_t}) and (\ref{eq:dot_Y_t}) in the interval $0\leq t \leq t_R$ with $\alpha(t) = V^{(n)}$, we find
\begin{widetext}
\begin{align}
    \label{eq:transfer_matrix}
& \begin{pmatrix}    \chi\sn(t_R) \\ \mathcal{X}\sn(t_R) \end{pmatrix} = \hat T(V^{(n)},t_R) \begin{pmatrix}    \chi\sn(0) \\ \mathcal{X}\sn(0) \end{pmatrix}, 
  \qquad \begin{pmatrix} \chi\sn(0)\\   \mathcal{X}\sn(0)\end{pmatrix} = 
 \begin{pmatrix} 1\\  \Delta W^{(n)} / W^{(n)} \end{pmatrix},\\
  %
    \label{eq:def_T_matrix}
 &   \hat T(V^{(n)},t_R)  =\frac{1}{\gamma\sn} \exp[-W^{(n)} t_R/2] \nonumber\\
&\times \begin{pmatrix} \frac{1}{2}W^{(n)} \sinh(\gamma^{(n)} t_R)+ \gamma\sn\cosh(\gamma^{(n)} t_R) & iV^{(n)}\sinh(\gamma^{(n)} t_R) \\ (iV^{(n)}+\Delta W^{(n)} )\sinh(\gamma^{(n)} t_R) & \gamma\sn\cosh(\gamma^{(n)} t_R) - \frac{1}{2}W^{(n)}\sinh(\gamma^{(n)} t_R)    \end{pmatrix}.
\end{align}
\end{widetext}
Here we have introduced the transfer matrix $\hat T(V^{(n)},t_R)$ which will be useful below. Carrying out the matrix multiplication in Eq.~(\ref{eq:transfer_matrix}), we obtain 
\begin{align}
\label{eq:chi_tR}
    &\chi\sn(t_R) = \Xi^{(n)}(t_R) 
\end{align}
where $\Xi\sn(t_R)$ is given by Eq.~(\ref{eq:Xi_n}). Equations (\ref{eq:r1_general}), (\ref{eq:Phi_one_TLS}), and (\ref{eq:chi_tR}) give the expression for $r_1$, which coincides with Eq.~(\ref{eq:Pi_1_multi_TLS_explicit}) for the case of one TLS.


\subsubsection{Two-time characteristic function}

We now evaluate  $\Phi(q_0=\pm1,q_k=1,q_{k'\neq k,0})$ in Eq.~(\ref{eq:r2_general}), which we refer to as the two-time characteristic functions. Again, we start with the case of the coupling to one TLS. 
To this end, we need to solve Eqs.~(\ref{eq:dot_chi_t}) and (\ref{eq:dot_Y_t}) for  $\alpha(t)$ of the form: 
\begin{align}
    \alpha (t) &=  \pm V^{(n)}, \quad 0 \leq t \leq t_R, \nonumber \\
    \alpha (t) &=  0, \quad t_R < t < kt_{\rm seq}, \nonumber \\
    \alpha (t) &=  V^{(n)},  \quad kt_{\rm seq}\leq t \leq k t_{\rm seq}+t_R.
    \label{eq:alpha_t_two_time}
\end{align}
The relevant two-time characteristic functions in Eq.~(\ref{eq:r2_general}) are then expressed in terms of $\chi\sn(t)$ as
\begin{align}
\label{eq:Phi_to_chi_pm}
    \Phi_0(q_0=\pm 1, q_k=1,q_{k'\neq k,0}) = \chi\sn_{\pm} (k t_{\rm seq}+t_R),
\end{align}
where the subscript $\pm$ corresponds to $\chi\sn(t)$ calculated for $\alpha (t) = \pm V^{(n)}$ in the time interval $0 \leq t \leq t_R$, respectively. 

The function $\chi\sn_\pm (kt_{\rm seq} + t_R)$ can be found using the transfer matrix in Eq.~(\ref{eq:transfer_matrix}) to connect the solutions of $\chi\sn(t)$ in different regions where $\alpha(t)$ is a constant. The solution is reduced just to matrix multiplication, 
\begin{widetext}
\begin{align}
    \begin{pmatrix}    \chi\sn_{\pm}(kt_{\rm seq}+t_R) \\\mathcal{X}\sn(kt_{\rm seq}+t_R)) \end{pmatrix} &= \hat T(V^{(n)},t_R)\hat T(0,kt_{\rm seq}) \hat T(\pm V^{(n)},t_R) \begin{pmatrix}    \chi\sn(0) \\\mathcal{X}\sn(0)  \end{pmatrix}.
\end{align}
\end{widetext}
This gives 
\begin{align}
\label{eq:chi_pm}
&    \chi\sn_+(kt_{\rm seq}+t_R)  = [\Xi\sn(t_R)]^2 + [\xi_k\sn (t_R)]^2,\nonumber\\
&        \chi\sn_-(kt_{\rm seq}+t_R)  = |\Xi\sn(t_R)|^2 + |\xi_k\sn (t_R)|^2,
\end{align}
where $\xi_k\sn(t_R)$ is given in Eq.~(\ref{eq:pair_full_1}). Substituting Eqs.~(\ref{eq:Phi_to_chi_pm}) and (\ref{eq:chi_pm}) into Eq.~(\ref{eq:r2_general}) we obtain for the pair correlator $r_2(k)$ the same expression as Eq.~(\ref{eq:pair_full_1}) written for the case of  coupling to one TLS.


\subsection{Characteristic function in the presence of multiple TLSs}

Having found the characteristic function in the presence of one TLS, let us consider multiple independent TLSs. Qubit frequency noise is now a sum over the TLSs, $\delta \omega_\mathrm{q} = \sum_n V\sn \tau_z\sn (t)$.

A key advantage of using the characteristic function is that, in the presence of many independent TLSs coupled to the qubit, it factors into a product of the characteristic functions for individual TLSs. Specifically, the one-time characteristic function now becomes
\begin{align}
\label{eq:one-point-Phi}
    \Phi(q_0=1,q_{k\neq 0}=0) = \prod_n \chi\sn (t_R),
\end{align}
with $\chi\sn (t_R)$ given in Eq.~(\ref{eq:chi_tR}). Similarly, the expression for the two-time characteristic function reads
\begin{align}
\label{eq:two-point-Phi}
    \Phi(q_0=\pm 1, q_k = 1,q_{k'\neq k,0}) = \prod_n \chi_{\pm}\sn (k \ts + t_R),
\end{align}
with $\chi_{\pm}\sn (k \ts + t_R)$ given in Eq.~(\ref{eq:chi_pm}). 

Substituting Eqs.~(\ref{eq:one-point-Phi}) and (\ref{eq:two-point-Phi}) into Eqs.~(\ref{eq:r1_general}) and (\ref{eq:r2_general}), we immediately obtain the same expressions for $r_1$ and $\tilde r_2(k)$ as  Eqs.~(\ref{eq:Pi_1_multi_TLS_explicit})  and (\ref{eq:pair_full_1}).

\section{Simulations}
\label{sec:simulation_algorithms}

In this section we described the algorithms used in the simulations of qubit frequency noise induced by two-level systems and of Gaussian qubit frequency noise.

\subsection{Simulating noise from the TLSs}
\label{subsec:simulations_TLSs}

Noise from the TLSs was simulated as a sum of telegraph noises produced by each TLS independently. An $n$th TLS has two states,  $\ket{0}\sn$ and $\ket{1}\sn$, in which its contributions to noise are $1$ and $-1$, respectively. These contributions are multiplied by the parameter $V\sn$ of the coupling to the qubit to obtain the qubit frequency shift. The time was discretized with the same step $\delta t$ for all TLSs.  In the simulations we used $\delta t/t_R =0.1$, where $t_R$ is the duration of the Ramsey measurement. As everywhere else, we used the relative length of the cycle $\ts/t_R=3$

At each time step a TLS can switch between its states $\ket{0}\sn$ and $\ket{1}\sn$. The switching probabilities are   
\begin{align}
\label{eq:probab_numerical_TLS}
    p_{01}\sn = W_{01}\sn \delta t, \qquad p_{10}\sn = W_{10}\sn \delta t\ .
\end{align}
where $W_{01}\sn$ and $W_{10}\sn$ are the switching rates. We determined numerically whether the TLS switches or not in a standard way by comparing $W_{ij}\sn \delta t$ with a random number from the uniform distribution $U(0,1)$. 

The generated states of the TLS  produce an array $d\sn(m)$ of random numbers that take values $\pm 1$. Here $m$ enumerates the time steps. We collect $N=10^5$ outcomes of the simulated Ramsey measurements , which means that we use $\tilde{N}=(\ts/\delta t)\times 10^5$ samples, that is, $1\leq m\leq \tilde{N}$. A $k$th Ramsey measurement, which is done in the  time interval $k\ts \leq t \leq k\ts + t_R$, corresponds to the range of steps $k(\ts/\delta t) < m  \leq k(\ts/\delta t) + (t_R/\delta t)$. 
For our discretized time sequence, the random phase accumulated by the qubit in this time interval is  
\begin{align}
    \th_k = \sum_{m=1}^{\lceil t_R/\delta t\rceil}\sum_{n=1}^{\NTLS} V\sn d\sn(\lfloor k\ts/\delta t\rfloor+m).
\end{align}

We used the probability of obtaining ``1'' in a $k$th measurement $p(\theta_k)$ as given by Eq.~(\ref{eq:standard_probability}). We compared $p(\theta_k)$ with a random numbers $o_k$ from $U(0,1)$. If $p(\theta_k)>o_k$, we set the outcome of the $k$th Ramsey measurement to $x_k=1$, otherwise we set $x_k=0$. 
The whole procedure was independently repeated 300 times for statistical averaging. This allowed us to numerically analyze and compare  the parameters $r_1, r_2(k), r_3(k,l)$ with the theory, as well as to investigate other parameters of interest, as discussed in the main text. 

Of primary interest to us was the analysis of the TLSs that produce noise of the $1/f$ type in a reasonably broad frequency range. There are many ways to obtain such noise. The results presented in the main text refer to the noise in which the coupling of the TLSs to the qubit was the same for all TLSs, $V\sn = V$, but the distribution of the switching rates $W\sn = W_{01}\sn + W_{10}\sn$ was log-normal. We also simulated other types of the TLSs. They lead to qualitatively similar results.


\subsection{Simulating Gaussian noise}
\label{simulations_Gaussian}

The effect of a Gaussian noise on the outcomes of Ramsey measurements is fully characterized by the correlation function $f_k = \langle \theta_n\theta_{n+k}\rangle$ of the phases $\th_k$ acquired by the qubit in the measurements. Since the measurements are periodically repeated, one has to  sample $\th_k$ for successive $k$. The probability to have a given  phase $\theta_k$ depends on the entire history of the previously ``observed'' phases $\theta_{k'}$ with $k'<k$. This means that the quantity of interest is the conditional probability 
\begin{align}
\label{eq:phase_conditional}
& P(\th_k|\th_0,...,\th_{k-1}) = P(\th_0,...,\th_k)\nonumber\\
&\times [P(\th_0,...,\th_{k-1})]^{-1}
\end{align}
where $\theta_0$ is the outcome of the first measurement.
The probability (\ref{eq:phase_conditional}) has to be evaluated recursively starting with the probability of $\theta_0$. We will use that the distribution of the phases is stationary, 
\[P(\{\theta\}) = Z^{-1}\exp\left[-\frac{1}{2}\sum_{k,k'}(\hat f{}^{-1})_{k-k'}\theta_k\theta_{k'}\right],\]
where $Z$ is the normalization constant and $\hat f{}^{-1}$ is the matrix reciprocal to the matrix $f_{kk'}$; we note that the matrix elements of the latter matrix are $f_{kk'}= f_{|k-k'|}$, cf. Eq.~(\ref{eq:phase_distribution}). 

In evaluating the conditional probability given by Eq.~(\ref{eq:phase_conditional}) one should keep in mind that the values of $\th_k$ are correlated at a finite distance $k_\mathrm{corr}$, which is determined by the relation between the correlation time of the underlying noise and the period of the sequence $\ts$. In other words, it means that, to a good approximation (which needs to be checked)  $|f_k| $ can be set equal to zero for $k>k_\mathrm{corr}$. Then for $k > k_\mathrm{corr}$ one can approximate the conditional probability $ P(\th_k|\th_0,...,\th_{k-1})$ with $P(\th_k|\th_{k-k_\mathrm{corr}}, \th_{k-k_\mathrm{corr}+1},...,\th_{k-1})$, i.e., instead of the probability (\ref{eq:phase_conditional}) we have to calculate
%
\begin{align}
\label{eq:drawing_theta_distribution}
&P(\th_k|\th_{k-k_\mathrm{corr}},...,\th_{k-1}) = P(\th_{k-k_\mathrm{corr}},...,\th_k) \nonumber\\
&\times[P(\th_{k-k_\mathrm{corr}},...,\th_{k-1})]^{-1}.
\end{align}

It is important that all conditional probabilities in Eq.~(\ref{eq:drawing_theta_distribution}) have a Gaussian form, albeit they are not zero-mean, because of the correlations. As we now show, we can sample each $\th_k$ from $\mathcal{N}(\mu_k, \sigma_k)$, where $\mu_k$ and $\sigma_k$ are respectively the mean value and the standard deviation for $\theta_k$. They depend on the values of $\theta_{k'}$ with $k'<k$.

The first phase to be sampled, $\th_0$, is sampled with $\mu_0 = 0$ and $\sigma_0 = 1/\sqrt{f_0}$. To find the distribution of the phases $\th_k$ with $k>0$ we note that, when calculating $P(\th_{k-k_\mathrm{corr}},...,\th_{k})$, rather than using the full reciprocal matrix $(\hat f{}^{-1})_{|k-k'|}$ we should use a $(k_\mathrm{corr}+1)\times (k_\mathrm{corr}+1)$ matrix reciprocal to the $(k_\mathrm{corr}+1)\times (k_\mathrm{corr}+1)$ part of the matrix $f_{kk'}$. This matrix $\Psi_{kk'}$ is defined by the equation 
\[\sum_{m=k-k_\mathrm{corr}}^{k} \Psi_{km}f_{mk'} = \delta_{kk'},\]
Along with $\Psi_{kk'}$ we need the matrix $\psi_{kk'}$, which is the reciprocal of  the  $k_\mathrm{corr}\times k_\mathrm{corr}$ part of $f_{kk'}$, 
\[\sum_{m=k-k_\mathrm{corr}}^{k-1} \psi_{km}f_{mk'} = \delta_{kk'}.\]

The matrices $\Psi_{kk'}$ and $\psi_{kk'}$ are symmetric. However,  even though $f_{kk}=f_{0}$ is independent of $k$, the diagonal matrix elements of the matrices  $\Psi_{kk}$  and $\psi_{kk}$ depend on $k$.

There is an important relation between the matrices $\hat\psi$ and $\hat\Psi$,
\begin{align}
\label{eq:reciprocal_relation}
&\psi_{kk'} =\Psi_{kk'} - (\Psi_{mk}\Psi_{mk'}/\Psi_{mm})
\nonumber\\
&k,k'=m-k_\mathrm{corr}, \ldots, m-1;
\end{align}
This relation can be checked by multiplying from the left by $f_{k_1k}$ and summing over $k=m-k_\mathrm{corr},\ldots,m-1$.

Taking the relation (\ref{eq:reciprocal_relation}) into account, we can write the exponential in $P(\th_{k-k_\mathrm{corr}},...,\th_k)$ as
\begin{align*}
&\exp\left[-\frac{1}{2}\Psi_{kk}(\th_k-\mu_k)^2 \right.\\
&\left.- \frac{1}{2}\sum_{m,m'=k-k_\mathrm{corr}}^{k-1}\psi_{mm'}\th_m\th_{m'}\right],\nonumber\\
\end{align*}
where
\begin{align}
\label{eq:mean_phase_picked}
\mu_k= -\Psi_{kk}^{-1}\sum_{m=k-k_\mathrm{corr}}^{k-1}\Psi_{km}\th_m
\end{align}
is the phase accumulated over $k_\mathrm{corr}$ steps that preceded the $k$th step.

Ultimately, we have for the conditional probability of $\th_k$  the distribution
\begin{align}
\label{eq:full_drawing_theta}
&P(\th_k|\th_{k-k_\mathrm{corr}}, \th_{k-k_\mathrm{corr}+1},...,\th_{k-1}) \nonumber\\
&=(\Psi_{kk}/2\pi)^{1/2} \exp\left[-\frac{1}{2}\Psi_{kk}(\th_k-\mu_k)^2\right]
\end{align}
We have also used here the Cramer rule that relates the matrix element $\Psi_{kk}$ to the ratio of the determinants of the matrices $\hat \Psi$ and $\hat \psi$.

The above prescription allowed us to sample a sequence of random phases $\th_k$. Each obtained $\th_k$ was used to determine whether the outcome of the simulated Ramsey measurement gives ``0'' or ``1'' based on the probability \eqref{eq:standard_probability}. From the observed outcomes, we could calculate $r_1, r_2(k)$, and $r_3(k,l)$ as well as other statistical characteristics of the simulated sequence of periodic Ramsey measurements. 

The value of $k_\mathrm{corr}$ depends on a particular type of noise. We chose it in such a way that the results became virtually independent of the choice. For exponentially correlated frequency noise with the correlation time $\tau_\mathrm{corr}$ one can choose $k_\mathrm{corr}=a_\mathrm{corr} \tau_\mathrm{corr}/\ts$ with a sufficiently large  $a_\mathrm{corr}$. For the $1/f$-type noise we studied, with the characteristic minimal frequency $\omega_{\min}$, one can  choose $k_\mathrm{corr}=a_{1/f}( \omega_{\min}\ts)^{-1}$ with a sufficiently large $a_{1/f}$. However, for the parameters used in the simulations it was checked that the results become independent of  $k_\mathrm{corr}$ once one sets  $k_\corr\approx 10$ for the exponentially correlated noise and  $k_\corr=400$ for the $1/f$ noise.


\section{Power spectrum of the TLS-induced frequency fluctuations}\label{sec:TLS_power_spectrum}

The power spectrum of the qubit frequency fluctuations due to the coupling to TLSs becomes particularly simple if the TLSs are symmetric, the coupling is the same for all TLSs, and the distribution of the switching rates is log-normal, i.e., the probability density of having a given switching rate $W$ is $\propto 1/W$. Such a distribution is natural if one thinks of a uniform distribution of the barriers which the TLSs have to overcome in switching. For a finite set of the TLSs the log-normal distribution is mimicked by the switching probabilities of the form
\begin{equation}
\label{eq:exponential_rates}
    W^{\sn}t_R = \exp\left(-\alpha (n+n_0)\right),
\end{equation}
where $\alpha$ and $n_0$ are control parameters. In the simulations  we used $\alpha =0.75$ and considered $n_0=0, 5$ or $7$.

\begin{figure}
    \centering
    \includegraphics[width=0.47\textwidth]{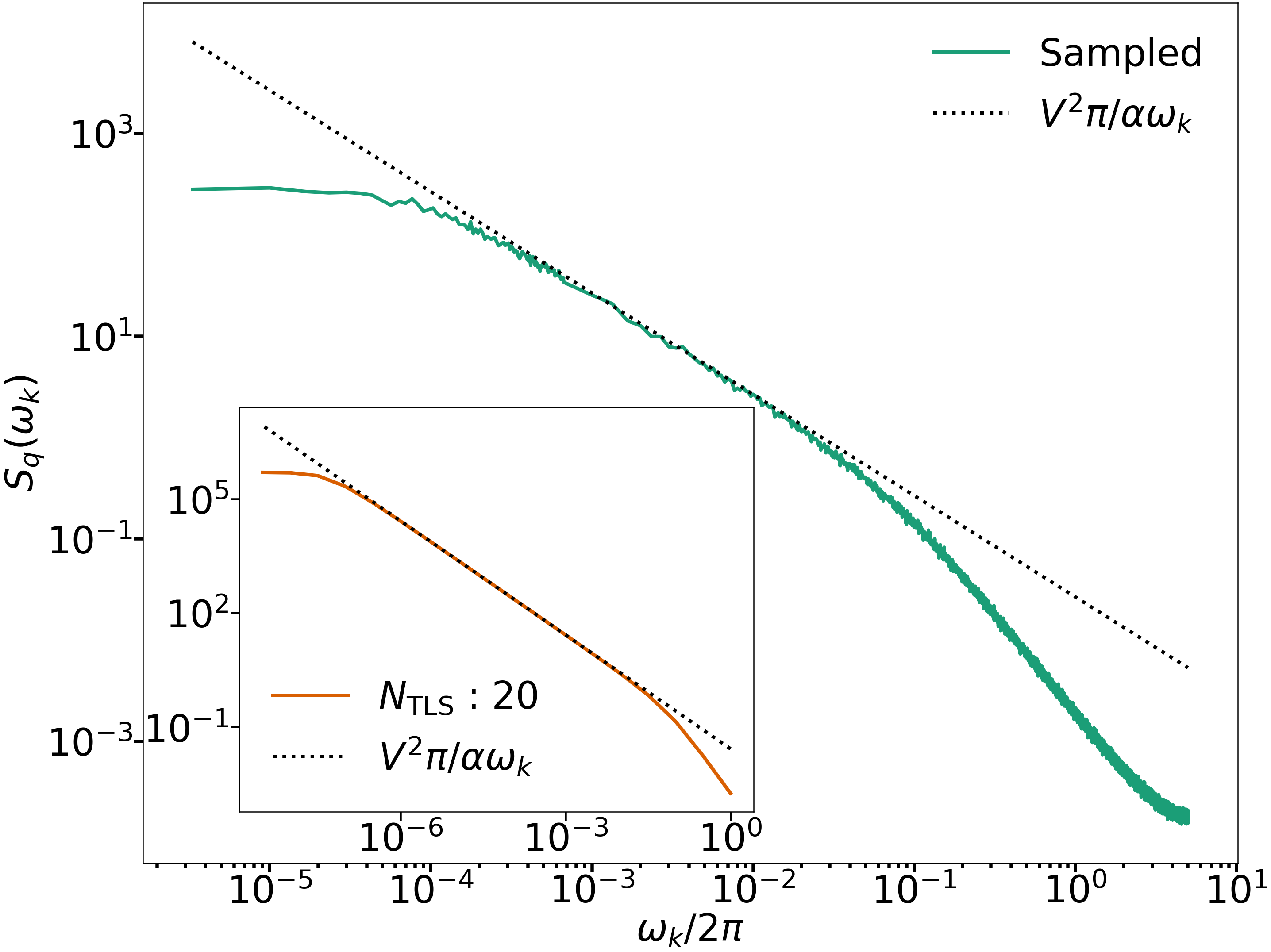}
    \caption{Simulated power spectrum of 10 symmetric TLSs with $V\sn=V=0.2/t_R$ and $W_{01}^{\sn}=W_{10}^{\sn}=\exp(-\alpha n)/t_R$ for $n=1,2,\ldots, 10$ and $\alpha = 0.75$ (green data points). The spectrum displays $1/f$ character over more than two decades. The results agree with the analytical result (\ref{eq:TLS_power_spectrum}). Inset: the power spectrum for 20 TLSs, $n=1,...,20$ with the same $V$ and $\alpha$ as given by Eq.~(\ref{eq:TLS_power_spectrum}).  In this case the spectrum displays more than five decades of $1/f$ behavior.  The dashed lines in the main plot and in the inset show $1/f$ noise.}
    \label{fig:PS_10TLS}
\end{figure}

The selected switching rates correspond to the power spectrum $S_q(\omega)$,  which is a superposition of Lorentzian peaks given by Eq.~\eqref{eq:TLS_power_spectrum}. This superposition provides a $1/f$-type behavior. It is clearly  seen in Fig.~\ref{fig:PS_10TLS} already for 10 TLSs, in which case the $1/f$ spectrum spreads over two decades in  frequency. The range of $1/f$-type behavior can be extended by altering $\alpha$ and $n_0$  or by including more TLSs. The inset  in Fig.~\ref{fig:PS_10TLS} shows that, for 20 TLSs, $1/f$-behavior spreads over five decades.

The spectrum of the main plot in~Fig.~\ref{fig:PS_10TLS} is obtained via sampling the TLSs states $d^{\sn}(m)=\pm 1$ with a time step $\delta t$. At each time step $m$ the outcome is 
\begin{equation}
    D(m) = \sum_{n=1}^{\NTLS}V^{\sn} t_R d^{\sn}(m) .
\end{equation}
The fast Fourier transform of $D(m)$ is calculated in the standard way as
\begin{align}
\label{eq:FFT_of_D}
\tilde D(\omega_k) = \left|\sum_m \exp(-2\pi i m k/\tilde N)D(m)\right|^2/\tilde N
\end{align}
where $\omega_k=2\pi k/\tilde N$ and, as before, $\tilde N = N\ts/\delta t$ with $N=10^5$. 

To obtain the power spectrum $S_q(\omega_k) = \langle \tilde D(\omega_k)\rangle$, this procedure is repeated 300 times to get sufficient statistics for averaging. 
The result is shown in the main plot in~Fig.~\ref{fig:PS_10TLS}. For weak couplng to the TLSs the spectrum $S_q(\omega)$ is immediately related to the centered correlator $\tilde r_2(k)$.


\section{Broadening of the distribution by the noise correlations}
\label{sec:distribution_broadening}

Noise correlations lead to broadening of the distribution of the measurement outcomes compared to the standard binomial distribution. Here we illustrate this well-known effect for our setting. For a random variable $x_n$ that takes on the values 0 and 1, the probability  of observing $m$ ones in $M$ measurements is
\begin{equation}
    \langle m/M\rangle \equiv M^{-1}\mathbb{E}\left[\sum_{n=1}^M x_n\right]  = r_1,
\end{equation}
whereas the variance is 
\begin{align}
\label{eq:variance_correlated}
    &\sigma_M^2 = \langle (m/M)^2\rangle - r_1^2 \equiv
M^{-2}\mathbb{E}\left[\left(\sum_{n=1}^M x_n\right)^2\right]- r_1^2
 \nonumber\\
& = M^{-1}r_1(1-r_1) 
+ 2M^{-2}\sum_{k=1}^{M-1}(M-k)\tilde{r}_2(k).
\end{align}
For the case where $M\ts$ is much larger  than the correlation time of the noise one should replace in the second line $M-k$ with $M$. 

The term $r_1(1-r_1)/M$ in Eq.~(\ref{eq:variance_correlated}) is the standard variance of the binomial distribution. It falls off as $M^{-1}$ with the increasing number of measurements $M$. Equation~(\ref{eq:variance_correlated})  shows that the correction to the binomial distribution due to the noise correlations, which is determined by the centered correlator $\tilde r_2(k)$, also scales as $M^{-1}$. This correction can be significant. The change of the standard deviation due to the noise correlation for several types of noise that we studied is seen from Table~\ref{table:broadening}.

\begin{table}[h]
    \begin{tabular}{llcc}
   Noise type & Setup &   $M^{1/2}\sigma_\mathrm{simul} $ & $M^{1/2}\sigma_\mathrm{binom}$ \tabularnewline
      
    \hline 
    \hline 
  TLS&  $n_0=5$ &   1.283  & 0.379  \tabularnewline
  TLS&  $n_0=7$ &   1.696  & 0.379 \tabularnewline
    Gaussian EC  &    $\tau_{\corr}/t_R= 20$ &   0.605 & 0.379 \tabularnewline
  Gaussian EC & $\tau_{\corr}/t_R= 100$ &   1.120 & 0.379 \tabularnewline
   Gaussian $1/f$ &  $\omegaMin t_R= 1\times 10^{-3}$ &   0.956 & 0.379 \tabularnewline
  Gaussian $1/f$ & $\omegaMin t_R= 1\times 10^{-5}$ &   1.378 & 0.379 \tabularnewline
    \end{tabular}
    \caption{Standard deviations of the  simulated probabilities of the measurement outcomes scaled by the number of measurements $M^{1/2}\sigma_\mathrm{simul}$ and of the binomial distributions $M^{1/2}\sigma_\mathrm{binom}$ for several types of noise. The results for the noise from the TLS refer to 4 symmetric TLSs with scaled coupling $V\sn t_R =Vt_R=0.2$ and the switching rates $W^{\sn}t_R=\exp[-0.75(n+n_0)]$ ($n=1,2,3,4$). The results for the Gaussian exponentially correlated noise (Gaussian EC) and the $1/f$ noise with the spectrum (\ref{eq:1_f_spectrum}) refer to the noise intensities that give the phase variance  $\langle \theta_0^2\rangle \equiv f_0=0.16$.  
   }
   \label{table:broadening}
\end{table}

\begin{figure}
    \centering
    \includegraphics[width=0.45\textwidth]{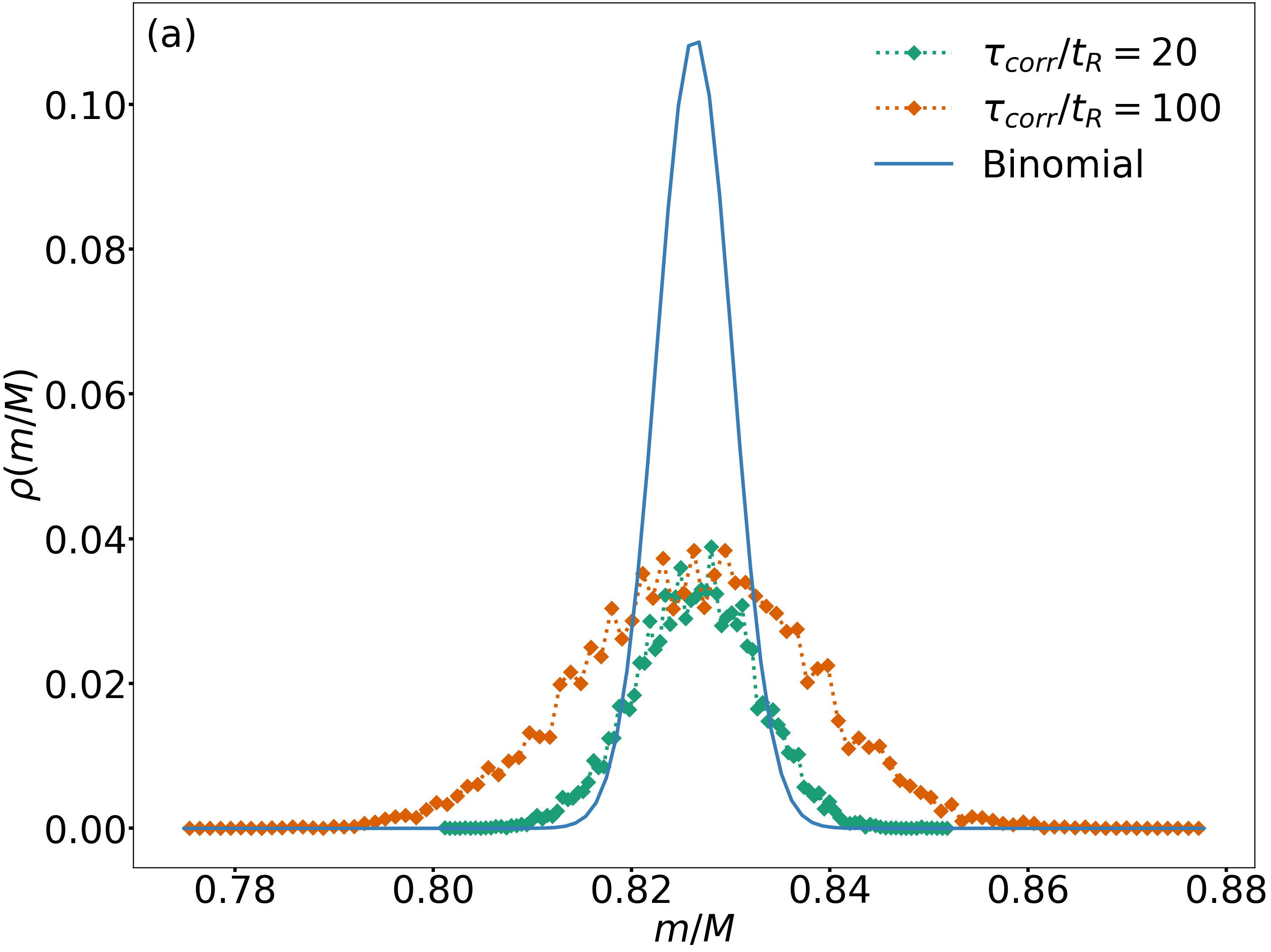}
        \includegraphics[width=0.45\textwidth]{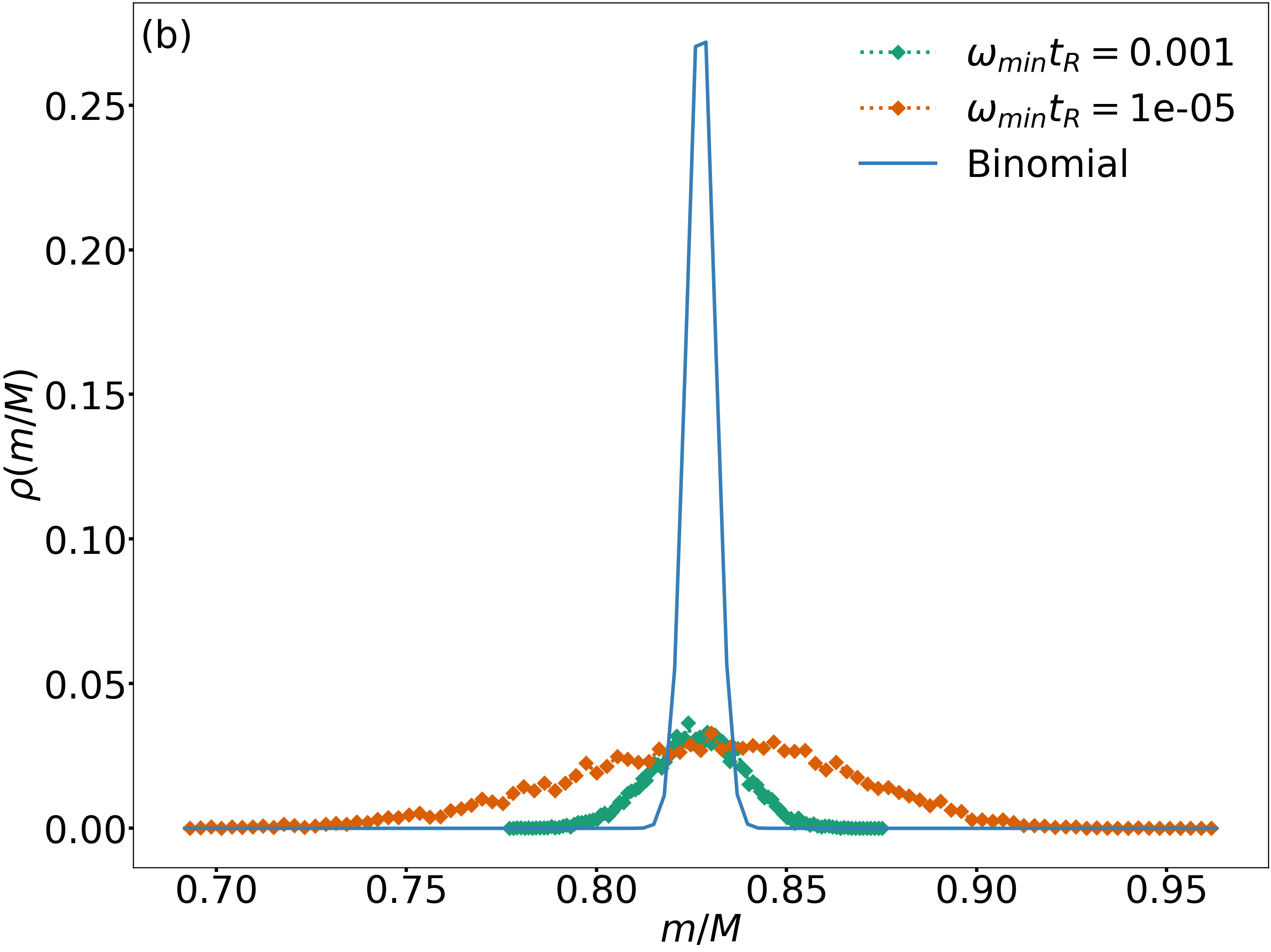}
    \includegraphics[width=0.45\textwidth]{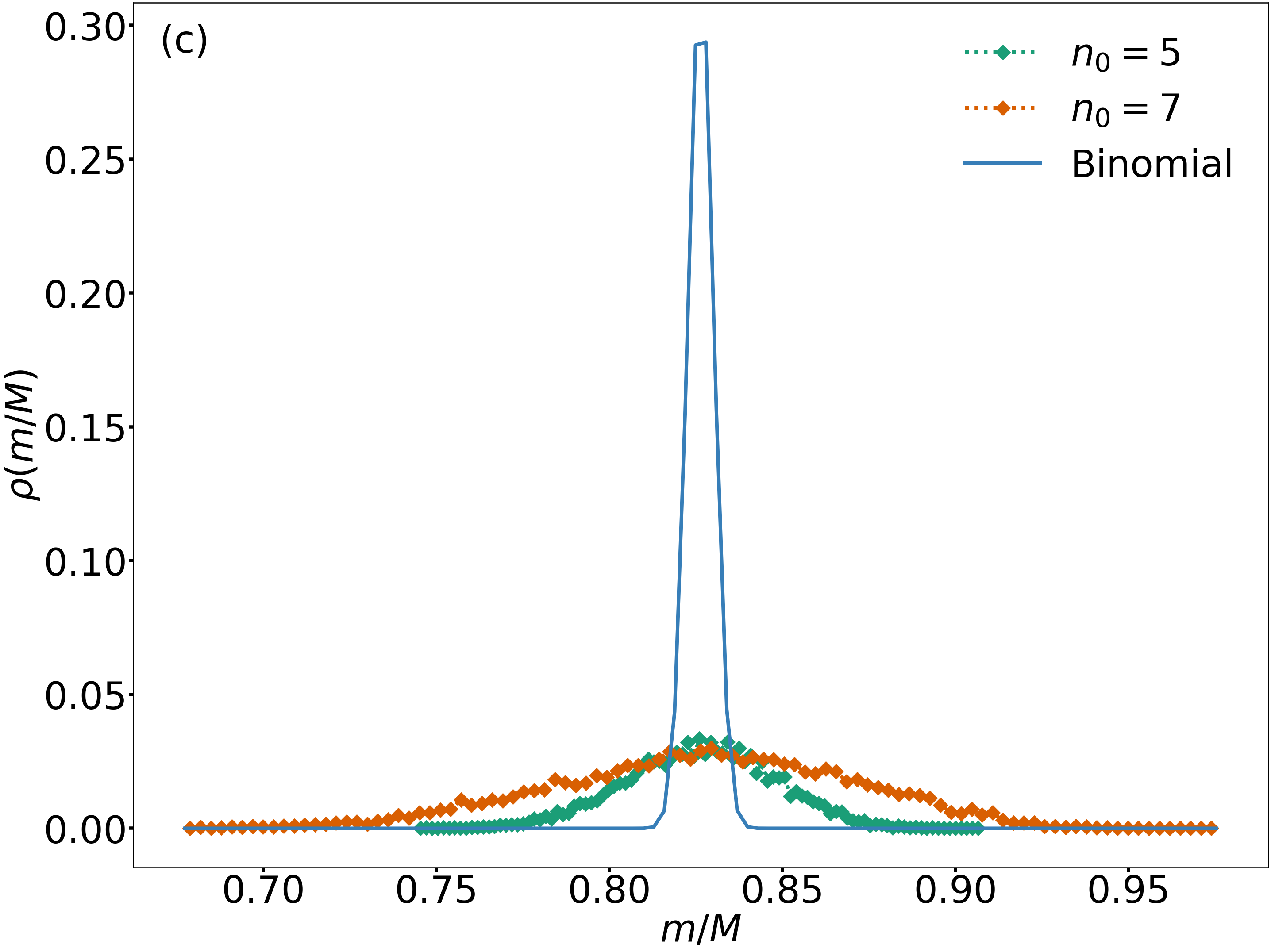}
    \caption{Broadening of the distribution of the outcomes of $M=10^4$ measurements. Panels (a) and (b) refer to the exponentially correlated and $1/f$ noise, respectively; the noise intensity is chosen such that the phase variance is $\langle \theta_0^2\rangle \equiv f_0=0.16$. (c) noise from 4 TLSs with the same parameters as in Table \ref{table:broadening}. Solid lines show the Gaussian distribution, which approximates the binomial distribution with  $r_1=0.826$ and $M=10^4$. Each histogram is aggregated into 100 bins, where the lower and upper ends are determined by the minimal and maximal sampled values, respectively. The theoretical curves use the same binning.}
    \label{fig:broadening_all}
\end{figure}

The effect of the correlations on the variance can be studied by evaluating the probability distribution $\rho(m|M)$ of having $m$ ones in $M$ measurements. This distribution for $M=10^4$ is shown in Fig.~\ref{fig:broadening_all} for several types of noise. It was obtained by repeating $10^4$ measurements $10^5$ times to have a good statistics. The distribution is compared with the binomial distribution for the same $M=10^4$, which is very well approximated by a Gaussian distribution $\propto \exp[-(m-Mr_1)^2/2Mr_1(1-r_1)]$.


\section{TLS dynamics during the Ramsey measurement}
\label{sec:C_eqns}

We consider here the dynamics of the TLS-dependent components of the density matrix during the Ramsey measurement, i.e., in the time intervals $n\ts<t< n\ts+t_R$. These components, $\rho_\lambda$ with $\lambda = I,z,\pm$, are defined in Eqs.~(\ref{eq:q_TLS_full}) and (\ref{eq:many_TLS_general}). As indicated in the main text, the equations for $\rho_\lambda$ with different $\lambda$ are obtained by substituting Eq.~(\ref{eq:many_TLS_general}) into the full master equation (\ref{eq:master_TLS}), multiplying the left- and right-hand sides by $\hI, \hI_q-\sigma_z, \sigma_\pm$, and taking trace over the qubit states. The left-hand side of the  resulting equation for $\rho_\lambda$ is
\begin{align}
\label{eq:rho_lambda_dot}
\partial_t\rho_\lambda =\sum_n \sum_{\varkappa=0,z}\dot C_{\lambda\varkappa}\sn{}\tT_\varkappa\sn\prod_{m\neq n} \sum_{\varkappa'=0,z}C_{\lambda\varkappa'}^{(m)}\tT_{\varkappa'}\sm.
\end{align}
As seen from Eq.~(\ref{eq:master_TLS}, the term  $\sum_n \mathcal{L}\sn \rho_\lambda$ in the right-hand side of the equation for $\partial_t\rho_\lambda$  has the same structure as Eq.~(\ref{eq:rho_lambda_dot}): a sum over $n$ multiplied by the product over $m\neq n$ of $\sum_{\varkappa'=0,z}C_{\lambda\varkappa'}^{(m)}\tT_{\varkappa'}\sm$. The commutator $[\rho, H_\mathrm{q-TLS}]$ also has the same structure. One can divide the both sides of the equation for $\partial_t\rho_\lambda$ by $\rho_\lambda$, reminiscent of the standard trick of separation of variables in a differential equation. This gives equations (\ref{eq:single_TLS_decay}).

The equations for $C_{\lambda\varkappa}\sn$ are split into sets of pairs of coupled equations.  For $\lambda=0,z$ we have
\begin{align}
\label{eq:eom_many_TLS_0}
&\dot C_{\lambda 0}\sn=0, \quad \dot C_{\lambda z}\sn = \Delta W\sn{} C_{\lambda 0}\sn{} - W\sn{}C_{\lambda z}\sn{}\ ,\nonumber\\
&\Delta W\sn{} = W_{10}\sn{} - W_{01}\sn{}\ .
\end{align}
The parameter
$\Delta W\sn{}$ characterizes the asymmetry of the TLS, whereas $W\sn{} = W_{10}\sn{} + W_{01}\sn{}$ is the TLS relaxation rate.

If $\lambda=+$, Eq.~(\ref{eq:single_TLS_decay}) for $C_{\lambda\varkappa}\sn$ reads
\begin{align}
\label{eq:dot_C_pm_equation}
&\dot C_{+0}\sn{} = iV\sn{} C_{+z}\sn{},\nonumber\\
&\dot C_{+z}\sn{} = (\Delta W\sn{}+iV\sn{} )C_{+0}\sn{} - W\sn{}C_{+z}\sn{}\ ,
\end{align}
and by construction, $C_{-\varkappa}\sn = C_{+\varkappa}\sn{}^*$.

Solving these equations, we obtain for $\lambda = I,z$
\begin{align}
\label{eq:solution_many_TLS_0}
& C_{\lambda z}\sn{}(t) =
\left[C_{\lambda z}\sn{}(0^+) - \frac{\Delta W\sn{}}{W\sn{}}C_{\lambda 0}\sn{}(0^+)\right]e^{-W\sn{}t}
\nonumber\\ 
&+ \frac{\Delta W\sn{}}{W\sn{}}C_{\lambda 0}\sn{}(0^+), \quad C_{\lambda 0}\sn{}(t)=C_{\lambda 0}\sn{}(0^+)\ ,
\end{align}
whereas
\begin{align}
\label{eq:sigma_plus_many}
&C_{+ \varkappa}\sn{}(t)=\sum_{k=1,2}B_\varkappa^{(k; n)}\exp(\nu_k\sn{}t),\nonumber\\
&\nu_{1,2}\sn{}=-\frac{1}{2}W\sn{} \pm \gamma\sn,\nonumber\\
&\gamma\sn=\frac{1}{2}\left[(W\sn{})^2 +4iV\sn{} (\Delta W\sn{}+iV\sn{} )\right]^{1/2}\ ,
\end{align}
here
\begin{align}
\label{eq:initial_cond_for_B}
&B_0^{(k;n)}= [\nu_{3-k}\sn C_{+0}\sn{}(0^+)-iV\sn{}  C_{+z}\sn{}(0^+)]\nonumber\\
&\times (\nu_{3-k}\sn{} - \nu_k\sn{})^{-1},\nonumber\\
&B_z^{(k;n)} = -i(\nu_k\sn{}/V\sn{} )B_0^{(k;n)} \quad (k=1,2).
\end{align}
These expressions are used in the main text to obtain $C_{\lambda\varkappa}\sn(t)$ in the explicit form.

To find the initial conditions for the equations for $C_{\lambda\varkappa}\sn$ in the above expressions we take into account that at the instant $t=0^+$  the qubit is in the state $(\ket{0} + \ket{1})/\sqrt{2}$, whereas the stationary TLS populations are given by Eq.~(\ref{eq:TLS_populations}). Therefore
\begin{align}
\label{eq:after_pulse_multi_TLS}
&C_{Iz}\sn{}(0^+) =C_{\pm z}(0^+) =- C_{zz}\sn{}(0^+)= \Delta W\sn/2W\sn\ ,\nonumber\\
&C_{I0}\sn{}(0^+)=C_{\pm 0}(0^+)= -C_{z0}\sn{}(0^+)=\frac{1}{2},
\end{align}
With these initial conditions, we have in particular 
\begin{align}
\label{eq:C_coeff_explicit}
&C_{Iz}\sn{}(t_R) = - C_{zz}\sn{}(t_R)=
\Delta W\sn{}/2W\sn{}\ , \nonumber\\
&C_{I0}\sn{}(t_R)= -C_{z0}\sn{}(t_R)=\frac{1}{2}\ ,  
\end{align}
and 
\begin{align}
\label{eq:C_+z}
&C_{+z}\sn{}(t_R) =\frac{1}{2} e^{-W\sn{}t_R/2}\left[\frac{\Delta W\sn{}}{W\sn{}}\cosh(\gamma\sn t_R)
\right. \nonumber\\
&\left. +\left(i\frac{V\sn{} }{\gamma\sn}+\frac{\Delta W\sn{}}{2\gamma\sn}\right)\sinh(\gamma\sn t_R) \right];
\end{align}
the expression for $C_{+0}\sn(t_R)$ is given in the main text, Eq.~(\ref{eq:Xi_n}).


\section{Time evolution of the pair correlator}
\label{sec:K_term_evolution}

The operator $\Rcorr$ defined in Eq.~(\ref{eq:Rcorr}) describes the decay of the pair correlations function of Ramsey measurement. The parameters $\mK$ in the expression for $\Rcorr$ read 
\begin{align}
\label{eq:K_coefficients}
&\mathbb{K}_{\{m\}_s}=\frac{1}{2}e^{-t_R/T_2}\mathrm{Re}\Bigl[e^{i\tilde\phi_R} \mathcal{J}^{(m_1)}...\mathcal{J}^{(m_s)} \nonumber\\
&\times \prod_{n\neq m_1,...,m_s}e^{-W\sn t_R/2}\Xi\sn(t_R)\Bigr]
\end{align}
where
\begin{align}
\label{eq:L_many}
&\mathcal{J}\sn{}= C_{+z}\sn{}(t_R) - \frac{\Delta W\sn{}}{W\sn{}}C_{+0}\sn{}(t_R),\nonumber\\
&=2 i\frac{V\sn{} }{\gamma\sn}\,\frac{W_{01}\sn{} W_{10}\sn{}}{W\sn{}{}^2}e^{-W\sn{}t_R/2}\sinh \gamma\sn t_R
\end{align} 
We use here the Hamiltonian $H_R$ in which $\phi_R$ was replaced by the auxiliary phase $\tilde\phi_R$ to compensate the coupling-induced shift of the average qubit frequency.


\subsection{Time evolution of the terms $\propto \sigma_\pm$ in $\Rcorr$}

 As indicated in the main text, the evolution of the terms $\propto \sigma_\pm$ in $\Rcorr(t)$ in the time interval $k\ts < t < k\ts+t_R$  have to be studied separately for each term in the sum over $m_1,..., m_s$ in $\Rcorr(t)$. We denote the corresponding terms as $\Rcorr(t|\{m\}_s)$,
 \[\Rcorr(t) = \sum_{s\geq 1}\sum_{\{m\}_s}\Rcorr(t|\{m\}_s).
 \]
 Similar to Eqs.~(\ref{eq:many_TLS_general}) and (\ref{eq:many_TLS_Schmidt}), we seek $\Rcorr(t|\{m\}_s)$ in the form  
\begin{align}
\label{eq:many_rho_relevant_m_term}
&\Rcorr(t|\{m\}_s) = \frac{1}{4}e^{-(t-k\ts)/T_2}
\mathbb{K}_{\{m\}_s}
\sum_{\alpha=\pm}
\sigma_\alpha \nonumber\\
&\times\prod_n\sum_{\varkappa=0,z} \tilde C_{\alpha\varkappa}\sn(t|\{m\}_s)\tT_\varkappa\sn\ .
\end{align}
The equations for the coefficients $\tilde C_{\alpha\varkappa}\sn{}(t|\{m\}_s)$ have the same form as Eqs.~(\ref{eq:dot_C_pm_equation}) for $C_{\alpha\varkappa}\sn(t)$. The initial conditions for these equations are set at $t=\ts^+$. They follow from the expression (\ref{eq:Rcorr_prime}) for $\Rcorr(k\ts)$ and Eq.~(\ref{eq:transform_qubit}). They are different for the values of $n$ that coincide with one of the components $m_i$ of the vector $\{m\}_s$ and for those values of $n$ that differ from the components of $\{m\}_s$. For the first group we have 
\begin{align}
\label{eq:C_initial_mi}
\tilde C_{+0}^{(m_i)}(k\ts^+|\{m\}_s) =0, \qquad 
\tilde C_{+z}^{(m_i)}(k\ts^+|\{m\}_s)\nonumber\\
=\exp [-W^{(m_i)}(k\ts - t_R)], \quad m_i\in \{m\}_s\ .
\end{align}
In contrast, for the second group we have
\begin{align}
\label{eq:C_initial_non_mi}
&\tilde C_{+0}\sn{}(k\ts^+|\{m\}_s) =1/2, \quad \tilde C_{+z}\sn{}(k\ts^+|\{m\}_s)\nonumber\\
&= \Delta W\sn{}/2W\sn{} \quad n\neq m_1,...,m_s\ .
\end{align}
It then follows from Eqs.~(\ref{eq:solution_many_TLS_0}) - (\ref{eq:initial_cond_for_B}) that
\begin{align}
\label{eq:coeff_at_tau_z}
&\tilde C_{+0}^{(m_j)}(k\ts +t_R|\{m\}_s) =i(V_{m_j}/\mu^{(m_j)})\sinh \mu^{(m_j)}t_R\nonumber\\
&\times \exp\left[-W^{(m_j)}(k\ts -t_R /2)\right]  , \quad m_j\in\{m\}_s\ ,
\end{align} 
whereas 
\begin{align}
\label{eq:coeff_at_I_TLS}
&\tilde C_{+0}\sn{}(k\ts + t_R|\{m\}_s)=\frac{1}{2}\Xi\sn(t_R)\ , \nonumber\\
& n\notin\{m\}_s\ ,
\end{align}
where $\Xi\sn(t_R)$ is given by Eq.~(\ref{eq:Xi_n}). 

At time $k\ts + t_R$ the qubit undergoes a rotation about the $z$-axis by the angle $\tilde\phi_R$ followed by a $\pi/2$ rotation about the $y$-axis, as seen from Eqs.~(\ref{eq:Hamiltonian}) and (\ref{eq:H_R}). As a result the matrices $\sigma_\alpha$ in the operator $\Rcorr(k\ts+t_R|\{m\}_s)$, Eq.~(\ref{eq:many_rho_relevant_m_term}), transform as 
\[\sigma_\alpha \to e^{i\alpha\tilde\phi_R} (i\alpha\sigma_y-\sigma_z)\ .
\]
This transformation corresponds to the transformation of $\Rcorr(k\ts+t_R|\{m\}_s)$ into $\Rcorr(k\ts+t_R^+|\{m\}_s)$. As indicated in the main text, if the qubit is reset at $k\ts+t_R^+$, then $\Rcorr(k\ts+t_R^+|\{m\}_s)$ decays. However, if the measurement is performed at time $k\ts+t_R^+$, i.e., $\Rcorr(k\ts+t_R^+|\{m\}_s)$ is multiplied by $\hat\pi= (\hI_q-\sigma_z)/2$ and the trace is taken over the qubit and the TLSs, the contribution of $\Rcorr(k\ts+t_R^+|\{m\}_s)$ to the result of the measurement is
\begin{align}
\label{eq:r_2_tilde_C}
&\tilde r_2(k)= \frac{1}{2}e^{-t_R/T_2}\sum_s\sum_{\{m\}_s}
\mathbb{K}_{\{m\}_s}
\nonumber\\
&\times\mathrm{Re}\left[e^{i\tilde\phi_R}\prod_n 2\tilde C_{+0}\sn(k\ts+t_R|\{m\}_s)\right]\ .
\end{align}
This result depends only on the coefficients $\tilde C_{+\varkappa}\sn$ with $\varkappa=0$, which describe the contribution of the unit operators of the TLSs; the trace of the terms proportional to $\tT_z\sn$ is zero.



\section{Explicit expressions for the phase correlators for $1/f$ noise}
\label{sec:1_f_integrals}

For completeness, here we provide the explicit expressions for the correlators $f_k= \langle \theta_0 \theta_k\rangle$ of the phases accumulated by the qubit during Ramsey measurements separated by $k$ cycles, i.e., separated by the time $k\ts$. The expressions are related to  Gaussian noise $\omq(t)$ produced by a large number of TLSs with the same coupling to the qubit and with the log-normal distribution of the switching rates.  The power spectrum $S_q(\omega)$ of $\omq(t)$ in this case is given by Eq.~(\ref{eq:1_f_spectrum}).

The integral over $\omega$, Eq.~(\ref{eq:phase_correlation}), that relates $f_k$ to $S_q(\omega)$, can be  expressed in terms of the exponential integral and hyperbolic sine and cosine integral functions $\mathrm{Ei}(z), \mathrm{shi}(z)$, and $\mathrm{chi}(z)$ as follows
\begin{widetext}
\begin{align}
\label{eq:Yaxing_1_f_k}
&  f_{k} = \frac{ D_{\rm fl }t_R^2}{2\pi}\{2e^{-a_k b_{\min}}[(1-a_kb_{\min})[\cosh b_{\min}-1]
 +b_{\min}\sinh b_{\min}]/b_{\min}^2   \nonumber\\
    & + 2a_k^2 \mathrm{Ei}(-a_kb_{\min})  - (a_k+1)^2 \mathrm{Ei}(-a_k b_{\min}-b_{\min}) - (a_k-1)^2\mathrm{Ei}(b_{\min}-a_k b_{\min})  \}, \quad k > 0\ ,
 \end{align}
 and
 \begin{align}
 \label{eq:Yaxing_1_f_0}
& f_{0} =\frac{D_{\rm fl}t_R^2}{\pi}[-\frac{1}{b_{\min}^2}+\frac{2}{b_{\min}}-(b_{\min}-1)\exp(-b_{\min})/b_{\min}^2 - \mathrm{chi}(b_{\min}) + \mathrm{shi}(b_{\min})]\ .
\end{align}
\end{widetext}
These expressions depend on two dimensionless parameters 
\[a_k = \frac{k\ts}{t_R}\ , \qquad b_{\min} = \omega_{\min}t_R\ , \]
that are convenient for a numerical evaluation of the correlators $f_k$.


%


\begin{thebibliography}{45}%
\makeatletter
\providecommand \@ifxundefined [1]{%
 \@ifx{#1\undefined}
}%
\providecommand \@ifnum [1]{%
 \ifnum #1\expandafter \@firstoftwo
 \else \expandafter \@secondoftwo
 \fi
}%
\providecommand \@ifx [1]{%
 \ifx #1\expandafter \@firstoftwo
 \else \expandafter \@secondoftwo
 \fi
}%
\providecommand \natexlab [1]{#1}%
\providecommand \enquote  [1]{``#1''}%
\providecommand \bibnamefont  [1]{#1}%
\providecommand \bibfnamefont [1]{#1}%
\providecommand \citenamefont [1]{#1}%
\providecommand \href@noop [0]{\@secondoftwo}%
\providecommand \href [0]{\begingroup \@sanitize@url \@href}%
\providecommand \@href[1]{\@@startlink{#1}\@@href}%
\providecommand \@@href[1]{\endgroup#1\@@endlink}%
\providecommand \@sanitize@url [0]{\catcode `\\12\catcode `\$12\catcode
  `\&12\catcode `\#12\catcode `\^12\catcode `\_12\catcode `\%12\relax}%
\providecommand \@@startlink[1]{}%
\providecommand \@@endlink[0]{}%
\providecommand \url  [0]{\begingroup\@sanitize@url \@url }%
\providecommand \@url [1]{\endgroup\@href {#1}{\urlprefix }}%
\providecommand \urlprefix  [0]{URL }%
\providecommand \Eprint [0]{\href }%
\providecommand \doibase [0]{https://doi.org/}%
\providecommand \selectlanguage [0]{\@gobble}%
\providecommand \bibinfo  [0]{\@secondoftwo}%
\providecommand \bibfield  [0]{\@secondoftwo}%
\providecommand \translation [1]{[#1]}%
\providecommand \BibitemOpen [0]{}%
\providecommand \bibitemStop [0]{}%
\providecommand \bibitemNoStop [0]{.\EOS\space}%
\providecommand \EOS [0]{\spacefactor3000\relax}%
\providecommand \BibitemShut  [1]{\csname bibitem#1\endcsname}%
\let\auto@bib@innerbib\@empty
\bibitem [{\citenamefont {{\'A}lvarez}\ and\ \citenamefont
  {Suter}(2011)}]{Alvarez2011}%
  \BibitemOpen
  \bibfield  {author} {\bibinfo {author} {\bibfnamefont {G.~A.}\ \bibnamefont
  {{\'A}lvarez}}\ and\ \bibinfo {author} {\bibfnamefont {D.}~\bibnamefont
  {Suter}},\ }\bibfield  {title} {\bibinfo {title} {Measuring the {{Spectrum}}
  of {{Colored Noise}} by {{Dynamical Decoupling}}},\ }\href
  {https://doi.org/10.1103/PhysRevLett.107.230501} {\bibfield  {journal}
  {\bibinfo  {journal} {Phys. Rev. Lett.}\ }\textbf {\bibinfo {volume} {107}},\
  \bibinfo {pages} {230501} (\bibinfo {year} {2011})}\BibitemShut {NoStop}%
\bibitem [{\citenamefont {Bylander}\ \emph {et~al.}(2011)\citenamefont
  {Bylander}, \citenamefont {Gustavsson}, \citenamefont {Yan}, \citenamefont
  {Yoshihara}, \citenamefont {Harrabi}, \citenamefont {Fitch}, \citenamefont
  {Cory}, \citenamefont {Nakamura}, \citenamefont {Tsai},\ and\ \citenamefont
  {Oliver}}]{Bylander2011}%
  \BibitemOpen
  \bibfield  {author} {\bibinfo {author} {\bibfnamefont {J.}~\bibnamefont
  {Bylander}}, \bibinfo {author} {\bibfnamefont {S.}~\bibnamefont
  {Gustavsson}}, \bibinfo {author} {\bibfnamefont {F.}~\bibnamefont {Yan}},
  \bibinfo {author} {\bibfnamefont {F.}~\bibnamefont {Yoshihara}}, \bibinfo
  {author} {\bibfnamefont {K.}~\bibnamefont {Harrabi}}, \bibinfo {author}
  {\bibfnamefont {G.}~\bibnamefont {Fitch}}, \bibinfo {author} {\bibfnamefont
  {D.~G.}\ \bibnamefont {Cory}}, \bibinfo {author} {\bibfnamefont
  {Y.}~\bibnamefont {Nakamura}}, \bibinfo {author} {\bibfnamefont {J.-S.}\
  \bibnamefont {Tsai}},\ and\ \bibinfo {author} {\bibfnamefont {W.~D.}\
  \bibnamefont {Oliver}},\ }\bibfield  {title} {\bibinfo {title} {Noise
  spectroscopy through dynamical decoupling with a superconducting flux
  qubit},\ }\href {https://doi.org/10.1038/nphys1994} {\bibfield  {journal}
  {\bibinfo  {journal} {Nat. Phys.}\ }\textbf {\bibinfo {volume} {7}},\
  \bibinfo {pages} {565} (\bibinfo {year} {2011})}\BibitemShut {NoStop}%
\bibitem [{\citenamefont {Sank}\ \emph {et~al.}(2012)\citenamefont {Sank},
  \citenamefont {Barends}, \citenamefont {Bialczak}, \citenamefont {Chen},
  \citenamefont {Kelly}, \citenamefont {Lenander}, \citenamefont {Lucero},
  \citenamefont {Mariantoni}, \citenamefont {Megrant}, \citenamefont {Neeley},
  \citenamefont {O'Malley}, \citenamefont {Vainsencher}, \citenamefont {Wang},
  \citenamefont {Wenner}, \citenamefont {White}, \citenamefont {Yamamoto},
  \citenamefont {Yin}, \citenamefont {Cleland},\ and\ \citenamefont
  {Martinis}}]{Sank2012}%
  \BibitemOpen
  \bibfield  {author} {\bibinfo {author} {\bibfnamefont {D.}~\bibnamefont
  {Sank}}, \bibinfo {author} {\bibfnamefont {R.}~\bibnamefont {Barends}},
  \bibinfo {author} {\bibfnamefont {R.~C.}\ \bibnamefont {Bialczak}}, \bibinfo
  {author} {\bibfnamefont {Y.}~\bibnamefont {Chen}}, \bibinfo {author}
  {\bibfnamefont {J.}~\bibnamefont {Kelly}}, \bibinfo {author} {\bibfnamefont
  {M.}~\bibnamefont {Lenander}}, \bibinfo {author} {\bibfnamefont
  {E.}~\bibnamefont {Lucero}}, \bibinfo {author} {\bibfnamefont
  {M.}~\bibnamefont {Mariantoni}}, \bibinfo {author} {\bibfnamefont
  {A.}~\bibnamefont {Megrant}}, \bibinfo {author} {\bibfnamefont
  {M.}~\bibnamefont {Neeley}}, \bibinfo {author} {\bibfnamefont {P.~J.~J.}\
  \bibnamefont {O'Malley}}, \bibinfo {author} {\bibfnamefont {A.}~\bibnamefont
  {Vainsencher}}, \bibinfo {author} {\bibfnamefont {H.}~\bibnamefont {Wang}},
  \bibinfo {author} {\bibfnamefont {J.}~\bibnamefont {Wenner}}, \bibinfo
  {author} {\bibfnamefont {T.~C.}\ \bibnamefont {White}}, \bibinfo {author}
  {\bibfnamefont {T.}~\bibnamefont {Yamamoto}}, \bibinfo {author}
  {\bibfnamefont {Y.}~\bibnamefont {Yin}}, \bibinfo {author} {\bibfnamefont
  {A.~N.}\ \bibnamefont {Cleland}},\ and\ \bibinfo {author} {\bibfnamefont
  {J.~M.}\ \bibnamefont {Martinis}},\ }\bibfield  {title} {\bibinfo {title}
  {Flux {{Noise Probed}} with {{Real Time Qubit Tomography}} in a {{Josephson
  Phase Qubit}}},\ }\href {https://doi.org/10.1103/PhysRevLett.109.067001}
  {\bibfield  {journal} {\bibinfo  {journal} {Phys. Rev. Lett.}\ }\textbf
  {\bibinfo {volume} {109}},\ \bibinfo {pages} {067001} (\bibinfo {year}
  {2012})}\BibitemShut {NoStop}%
\bibitem [{\citenamefont {Yan}\ \emph {et~al.}(2012)\citenamefont {Yan},
  \citenamefont {Bylander}, \citenamefont {Gustavsson}, \citenamefont
  {Yoshihara}, \citenamefont {Harrabi}, \citenamefont {Cory}, \citenamefont
  {Orlando}, \citenamefont {Nakamura}, \citenamefont {Tsai},\ and\
  \citenamefont {Oliver}}]{Yan2012}%
  \BibitemOpen
  \bibfield  {author} {\bibinfo {author} {\bibfnamefont {F.}~\bibnamefont
  {Yan}}, \bibinfo {author} {\bibfnamefont {J.}~\bibnamefont {Bylander}},
  \bibinfo {author} {\bibfnamefont {S.}~\bibnamefont {Gustavsson}}, \bibinfo
  {author} {\bibfnamefont {F.}~\bibnamefont {Yoshihara}}, \bibinfo {author}
  {\bibfnamefont {K.}~\bibnamefont {Harrabi}}, \bibinfo {author} {\bibfnamefont
  {D.~G.}\ \bibnamefont {Cory}}, \bibinfo {author} {\bibfnamefont {T.~P.}\
  \bibnamefont {Orlando}}, \bibinfo {author} {\bibfnamefont {Y.}~\bibnamefont
  {Nakamura}}, \bibinfo {author} {\bibfnamefont {J.-S.}\ \bibnamefont {Tsai}},\
  and\ \bibinfo {author} {\bibfnamefont {W.~D.}\ \bibnamefont {Oliver}},\
  }\bibfield  {title} {\bibinfo {title} {Spectroscopy of low-frequency noise
  and its temperature dependence in a superconducting qubit},\ }\href
  {https://doi.org/10.1103/PhysRevB.85.174521} {\bibfield  {journal} {\bibinfo
  {journal} {Phys. Rev. B}\ }\textbf {\bibinfo {volume} {85}},\ \bibinfo
  {pages} {174521} (\bibinfo {year} {2012})}\BibitemShut {NoStop}%
\bibitem [{\citenamefont {{Paz-Silva}}\ and\ \citenamefont
  {Viola}(2014)}]{Paz-Silva2014}%
  \BibitemOpen
  \bibfield  {author} {\bibinfo {author} {\bibfnamefont {G.~A.}\ \bibnamefont
  {{Paz-Silva}}}\ and\ \bibinfo {author} {\bibfnamefont {L.}~\bibnamefont
  {Viola}},\ }\bibfield  {title} {\bibinfo {title} {General {{Transfer-Function
  Approach}} to {{Noise Filtering}} in {{Open-Loop Quantum Control}}},\ }\href
  {https://doi.org/10.1103/PhysRevLett.113.250501} {\bibfield  {journal}
  {\bibinfo  {journal} {Phys. Rev. Lett.}\ }\textbf {\bibinfo {volume} {113}},\
  \bibinfo {pages} {250501} (\bibinfo {year} {2014})}\BibitemShut {NoStop}%
\bibitem [{\citenamefont {Yoshihara}\ \emph {et~al.}(2014)\citenamefont
  {Yoshihara}, \citenamefont {Nakamura}, \citenamefont {Yan}, \citenamefont
  {Gustavsson}, \citenamefont {Bylander}, \citenamefont {Oliver},\ and\
  \citenamefont {Tsai}}]{Yoshihara2014}%
  \BibitemOpen
  \bibfield  {author} {\bibinfo {author} {\bibfnamefont {F.}~\bibnamefont
  {Yoshihara}}, \bibinfo {author} {\bibfnamefont {Y.}~\bibnamefont {Nakamura}},
  \bibinfo {author} {\bibfnamefont {F.}~\bibnamefont {Yan}}, \bibinfo {author}
  {\bibfnamefont {S.}~\bibnamefont {Gustavsson}}, \bibinfo {author}
  {\bibfnamefont {J.}~\bibnamefont {Bylander}}, \bibinfo {author}
  {\bibfnamefont {W.~D.}\ \bibnamefont {Oliver}},\ and\ \bibinfo {author}
  {\bibfnamefont {J.-S.}\ \bibnamefont {Tsai}},\ }\bibfield  {title} {\bibinfo
  {title} {Flux qubit noise spectroscopy using {{Rabi}} oscillations under
  strong driving conditions},\ }\href
  {https://doi.org/10.1103/PhysRevB.89.020503} {\bibfield  {journal} {\bibinfo
  {journal} {Phys. Rev. B}\ }\textbf {\bibinfo {volume} {89}},\ \bibinfo
  {pages} {020503} (\bibinfo {year} {2014})}\BibitemShut {NoStop}%
\bibitem [{\citenamefont {Kim}\ \emph {et~al.}(2015)\citenamefont {Kim},
  \citenamefont {Mamin}, \citenamefont {Sherwood}, \citenamefont {Ohno},
  \citenamefont {Awschalom},\ and\ \citenamefont {Rugar}}]{Kim2015}%
  \BibitemOpen
  \bibfield  {author} {\bibinfo {author} {\bibfnamefont {M.}~\bibnamefont
  {Kim}}, \bibinfo {author} {\bibfnamefont {H.~J.}\ \bibnamefont {Mamin}},
  \bibinfo {author} {\bibfnamefont {M.~H.}\ \bibnamefont {Sherwood}}, \bibinfo
  {author} {\bibfnamefont {K.}~\bibnamefont {Ohno}}, \bibinfo {author}
  {\bibfnamefont {D.~D.}\ \bibnamefont {Awschalom}},\ and\ \bibinfo {author}
  {\bibfnamefont {D.}~\bibnamefont {Rugar}},\ }\bibfield  {title} {\bibinfo
  {title} {Decoherence of {{Near-Surface Nitrogen-Vacancy Centers Due}} to
  {{Electric Field Noise}}},\ }\href
  {https://doi.org/10.1103/PhysRevLett.115.087602} {\bibfield  {journal}
  {\bibinfo  {journal} {Phys. Rev. Lett.}\ }\textbf {\bibinfo {volume} {115}},\
  \bibinfo {pages} {087602} (\bibinfo {year} {2015})}\BibitemShut {NoStop}%
\bibitem [{\citenamefont {Brownnutt}\ \emph {et~al.}(2015)\citenamefont
  {Brownnutt}, \citenamefont {Kumph}, \citenamefont {Rabl},\ and\ \citenamefont
  {Blatt}}]{Brownnutt2015}%
  \BibitemOpen
  \bibfield  {author} {\bibinfo {author} {\bibfnamefont {M.}~\bibnamefont
  {Brownnutt}}, \bibinfo {author} {\bibfnamefont {M.}~\bibnamefont {Kumph}},
  \bibinfo {author} {\bibfnamefont {P.}~\bibnamefont {Rabl}},\ and\ \bibinfo
  {author} {\bibfnamefont {R.}~\bibnamefont {Blatt}},\ }\bibfield  {title}
  {\bibinfo {title} {Ion-trap measurements of electric-field noise near
  surfaces},\ }\href {https://doi.org/10.1103/RevModPhys.87.1419} {\bibfield
  {journal} {\bibinfo  {journal} {Rev. Mod. Phys.}\ }\textbf {\bibinfo {volume}
  {87}},\ \bibinfo {pages} {1419} (\bibinfo {year} {2015})}\BibitemShut
  {NoStop}%
\bibitem [{\citenamefont {O'Malley}\ \emph {et~al.}(2015)\citenamefont
  {O'Malley}, \citenamefont {Kelly}, \citenamefont {Barends}, \citenamefont
  {Campbell}, \citenamefont {Chen}, \citenamefont {Chen}, \citenamefont
  {Chiaro}, \citenamefont {Dunsworth}, \citenamefont {Fowler}, \citenamefont
  {Hoi}, \citenamefont {Jeffrey}, \citenamefont {Megrant}, \citenamefont
  {Mutus}, \citenamefont {Neill}, \citenamefont {Quintana}, \citenamefont
  {Roushan}, \citenamefont {Sank}, \citenamefont {Vainsencher}, \citenamefont
  {Wenner}, \citenamefont {White}, \citenamefont {Korotkov}, \citenamefont
  {Cleland},\ and\ \citenamefont {Martinis}}]{O'Malley2015}%
  \BibitemOpen
  \bibfield  {author} {\bibinfo {author} {\bibfnamefont {P.~J.~J.}\
  \bibnamefont {O'Malley}}, \bibinfo {author} {\bibfnamefont {J.}~\bibnamefont
  {Kelly}}, \bibinfo {author} {\bibfnamefont {R.}~\bibnamefont {Barends}},
  \bibinfo {author} {\bibfnamefont {B.}~\bibnamefont {Campbell}}, \bibinfo
  {author} {\bibfnamefont {Y.}~\bibnamefont {Chen}}, \bibinfo {author}
  {\bibfnamefont {Z.}~\bibnamefont {Chen}}, \bibinfo {author} {\bibfnamefont
  {B.}~\bibnamefont {Chiaro}}, \bibinfo {author} {\bibfnamefont
  {A.}~\bibnamefont {Dunsworth}}, \bibinfo {author} {\bibfnamefont {A.~G.}\
  \bibnamefont {Fowler}}, \bibinfo {author} {\bibfnamefont {I.-C.}\
  \bibnamefont {Hoi}}, \bibinfo {author} {\bibfnamefont {E.}~\bibnamefont
  {Jeffrey}}, \bibinfo {author} {\bibfnamefont {A.}~\bibnamefont {Megrant}},
  \bibinfo {author} {\bibfnamefont {J.}~\bibnamefont {Mutus}}, \bibinfo
  {author} {\bibfnamefont {C.}~\bibnamefont {Neill}}, \bibinfo {author}
  {\bibfnamefont {C.}~\bibnamefont {Quintana}}, \bibinfo {author}
  {\bibfnamefont {P.}~\bibnamefont {Roushan}}, \bibinfo {author} {\bibfnamefont
  {D.}~\bibnamefont {Sank}}, \bibinfo {author} {\bibfnamefont {A.}~\bibnamefont
  {Vainsencher}}, \bibinfo {author} {\bibfnamefont {J.}~\bibnamefont {Wenner}},
  \bibinfo {author} {\bibfnamefont {T.~C.}\ \bibnamefont {White}}, \bibinfo
  {author} {\bibfnamefont {A.~N.}\ \bibnamefont {Korotkov}}, \bibinfo {author}
  {\bibfnamefont {A.~N.}\ \bibnamefont {Cleland}},\ and\ \bibinfo {author}
  {\bibfnamefont {J.~M.}\ \bibnamefont {Martinis}},\ }\bibfield  {title}
  {\bibinfo {title} {Qubit {{Metrology}} of {{Ultralow Phase Noise Using
  Randomized Benchmarking}}},\ }\href
  {https://doi.org/10.1103/PhysRevApplied.3.044009} {\bibfield  {journal}
  {\bibinfo  {journal} {Phys. Rev. Applied}\ }\textbf {\bibinfo {volume} {3}},\
  \bibinfo {pages} {044009} (\bibinfo {year} {2015})}\BibitemShut {NoStop}%
\bibitem [{\citenamefont {Sza{\'n}kowski}\ \emph {et~al.}(2016)\citenamefont
  {Sza{\'n}kowski}, \citenamefont {Trippenbach},\ and\ \citenamefont
  {Cywi{\'n}ski}}]{Szankowski2016}%
  \BibitemOpen
  \bibfield  {author} {\bibinfo {author} {\bibfnamefont {P.}~\bibnamefont
  {Sza{\'n}kowski}}, \bibinfo {author} {\bibfnamefont {M.}~\bibnamefont
  {Trippenbach}},\ and\ \bibinfo {author} {\bibfnamefont {{\L}.}~\bibnamefont
  {Cywi{\'n}ski}},\ }\bibfield  {title} {\bibinfo {title} {Spectroscopy of
  cross correlations of environmental noises with two qubits},\ }\href
  {https://doi.org/10.1103/PhysRevA.94.012109} {\bibfield  {journal} {\bibinfo
  {journal} {Phys. Rev. A}\ }\textbf {\bibinfo {volume} {94}},\ \bibinfo
  {pages} {012109} (\bibinfo {year} {2016})}\BibitemShut {NoStop}%
\bibitem [{\citenamefont {Yan}\ \emph {et~al.}(2016)\citenamefont {Yan},
  \citenamefont {Gustavsson}, \citenamefont {Kamal}, \citenamefont {Birenbaum},
  \citenamefont {Sears}, \citenamefont {Hover}, \citenamefont {Gudmundsen},
  \citenamefont {Rosenberg}, \citenamefont {Samach}, \citenamefont {Weber},
  \citenamefont {Yoder}, \citenamefont {Orlando}, \citenamefont {Clarke},
  \citenamefont {Kerman},\ and\ \citenamefont {Oliver}}]{Yan2016}%
  \BibitemOpen
  \bibfield  {author} {\bibinfo {author} {\bibfnamefont {F.}~\bibnamefont
  {Yan}}, \bibinfo {author} {\bibfnamefont {S.}~\bibnamefont {Gustavsson}},
  \bibinfo {author} {\bibfnamefont {A.}~\bibnamefont {Kamal}}, \bibinfo
  {author} {\bibfnamefont {J.}~\bibnamefont {Birenbaum}}, \bibinfo {author}
  {\bibfnamefont {A.~P.}\ \bibnamefont {Sears}}, \bibinfo {author}
  {\bibfnamefont {D.}~\bibnamefont {Hover}}, \bibinfo {author} {\bibfnamefont
  {T.~J.}\ \bibnamefont {Gudmundsen}}, \bibinfo {author} {\bibfnamefont
  {D.}~\bibnamefont {Rosenberg}}, \bibinfo {author} {\bibfnamefont
  {G.}~\bibnamefont {Samach}}, \bibinfo {author} {\bibfnamefont
  {S.}~\bibnamefont {Weber}}, \bibinfo {author} {\bibfnamefont {J.~L.}\
  \bibnamefont {Yoder}}, \bibinfo {author} {\bibfnamefont {T.~P.}\ \bibnamefont
  {Orlando}}, \bibinfo {author} {\bibfnamefont {J.}~\bibnamefont {Clarke}},
  \bibinfo {author} {\bibfnamefont {A.~J.}\ \bibnamefont {Kerman}},\ and\
  \bibinfo {author} {\bibfnamefont {W.~D.}\ \bibnamefont {Oliver}},\ }\bibfield
   {title} {\bibinfo {title} {The flux qubit revisited to enhance coherence and
  reproducibility},\ }\href {https://doi.org/10.1038/ncomms12964} {\bibfield
  {journal} {\bibinfo  {journal} {Nat Commun}\ }\textbf {\bibinfo {volume}
  {7}},\ \bibinfo {pages} {1} (\bibinfo {year} {2016})}\BibitemShut {NoStop}%
\bibitem [{\citenamefont {Myers}\ \emph {et~al.}(2017)\citenamefont {Myers},
  \citenamefont {Ariyaratne},\ and\ \citenamefont {Jayich}}]{Myers2017}%
  \BibitemOpen
  \bibfield  {author} {\bibinfo {author} {\bibfnamefont {B.~A.}\ \bibnamefont
  {Myers}}, \bibinfo {author} {\bibfnamefont {A.}~\bibnamefont {Ariyaratne}},\
  and\ \bibinfo {author} {\bibfnamefont {A.~C.~B.}\ \bibnamefont {Jayich}},\
  }\bibfield  {title} {\bibinfo {title} {Double-{{Quantum Spin-Relaxation
  Limits}} to {{Coherence}} of {{Near-Surface Nitrogen-Vacancy Centers}}},\
  }\href {https://doi.org/10.1103/PhysRevLett.118.197201} {\bibfield  {journal}
  {\bibinfo  {journal} {Phys. Rev. Lett.}\ }\textbf {\bibinfo {volume} {118}},\
  \bibinfo {pages} {197201} (\bibinfo {year} {2017})}\BibitemShut {NoStop}%
\bibitem [{\citenamefont {Quintana}\ \emph {et~al.}(2017)\citenamefont
  {Quintana}, \citenamefont {Chen}, \citenamefont {Sank}, \citenamefont
  {Petukhov}, \citenamefont {White}, \citenamefont {Kafri}, \citenamefont
  {Chiaro}, \citenamefont {Megrant}, \citenamefont {Barends}, \citenamefont
  {Campbell}, \citenamefont {Chen}, \citenamefont {Dunsworth}, \citenamefont
  {Fowler}, \citenamefont {Graff}, \citenamefont {Jeffrey}, \citenamefont
  {Kelly}, \citenamefont {Lucero}, \citenamefont {Mutus}, \citenamefont
  {Neeley}, \citenamefont {Neill}, \citenamefont {O'Malley}, \citenamefont
  {Roushan}, \citenamefont {Shabani}, \citenamefont {Smelyanskiy},
  \citenamefont {Vainsencher}, \citenamefont {Wenner}, \citenamefont {Neven},\
  and\ \citenamefont {Martinis}}]{Quintana2017}%
  \BibitemOpen
  \bibfield  {author} {\bibinfo {author} {\bibfnamefont {C.~M.}\ \bibnamefont
  {Quintana}}, \bibinfo {author} {\bibfnamefont {Y.}~\bibnamefont {Chen}},
  \bibinfo {author} {\bibfnamefont {D.}~\bibnamefont {Sank}}, \bibinfo {author}
  {\bibfnamefont {A.~G.}\ \bibnamefont {Petukhov}}, \bibinfo {author}
  {\bibfnamefont {T.~C.}\ \bibnamefont {White}}, \bibinfo {author}
  {\bibfnamefont {D.}~\bibnamefont {Kafri}}, \bibinfo {author} {\bibfnamefont
  {B.}~\bibnamefont {Chiaro}}, \bibinfo {author} {\bibfnamefont
  {A.}~\bibnamefont {Megrant}}, \bibinfo {author} {\bibfnamefont
  {R.}~\bibnamefont {Barends}}, \bibinfo {author} {\bibfnamefont
  {B.}~\bibnamefont {Campbell}}, \bibinfo {author} {\bibfnamefont
  {Z.}~\bibnamefont {Chen}}, \bibinfo {author} {\bibfnamefont {A.}~\bibnamefont
  {Dunsworth}}, \bibinfo {author} {\bibfnamefont {A.~G.}\ \bibnamefont
  {Fowler}}, \bibinfo {author} {\bibfnamefont {R.}~\bibnamefont {Graff}},
  \bibinfo {author} {\bibfnamefont {E.}~\bibnamefont {Jeffrey}}, \bibinfo
  {author} {\bibfnamefont {J.}~\bibnamefont {Kelly}}, \bibinfo {author}
  {\bibfnamefont {E.}~\bibnamefont {Lucero}}, \bibinfo {author} {\bibfnamefont
  {J.~Y.}\ \bibnamefont {Mutus}}, \bibinfo {author} {\bibfnamefont
  {M.}~\bibnamefont {Neeley}}, \bibinfo {author} {\bibfnamefont
  {C.}~\bibnamefont {Neill}}, \bibinfo {author} {\bibfnamefont {P.~J.~J.}\
  \bibnamefont {O'Malley}}, \bibinfo {author} {\bibfnamefont {P.}~\bibnamefont
  {Roushan}}, \bibinfo {author} {\bibfnamefont {A.}~\bibnamefont {Shabani}},
  \bibinfo {author} {\bibfnamefont {V.~N.}\ \bibnamefont {Smelyanskiy}},
  \bibinfo {author} {\bibfnamefont {A.}~\bibnamefont {Vainsencher}}, \bibinfo
  {author} {\bibfnamefont {J.}~\bibnamefont {Wenner}}, \bibinfo {author}
  {\bibfnamefont {H.}~\bibnamefont {Neven}},\ and\ \bibinfo {author}
  {\bibfnamefont {J.~M.}\ \bibnamefont {Martinis}},\ }\bibfield  {title}
  {\bibinfo {title} {Observation of {{Classical-Quantum Crossover}} of $1/f$
  {{Flux Noise}} and {{Its Paramagnetic Temperature Dependence}}},\ }\href
  {https://doi.org/10.1103/PhysRevLett.118.057702} {\bibfield  {journal}
  {\bibinfo  {journal} {Phys. Rev. Lett.}\ }\textbf {\bibinfo {volume} {118}},\
  \bibinfo {pages} {057702} (\bibinfo {year} {2017})}\BibitemShut {NoStop}%
\bibitem [{\citenamefont {Ferrie}\ \emph {et~al.}(2018)\citenamefont {Ferrie},
  \citenamefont {Granade}, \citenamefont {{Paz-Silva}},\ and\ \citenamefont
  {Wiseman}}]{Ferrie2018}%
  \BibitemOpen
  \bibfield  {author} {\bibinfo {author} {\bibfnamefont {C.}~\bibnamefont
  {Ferrie}}, \bibinfo {author} {\bibfnamefont {C.}~\bibnamefont {Granade}},
  \bibinfo {author} {\bibfnamefont {G.}~\bibnamefont {{Paz-Silva}}},\ and\
  \bibinfo {author} {\bibfnamefont {H.~M.}\ \bibnamefont {Wiseman}},\
  }\bibfield  {title} {\bibinfo {title} {Bayesian quantum noise spectroscopy},\
  }\href {https://doi.org/10.1088/1367-2630/aaf207} {\bibfield  {journal}
  {\bibinfo  {journal} {New J. Phys.}\ }\textbf {\bibinfo {volume} {20}},\
  \bibinfo {pages} {123005} (\bibinfo {year} {2018})}\BibitemShut {NoStop}%
\bibitem [{\citenamefont {Noel}\ \emph {et~al.}(2019)\citenamefont {Noel},
  \citenamefont {{Berlin-Udi}}, \citenamefont {Matthiesen}, \citenamefont {Yu},
  \citenamefont {Zhou}, \citenamefont {Lordi},\ and\ \citenamefont
  {H{\"a}ffner}}]{Noel2019}%
  \BibitemOpen
  \bibfield  {author} {\bibinfo {author} {\bibfnamefont {C.}~\bibnamefont
  {Noel}}, \bibinfo {author} {\bibfnamefont {M.}~\bibnamefont {{Berlin-Udi}}},
  \bibinfo {author} {\bibfnamefont {C.}~\bibnamefont {Matthiesen}}, \bibinfo
  {author} {\bibfnamefont {J.}~\bibnamefont {Yu}}, \bibinfo {author}
  {\bibfnamefont {Y.}~\bibnamefont {Zhou}}, \bibinfo {author} {\bibfnamefont
  {V.}~\bibnamefont {Lordi}},\ and\ \bibinfo {author} {\bibfnamefont
  {H.}~\bibnamefont {H{\"a}ffner}},\ }\bibfield  {title} {\bibinfo {title}
  {Electric-field noise from thermally activated fluctuators in a surface ion
  trap},\ }\href {https://doi.org/10.1103/PhysRevA.99.063427} {\bibfield
  {journal} {\bibinfo  {journal} {Phys. Rev. A}\ }\textbf {\bibinfo {volume}
  {99}},\ \bibinfo {pages} {063427} (\bibinfo {year} {2019})}\BibitemShut
  {NoStop}%
\bibitem [{\citenamefont {{von L{\"u}pke}}\ \emph {et~al.}(2020)\citenamefont
  {{von L{\"u}pke}}, \citenamefont {Beaudoin}, \citenamefont {Norris},
  \citenamefont {Sung}, \citenamefont {Winik}, \citenamefont {Qiu},
  \citenamefont {Kjaergaard}, \citenamefont {Kim}, \citenamefont {Yoder},
  \citenamefont {Gustavsson}, \citenamefont {Viola},\ and\ \citenamefont
  {Oliver}}]{vonLupke2020}%
  \BibitemOpen
  \bibfield  {author} {\bibinfo {author} {\bibfnamefont {U.}~\bibnamefont {{von
  L{\"u}pke}}}, \bibinfo {author} {\bibfnamefont {F.}~\bibnamefont {Beaudoin}},
  \bibinfo {author} {\bibfnamefont {L.~M.}\ \bibnamefont {Norris}}, \bibinfo
  {author} {\bibfnamefont {Y.}~\bibnamefont {Sung}}, \bibinfo {author}
  {\bibfnamefont {R.}~\bibnamefont {Winik}}, \bibinfo {author} {\bibfnamefont
  {J.~Y.}\ \bibnamefont {Qiu}}, \bibinfo {author} {\bibfnamefont
  {M.}~\bibnamefont {Kjaergaard}}, \bibinfo {author} {\bibfnamefont
  {D.}~\bibnamefont {Kim}}, \bibinfo {author} {\bibfnamefont {J.}~\bibnamefont
  {Yoder}}, \bibinfo {author} {\bibfnamefont {S.}~\bibnamefont {Gustavsson}},
  \bibinfo {author} {\bibfnamefont {L.}~\bibnamefont {Viola}},\ and\ \bibinfo
  {author} {\bibfnamefont {W.~D.}\ \bibnamefont {Oliver}},\ }\bibfield  {title}
  {\bibinfo {title} {Two-{{Qubit Spectroscopy}} of {{Spatiotemporally
  Correlated Quantum Noise}} in {{Superconducting Qubits}}},\ }\href
  {https://doi.org/10.1103/PRXQuantum.1.010305} {\bibfield  {journal} {\bibinfo
   {journal} {PRX Quantum}\ }\textbf {\bibinfo {volume} {1}},\ \bibinfo {pages}
  {010305} (\bibinfo {year} {2020})}\BibitemShut {NoStop}%
\bibitem [{\citenamefont {Wolfowicz}\ \emph {et~al.}(2021)\citenamefont
  {Wolfowicz}, \citenamefont {Heremans}, \citenamefont {Anderson},
  \citenamefont {Kanai}, \citenamefont {Seo}, \citenamefont {Gali},
  \citenamefont {Galli},\ and\ \citenamefont {Awschalom}}]{Wolfowicz2021}%
  \BibitemOpen
  \bibfield  {author} {\bibinfo {author} {\bibfnamefont {G.}~\bibnamefont
  {Wolfowicz}}, \bibinfo {author} {\bibfnamefont {F.~J.}\ \bibnamefont
  {Heremans}}, \bibinfo {author} {\bibfnamefont {C.~P.}\ \bibnamefont
  {Anderson}}, \bibinfo {author} {\bibfnamefont {S.}~\bibnamefont {Kanai}},
  \bibinfo {author} {\bibfnamefont {H.}~\bibnamefont {Seo}}, \bibinfo {author}
  {\bibfnamefont {A.}~\bibnamefont {Gali}}, \bibinfo {author} {\bibfnamefont
  {G.}~\bibnamefont {Galli}},\ and\ \bibinfo {author} {\bibfnamefont {D.~D.}\
  \bibnamefont {Awschalom}},\ }\bibfield  {title} {\bibinfo {title} {Qubit
  guidelines for solid-state spin defects},\ }\href@noop {} {\bibfield
  {journal} {\bibinfo  {journal} {Nat. Rev. Mater.}\ }\textbf {\bibinfo
  {volume} {6}},\ \bibinfo {pages} {906} (\bibinfo {year} {2021})},\ \bibinfo
  {note} {comment: 40 pages, 7 figures, 259 references}\BibitemShut {NoStop}%
\bibitem [{\citenamefont {Wang}\ and\ \citenamefont {Clerk}(2021)}]{Wang2021}%
  \BibitemOpen
  \bibfield  {author} {\bibinfo {author} {\bibfnamefont {Y.-X.}\ \bibnamefont
  {Wang}}\ and\ \bibinfo {author} {\bibfnamefont {A.~A.}\ \bibnamefont
  {Clerk}},\ }\bibfield  {title} {\bibinfo {title} {Intrinsic and induced
  quantum quenches for enhancing qubit-based quantum noise spectroscopy},\
  }\href@noop {} {\bibfield  {journal} {\bibinfo  {journal} {Nat. Commun.}\
  }\textbf {\bibinfo {volume} {12}},\ \bibinfo {pages} {6528} (\bibinfo {year}
  {2021})}\BibitemShut {NoStop}%
\bibitem [{\citenamefont {Rist{\`e}}\ \emph {et~al.}(2013)\citenamefont
  {Rist{\`e}}, \citenamefont {Bultink}, \citenamefont {Tiggelman},
  \citenamefont {Schouten}, \citenamefont {Lehnert},\ and\ \citenamefont
  {DiCarlo}}]{Riste2013}%
  \BibitemOpen
  \bibfield  {author} {\bibinfo {author} {\bibfnamefont {D.}~\bibnamefont
  {Rist{\`e}}}, \bibinfo {author} {\bibfnamefont {C.~C.}\ \bibnamefont
  {Bultink}}, \bibinfo {author} {\bibfnamefont {M.~J.}\ \bibnamefont
  {Tiggelman}}, \bibinfo {author} {\bibfnamefont {R.~N.}\ \bibnamefont
  {Schouten}}, \bibinfo {author} {\bibfnamefont {K.~W.}\ \bibnamefont
  {Lehnert}},\ and\ \bibinfo {author} {\bibfnamefont {L.}~\bibnamefont
  {DiCarlo}},\ }\bibfield  {title} {\bibinfo {title} {Millisecond charge-parity
  fluctuations and induced decoherence in a superconducting transmon qubit},\
  }\href {https://doi.org/10.1038/ncomms2936} {\bibfield  {journal} {\bibinfo
  {journal} {Nat Commun}\ }\textbf {\bibinfo {volume} {4}},\ \bibinfo {pages}
  {1913} (\bibinfo {year} {2013})}\BibitemShut {NoStop}%
\bibitem [{\citenamefont {Serniak}\ \emph {et~al.}(2018)\citenamefont
  {Serniak}, \citenamefont {Hays}, \citenamefont {{de Lange}}, \citenamefont
  {Diamond}, \citenamefont {Shankar}, \citenamefont {Burkhart}, \citenamefont
  {Frunzio}, \citenamefont {Houzet},\ and\ \citenamefont
  {Devoret}}]{Serniak2018}%
  \BibitemOpen
  \bibfield  {author} {\bibinfo {author} {\bibfnamefont {K.}~\bibnamefont
  {Serniak}}, \bibinfo {author} {\bibfnamefont {M.}~\bibnamefont {Hays}},
  \bibinfo {author} {\bibfnamefont {G.}~\bibnamefont {{de Lange}}}, \bibinfo
  {author} {\bibfnamefont {S.}~\bibnamefont {Diamond}}, \bibinfo {author}
  {\bibfnamefont {S.}~\bibnamefont {Shankar}}, \bibinfo {author} {\bibfnamefont
  {L.~D.}\ \bibnamefont {Burkhart}}, \bibinfo {author} {\bibfnamefont
  {L.}~\bibnamefont {Frunzio}}, \bibinfo {author} {\bibfnamefont
  {M.}~\bibnamefont {Houzet}},\ and\ \bibinfo {author} {\bibfnamefont {M.~H.}\
  \bibnamefont {Devoret}},\ }\bibfield  {title} {\bibinfo {title} {Hot
  {{Nonequilibrium Quasiparticles}} in {{Transmon Qubits}}},\ }\href
  {https://doi.org/10.1103/PhysRevLett.121.157701} {\bibfield  {journal}
  {\bibinfo  {journal} {Phys. Rev. Lett.}\ }\textbf {\bibinfo {volume} {121}},\
  \bibinfo {pages} {157701} (\bibinfo {year} {2018})}\BibitemShut {NoStop}%
\bibitem [{\citenamefont {Christensen}\ \emph {et~al.}(2019)\citenamefont
  {Christensen}, \citenamefont {Wilen}, \citenamefont {Opremcak}, \citenamefont
  {Nelson}, \citenamefont {Schlenker}, \citenamefont {Zimonick}, \citenamefont
  {Faoro}, \citenamefont {Ioffe}, \citenamefont {Rosen}, \citenamefont
  {DuBois}, \citenamefont {Plourde},\ and\ \citenamefont
  {McDermott}}]{Christensen2019}%
  \BibitemOpen
  \bibfield  {author} {\bibinfo {author} {\bibfnamefont {B.~G.}\ \bibnamefont
  {Christensen}}, \bibinfo {author} {\bibfnamefont {C.~D.}\ \bibnamefont
  {Wilen}}, \bibinfo {author} {\bibfnamefont {A.}~\bibnamefont {Opremcak}},
  \bibinfo {author} {\bibfnamefont {J.}~\bibnamefont {Nelson}}, \bibinfo
  {author} {\bibfnamefont {F.}~\bibnamefont {Schlenker}}, \bibinfo {author}
  {\bibfnamefont {C.~H.}\ \bibnamefont {Zimonick}}, \bibinfo {author}
  {\bibfnamefont {L.}~\bibnamefont {Faoro}}, \bibinfo {author} {\bibfnamefont
  {L.~B.}\ \bibnamefont {Ioffe}}, \bibinfo {author} {\bibfnamefont {Y.~J.}\
  \bibnamefont {Rosen}}, \bibinfo {author} {\bibfnamefont {J.~L.}\ \bibnamefont
  {DuBois}}, \bibinfo {author} {\bibfnamefont {B.~L.~T.}\ \bibnamefont
  {Plourde}},\ and\ \bibinfo {author} {\bibfnamefont {R.}~\bibnamefont
  {McDermott}},\ }\bibfield  {title} {\bibinfo {title} {Anomalous {{Charge
  Noise}} in {{Superconducting Qubits}}},\ }\href@noop {} {\bibfield  {journal}
  {\bibinfo  {journal} {Phys. Rev. B}\ }\textbf {\bibinfo {volume} {100}},\
  \bibinfo {pages} {140503} (\bibinfo {year} {2019})},\ \bibinfo {note}
  {comment: 10 pages, 7 figures},\ \Eprint {https://arxiv.org/abs/1905.13712}
  {arXiv:1905.13712} \BibitemShut {NoStop}%
\bibitem [{\citenamefont {Schl{\"o}r}\ \emph {et~al.}(2019)\citenamefont
  {Schl{\"o}r}, \citenamefont {Lisenfeld}, \citenamefont {M{\"u}ller},
  \citenamefont {Bilmes}, \citenamefont {Schneider}, \citenamefont {Pappas},
  \citenamefont {Ustinov},\ and\ \citenamefont {Weides}}]{Schlor2019}%
  \BibitemOpen
  \bibfield  {author} {\bibinfo {author} {\bibfnamefont {S.}~\bibnamefont
  {Schl{\"o}r}}, \bibinfo {author} {\bibfnamefont {J.}~\bibnamefont
  {Lisenfeld}}, \bibinfo {author} {\bibfnamefont {C.}~\bibnamefont
  {M{\"u}ller}}, \bibinfo {author} {\bibfnamefont {A.}~\bibnamefont {Bilmes}},
  \bibinfo {author} {\bibfnamefont {A.}~\bibnamefont {Schneider}}, \bibinfo
  {author} {\bibfnamefont {D.~P.}\ \bibnamefont {Pappas}}, \bibinfo {author}
  {\bibfnamefont {A.~V.}\ \bibnamefont {Ustinov}},\ and\ \bibinfo {author}
  {\bibfnamefont {M.}~\bibnamefont {Weides}},\ }\bibfield  {title} {\bibinfo
  {title} {Correlating {{Decoherence}} in {{Transmon Qubits}}: {{Low Frequency
  Noise}} by {{Single Fluctuators}}},\ }\href
  {https://doi.org/10.1103/PhysRevLett.123.190502} {\bibfield  {journal}
  {\bibinfo  {journal} {Phys. Rev. Lett.}\ }\textbf {\bibinfo {volume} {123}},\
  \bibinfo {pages} {190502} (\bibinfo {year} {2019})}\BibitemShut {NoStop}%
\bibitem [{\citenamefont {Paladino}\ \emph {et~al.}(2002)\citenamefont
  {Paladino}, \citenamefont {Faoro}, \citenamefont {Falci},\ and\ \citenamefont
  {Fazio}}]{Paladino2002}%
  \BibitemOpen
  \bibfield  {author} {\bibinfo {author} {\bibfnamefont {E.}~\bibnamefont
  {Paladino}}, \bibinfo {author} {\bibfnamefont {L.}~\bibnamefont {Faoro}},
  \bibinfo {author} {\bibfnamefont {G.}~\bibnamefont {Falci}},\ and\ \bibinfo
  {author} {\bibfnamefont {R.}~\bibnamefont {Fazio}},\ }\bibfield  {title}
  {\bibinfo {title} {Decoherence and $1/f$ {{Noise}} in {{Josephson Qubits}}},\
  }\href {https://doi.org/10.1103/PhysRevLett.88.228304} {\bibfield  {journal}
  {\bibinfo  {journal} {Phys. Rev. Lett.}\ }\textbf {\bibinfo {volume} {88}},\
  \bibinfo {pages} {228304} (\bibinfo {year} {2002})}\BibitemShut {NoStop}%
\bibitem [{\citenamefont {Galperin}\ \emph {et~al.}(2004)\citenamefont
  {Galperin}, \citenamefont {Altshuler},\ and\ \citenamefont
  {Shantsev}}]{Galperin2004}%
  \BibitemOpen
  \bibfield  {author} {\bibinfo {author} {\bibfnamefont {Y.~M.}\ \bibnamefont
  {Galperin}}, \bibinfo {author} {\bibfnamefont {B.~L.}\ \bibnamefont
  {Altshuler}},\ and\ \bibinfo {author} {\bibfnamefont {D.~V.}\ \bibnamefont
  {Shantsev}},\ }\bibfield  {title} {\bibinfo {title} {Low-frequency noise as a
  source of dephasing of a qubit},\ }in\ \href@noop {} {\emph {\bibinfo
  {booktitle} {Fundamental {{Problems}} of {{Mesoscopic Physics}}}}},\ \bibinfo
  {editor} {edited by\ \bibinfo {editor} {\bibfnamefont {I.~V.}\ \bibnamefont
  {Lerner}}}\ (\bibinfo  {publisher} {{Kluwer Academic Publishing}},\ \bibinfo
  {address} {{The Netherlands}},\ \bibinfo {year} {2004})\ pp.\ \bibinfo
  {pages} {141--165},\ \bibinfo {note} {comment: 18 pages, 8 figures, Proc. of
  NATO/Euresco Conf. "Fundamental Problems of Mesoscopic Physics: Interactions
  and Decoherence", Granada, Spain, Sept.2003}\BibitemShut {NoStop}%
\bibitem [{\citenamefont {Galperin}\ \emph {et~al.}(2006)\citenamefont
  {Galperin}, \citenamefont {Altshuler}, \citenamefont {Bergli},\ and\
  \citenamefont {Shantsev}}]{Galperin2006a}%
  \BibitemOpen
  \bibfield  {author} {\bibinfo {author} {\bibfnamefont {Y.~M.}\ \bibnamefont
  {Galperin}}, \bibinfo {author} {\bibfnamefont {B.~L.}\ \bibnamefont
  {Altshuler}}, \bibinfo {author} {\bibfnamefont {J.}~\bibnamefont {Bergli}},\
  and\ \bibinfo {author} {\bibfnamefont {D.~V.}\ \bibnamefont {Shantsev}},\
  }\bibfield  {title} {\bibinfo {title} {Non-{{Gaussian Low-Frequency Noise}}
  as a {{Source}} of {{Qubit Decoherence}}},\ }\href
  {https://doi.org/10.1103/PhysRevLett.96.097009} {\bibfield  {journal}
  {\bibinfo  {journal} {Phys. Rev. Lett.}\ }\textbf {\bibinfo {volume} {96}},\
  \bibinfo {pages} {097009} (\bibinfo {year} {2006})}\BibitemShut {NoStop}%
\bibitem [{\citenamefont {Paladino}\ \emph {et~al.}(2014)\citenamefont
  {Paladino}, \citenamefont {Galperin}, \citenamefont {Falci},\ and\
  \citenamefont {Altshuler}}]{Paladino2014}%
  \BibitemOpen
  \bibfield  {author} {\bibinfo {author} {\bibfnamefont {E.}~\bibnamefont
  {Paladino}}, \bibinfo {author} {\bibfnamefont {Y.~M.}\ \bibnamefont
  {Galperin}}, \bibinfo {author} {\bibfnamefont {G.}~\bibnamefont {Falci}},\
  and\ \bibinfo {author} {\bibfnamefont {B.~L.}\ \bibnamefont {Altshuler}},\
  }\bibfield  {title} {\bibinfo {title} {$1/f$ noise: {{Implications}} for
  solid-state quantum information},\ }\href
  {https://doi.org/10.1103/RevModPhys.86.361} {\bibfield  {journal} {\bibinfo
  {journal} {Rev. Mod. Phys.}\ }\textbf {\bibinfo {volume} {86}},\ \bibinfo
  {pages} {361} (\bibinfo {year} {2014})}\BibitemShut {NoStop}%
\bibitem [{\citenamefont {M{\"u}ller}\ \emph {et~al.}(2019)\citenamefont
  {M{\"u}ller}, \citenamefont {Cole},\ and\ \citenamefont
  {Lisenfeld}}]{Muller2019}%
  \BibitemOpen
  \bibfield  {author} {\bibinfo {author} {\bibfnamefont {C.}~\bibnamefont
  {M{\"u}ller}}, \bibinfo {author} {\bibfnamefont {J.~H.}\ \bibnamefont
  {Cole}},\ and\ \bibinfo {author} {\bibfnamefont {J.}~\bibnamefont
  {Lisenfeld}},\ }\bibfield  {title} {\bibinfo {title} {Towards understanding
  two-level-systems in amorphous solids: Insights from quantum circuits},\
  }\href {https://doi.org/10.1088/1361-6633/ab3a7e} {\bibfield  {journal}
  {\bibinfo  {journal} {Rep. Prog. Phys.}\ }\textbf {\bibinfo {volume} {82}},\
  \bibinfo {pages} {124501} (\bibinfo {year} {2019})}\BibitemShut {NoStop}%
\bibitem [{\citenamefont {Herzog}\ and\ \citenamefont
  {Hahn}(1956)}]{Herzog1956}%
  \BibitemOpen
  \bibfield  {author} {\bibinfo {author} {\bibfnamefont {B.}~\bibnamefont
  {Herzog}}\ and\ \bibinfo {author} {\bibfnamefont {E.~L.}\ \bibnamefont
  {Hahn}},\ }\bibfield  {title} {\bibinfo {title} {Transient {{Nuclear
  Induction}} and {{Double Nuclear Resonance}} in {{Solids}}},\ }\href
  {https://doi.org/10.1103/PhysRev.103.148} {\bibfield  {journal} {\bibinfo
  {journal} {Phys. Rev.}\ }\textbf {\bibinfo {volume} {103}},\ \bibinfo {pages}
  {148} (\bibinfo {year} {1956})}\BibitemShut {NoStop}%
\bibitem [{\citenamefont {Klauder}\ and\ \citenamefont
  {Anderson}(1962)}]{Klauder1962}%
  \BibitemOpen
  \bibfield  {author} {\bibinfo {author} {\bibfnamefont {J.~R.}\ \bibnamefont
  {Klauder}}\ and\ \bibinfo {author} {\bibfnamefont {P.~W.}\ \bibnamefont
  {Anderson}},\ }\bibfield  {title} {\bibinfo {title} {Spectral {{Diffusion
  Decay}} in {{Spin Resonance Experiments}}},\ }\href
  {https://doi.org/10.1103/PhysRev.125.912} {\bibfield  {journal} {\bibinfo
  {journal} {Phys. Rev.}\ }\textbf {\bibinfo {volume} {125}},\ \bibinfo {pages}
  {912} (\bibinfo {year} {1962})}\BibitemShut {NoStop}%
\bibitem [{\citenamefont {Li}\ \emph {et~al.}(2013)\citenamefont {Li},
  \citenamefont {Saxena}, \citenamefont {Smith},\ and\ \citenamefont
  {Sinitsyn}}]{Li2013a}%
  \BibitemOpen
  \bibfield  {author} {\bibinfo {author} {\bibfnamefont {F.}~\bibnamefont
  {Li}}, \bibinfo {author} {\bibfnamefont {A.}~\bibnamefont {Saxena}}, \bibinfo
  {author} {\bibfnamefont {D.}~\bibnamefont {Smith}},\ and\ \bibinfo {author}
  {\bibfnamefont {N.~A.}\ \bibnamefont {Sinitsyn}},\ }\bibfield  {title}
  {\bibinfo {title} {Higher-order spin noise statistics},\ }\href
  {https://doi.org/10.1088/1367-2630/15/11/113038} {\bibfield  {journal}
  {\bibinfo  {journal} {New J. Phys.}\ }\textbf {\bibinfo {volume} {15}},\
  \bibinfo {pages} {113038} (\bibinfo {year} {2013})}\BibitemShut {NoStop}%
\bibitem [{\citenamefont {Norris}\ \emph {et~al.}(2016)\citenamefont {Norris},
  \citenamefont {{Paz-Silva}},\ and\ \citenamefont {Viola}}]{Norris2016}%
  \BibitemOpen
  \bibfield  {author} {\bibinfo {author} {\bibfnamefont {L.~M.}\ \bibnamefont
  {Norris}}, \bibinfo {author} {\bibfnamefont {G.~A.}\ \bibnamefont
  {{Paz-Silva}}},\ and\ \bibinfo {author} {\bibfnamefont {L.}~\bibnamefont
  {Viola}},\ }\bibfield  {title} {\bibinfo {title} {Qubit {{Noise
  Spectroscopy}} for {{Non-Gaussian Dephasing Environments}}},\ }\href
  {https://doi.org/10.1103/PhysRevLett.116.150503} {\bibfield  {journal}
  {\bibinfo  {journal} {Phys. Rev. Lett.}\ }\textbf {\bibinfo {volume} {116}},\
  \bibinfo {pages} {150503} (\bibinfo {year} {2016})}\BibitemShut {NoStop}%
\bibitem [{\citenamefont {Sza{\'n}kowski}\ \emph {et~al.}(2017)\citenamefont
  {Sza{\'n}kowski}, \citenamefont {Ramon}, \citenamefont {Krzywda},
  \citenamefont {Kwiatkowski},\ and\ \citenamefont
  {Cywi{\'n}ski}}]{Szankowski2017}%
  \BibitemOpen
  \bibfield  {author} {\bibinfo {author} {\bibfnamefont {P.}~\bibnamefont
  {Sza{\'n}kowski}}, \bibinfo {author} {\bibfnamefont {G.}~\bibnamefont
  {Ramon}}, \bibinfo {author} {\bibfnamefont {J.}~\bibnamefont {Krzywda}},
  \bibinfo {author} {\bibfnamefont {D.}~\bibnamefont {Kwiatkowski}},\ and\
  \bibinfo {author} {\bibfnamefont {L.}~\bibnamefont {Cywi{\'n}ski}},\
  }\bibfield  {title} {\bibinfo {title} {Environmental noise spectroscopy with
  qubits subjected to dynamical decoupling},\ }\href
  {https://doi.org/10.1088/1361-648X/aa7648} {\bibfield  {journal} {\bibinfo
  {journal} {J. Phys.: Condens. Matter}\ }\textbf {\bibinfo {volume} {29}},\
  \bibinfo {pages} {333001} (\bibinfo {year} {2017})}\BibitemShut {NoStop}%
\bibitem [{\citenamefont {Sung}\ \emph {et~al.}(2019)\citenamefont {Sung},
  \citenamefont {Beaudoin}, \citenamefont {Norris}, \citenamefont {Yan},
  \citenamefont {Kim}, \citenamefont {Qiu}, \citenamefont {{von L{\"u}pke}},
  \citenamefont {Yoder}, \citenamefont {Orlando}, \citenamefont {Gustavsson},
  \citenamefont {Viola},\ and\ \citenamefont {Oliver}}]{Sung2019}%
  \BibitemOpen
  \bibfield  {author} {\bibinfo {author} {\bibfnamefont {Y.}~\bibnamefont
  {Sung}}, \bibinfo {author} {\bibfnamefont {F.}~\bibnamefont {Beaudoin}},
  \bibinfo {author} {\bibfnamefont {L.~M.}\ \bibnamefont {Norris}}, \bibinfo
  {author} {\bibfnamefont {F.}~\bibnamefont {Yan}}, \bibinfo {author}
  {\bibfnamefont {D.~K.}\ \bibnamefont {Kim}}, \bibinfo {author} {\bibfnamefont
  {J.~Y.}\ \bibnamefont {Qiu}}, \bibinfo {author} {\bibfnamefont
  {U.}~\bibnamefont {{von L{\"u}pke}}}, \bibinfo {author} {\bibfnamefont
  {J.~L.}\ \bibnamefont {Yoder}}, \bibinfo {author} {\bibfnamefont {T.~P.}\
  \bibnamefont {Orlando}}, \bibinfo {author} {\bibfnamefont {S.}~\bibnamefont
  {Gustavsson}}, \bibinfo {author} {\bibfnamefont {L.}~\bibnamefont {Viola}},\
  and\ \bibinfo {author} {\bibfnamefont {W.~D.}\ \bibnamefont {Oliver}},\
  }\bibfield  {title} {\bibinfo {title} {Non-{{Gaussian}} noise spectroscopy
  with a superconducting qubit sensor},\ }\href
  {https://doi.org/10.1038/s41467-019-11699-4} {\bibfield  {journal} {\bibinfo
  {journal} {Nat. Commun.}\ }\textbf {\bibinfo {volume} {10}},\ \bibinfo
  {pages} {3715} (\bibinfo {year} {2019})}\BibitemShut {NoStop}%
\bibitem [{\citenamefont {Maizelis}\ \emph {et~al.}(2011)\citenamefont
  {Maizelis}, \citenamefont {Roukes},\ and\ \citenamefont
  {Dykman}}]{Maizelis2011}%
  \BibitemOpen
  \bibfield  {author} {\bibinfo {author} {\bibfnamefont {Z.~A.}\ \bibnamefont
  {Maizelis}}, \bibinfo {author} {\bibfnamefont {M.~L.}\ \bibnamefont
  {Roukes}},\ and\ \bibinfo {author} {\bibfnamefont {M.~I.}\ \bibnamefont
  {Dykman}},\ }\bibfield  {title} {\bibinfo {title} {Detecting and
  {{Characterizing Frequency Fluctuations}} of {{Vibrational Modes}}},\
  }\href@noop {} {\bibfield  {journal} {\bibinfo  {journal} {Phys. Rev. B}\
  }\textbf {\bibinfo {volume} {84}},\ \bibinfo {pages} {144301} (\bibinfo
  {year} {2011})}\BibitemShut {NoStop}%
\bibitem [{\citenamefont {Sun}\ \emph {et~al.}(2015)\citenamefont {Sun},
  \citenamefont {Zou}, \citenamefont {Maizelis},\ and\ \citenamefont
  {Chan}}]{Sun2015}%
  \BibitemOpen
  \bibfield  {author} {\bibinfo {author} {\bibfnamefont {F.}~\bibnamefont
  {Sun}}, \bibinfo {author} {\bibfnamefont {J.}~\bibnamefont {Zou}}, \bibinfo
  {author} {\bibfnamefont {Z.~A.}\ \bibnamefont {Maizelis}},\ and\ \bibinfo
  {author} {\bibfnamefont {H.~B.}\ \bibnamefont {Chan}},\ }\bibfield  {title}
  {\bibinfo {title} {Telegraph {{Frequency Noise}} in {{Electromechanical
  Resonators}}},\ }\href@noop {} {\bibfield  {journal} {\bibinfo  {journal}
  {Phys. Rev. B}\ }\textbf {\bibinfo {volume} {91}},\ \bibinfo {pages} {174102}
  (\bibinfo {year} {2015})}\BibitemShut {NoStop}%
\bibitem [{\citenamefont {Wilen}\ \emph {et~al.}(2021)\citenamefont {Wilen},
  \citenamefont {Abdullah}, \citenamefont {Kurinsky}, \citenamefont {Stanford},
  \citenamefont {Cardani}, \citenamefont {D'Imperio}, \citenamefont {Tomei},
  \citenamefont {Faoro}, \citenamefont {Ioffe}, \citenamefont {Liu},
  \citenamefont {Opremcak}, \citenamefont {Christensen}, \citenamefont
  {DuBois},\ and\ \citenamefont {McDermott}}]{Wilen2021}%
  \BibitemOpen
  \bibfield  {author} {\bibinfo {author} {\bibfnamefont {C.~D.}\ \bibnamefont
  {Wilen}}, \bibinfo {author} {\bibfnamefont {S.}~\bibnamefont {Abdullah}},
  \bibinfo {author} {\bibfnamefont {N.~A.}\ \bibnamefont {Kurinsky}}, \bibinfo
  {author} {\bibfnamefont {C.}~\bibnamefont {Stanford}}, \bibinfo {author}
  {\bibfnamefont {L.}~\bibnamefont {Cardani}}, \bibinfo {author} {\bibfnamefont
  {G.}~\bibnamefont {D'Imperio}}, \bibinfo {author} {\bibfnamefont
  {C.}~\bibnamefont {Tomei}}, \bibinfo {author} {\bibfnamefont
  {L.}~\bibnamefont {Faoro}}, \bibinfo {author} {\bibfnamefont {L.~B.}\
  \bibnamefont {Ioffe}}, \bibinfo {author} {\bibfnamefont {C.~H.}\ \bibnamefont
  {Liu}}, \bibinfo {author} {\bibfnamefont {A.}~\bibnamefont {Opremcak}},
  \bibinfo {author} {\bibfnamefont {B.~G.}\ \bibnamefont {Christensen}},
  \bibinfo {author} {\bibfnamefont {J.~L.}\ \bibnamefont {DuBois}},\ and\
  \bibinfo {author} {\bibfnamefont {R.}~\bibnamefont {McDermott}},\ }\bibfield
  {title} {\bibinfo {title} {Correlated charge noise and relaxation errors in
  superconducting qubits},\ }\href {https://doi.org/10.1038/s41586-021-03557-5}
  {\bibfield  {journal} {\bibinfo  {journal} {Nature}\ }\textbf {\bibinfo
  {volume} {594}},\ \bibinfo {pages} {369} (\bibinfo {year}
  {2021})}\BibitemShut {NoStop}%
\bibitem [{\citenamefont {Liu}\ \emph {et~al.}(2022)\citenamefont {Liu},
  \citenamefont {Harrison}, \citenamefont {Patel}, \citenamefont {Wilen},
  \citenamefont {Rafferty}, \citenamefont {Shearrow}, \citenamefont {Ballard},
  \citenamefont {Iaia}, \citenamefont {Ku}, \citenamefont {Plourde},\ and\
  \citenamefont {McDermott}}]{Liu2022a}%
  \BibitemOpen
  \bibfield  {author} {\bibinfo {author} {\bibfnamefont {C.-H.}\ \bibnamefont
  {Liu}}, \bibinfo {author} {\bibfnamefont {D.~C.}\ \bibnamefont {Harrison}},
  \bibinfo {author} {\bibfnamefont {S.}~\bibnamefont {Patel}}, \bibinfo
  {author} {\bibfnamefont {C.~D.}\ \bibnamefont {Wilen}}, \bibinfo {author}
  {\bibfnamefont {O.}~\bibnamefont {Rafferty}}, \bibinfo {author}
  {\bibfnamefont {A.}~\bibnamefont {Shearrow}}, \bibinfo {author}
  {\bibfnamefont {A.}~\bibnamefont {Ballard}}, \bibinfo {author} {\bibfnamefont
  {V.}~\bibnamefont {Iaia}}, \bibinfo {author} {\bibfnamefont {J.}~\bibnamefont
  {Ku}}, \bibinfo {author} {\bibfnamefont {B.~L.~T.}\ \bibnamefont {Plourde}},\
  and\ \bibinfo {author} {\bibfnamefont {R.}~\bibnamefont {McDermott}},\
  }\bibfield  {title} {\bibinfo {title} {Quasiparticle {{Poisoning}} of
  {{Superconducting Qubits}} from {{Resonant Absorption}} of {{Pair-breaking
  Photons}}},\ }\href@noop {} {\bibfield  {journal} {\bibinfo  {journal}
  {ArXiv220306577 Quant-Ph}\ } (\bibinfo {year} {2022})},\ \bibinfo {note}
  {comment: 13 pages, 11 figures, 3 tables},\ \Eprint
  {https://arxiv.org/abs/2203.06577} {arXiv:2203.06577 [quant-ph]} \BibitemShut
  {NoStop}%
\bibitem [{\citenamefont {Anderson}\ \emph {et~al.}(1972)\citenamefont
  {Anderson}, \citenamefont {Halperin},\ and\ \citenamefont
  {Varma}}]{Anderson1972}%
  \BibitemOpen
  \bibfield  {author} {\bibinfo {author} {\bibfnamefont {P.~W.}\ \bibnamefont
  {Anderson}}, \bibinfo {author} {\bibfnamefont {B.~I.}\ \bibnamefont
  {Halperin}},\ and\ \bibinfo {author} {\bibfnamefont {C.~M.}\ \bibnamefont
  {Varma}},\ }\bibfield  {title} {\bibinfo {title} {Anomalous low-temperature
  thermal properties of glasses and spin glasses},\ }\href
  {https://doi.org/10.1080/14786437208229210} {\bibfield  {journal} {\bibinfo
  {journal} {Philos. Mag. J. Theor. Exp. Appl. Phys.}\ }\textbf {\bibinfo
  {volume} {25}},\ \bibinfo {pages} {1} (\bibinfo {year} {1972})}\BibitemShut
  {NoStop}%
\bibitem [{\citenamefont {Phillips}(1972)}]{Phillips1972}%
  \BibitemOpen
  \bibfield  {author} {\bibinfo {author} {\bibfnamefont {W.~A.}\ \bibnamefont
  {Phillips}},\ }\bibfield  {title} {\bibinfo {title} {Tunneling states in
  amorphous solids},\ }\href {https://doi.org/10.1007/BF00660072} {\bibfield
  {journal} {\bibinfo  {journal} {J Low Temp Phys}\ }\textbf {\bibinfo {volume}
  {7}},\ \bibinfo {pages} {351} (\bibinfo {year} {1972})}\BibitemShut {NoStop}%
\bibitem [{\citenamefont {Phillips}(1987)}]{Phillips1987}%
  \BibitemOpen
  \bibfield  {author} {\bibinfo {author} {\bibfnamefont {W.~A.}\ \bibnamefont
  {Phillips}},\ }\bibfield  {title} {\bibinfo {title} {Two-{{Level States}} in
  {{Glasses}}},\ }\href@noop {} {\bibfield  {journal} {\bibinfo  {journal}
  {Rep. Prog. Phys.}\ }\textbf {\bibinfo {volume} {50}},\ \bibinfo {pages}
  {1657} (\bibinfo {year} {1987})}\BibitemShut {NoStop}%
\bibitem [{\citenamefont {Nielsen}\ and\ \citenamefont
  {Chuang}(2011)}]{Nielsen2011}%
  \BibitemOpen
  \bibfield  {author} {\bibinfo {author} {\bibfnamefont {M.~A.}\ \bibnamefont
  {Nielsen}}\ and\ \bibinfo {author} {\bibfnamefont {I.~L.}\ \bibnamefont
  {Chuang}},\ }\href@noop {} {\emph {\bibinfo {title} {Quantum {{Computation}}
  and {{Quantum Information}}: 10th {{Anniversary Edition}}}}},\ \bibinfo
  {edition} {1st}\ ed.\ (\bibinfo  {publisher} {{Cambridge University Press}},\
  \bibinfo {address} {{Cambridge ; New York}},\ \bibinfo {year}
  {2011})\BibitemShut {NoStop}%
\bibitem [{\citenamefont {You}\ \emph {et~al.}(2021)\citenamefont {You},
  \citenamefont {Clerk},\ and\ \citenamefont {Koch}}]{You2021}%
  \BibitemOpen
  \bibfield  {author} {\bibinfo {author} {\bibfnamefont {X.}~\bibnamefont
  {You}}, \bibinfo {author} {\bibfnamefont {A.~A.}\ \bibnamefont {Clerk}},\
  and\ \bibinfo {author} {\bibfnamefont {J.}~\bibnamefont {Koch}},\ }\bibfield
  {title} {\bibinfo {title} {Positive- and negative-frequency noise from an
  ensemble of two-level fluctuators},\ }\href@noop {} {\bibfield  {journal}
  {\bibinfo  {journal} {Phys Rev Res.}\ }\textbf {\bibinfo {volume} {3}},\
  \bibinfo {pages} {013045} (\bibinfo {year} {2021})},\ \bibinfo {note}
  {comment: 15 pages, 8 figures},\ \Eprint {https://arxiv.org/abs/2005.03591}
  {arXiv:2005.03591} \BibitemShut {NoStop}%
\bibitem [{\citenamefont {Ithier}\ \emph {et~al.}(2005)\citenamefont {Ithier},
  \citenamefont {Collin}, \citenamefont {Joyez}, \citenamefont {Meeson},
  \citenamefont {Vion}, \citenamefont {Esteve}, \citenamefont {Chiarello},
  \citenamefont {Shnirman}, \citenamefont {Makhlin}, \citenamefont {Schriefl},\
  and\ \citenamefont {Sch{\"o}n}}]{Ithier2005}%
  \BibitemOpen
  \bibfield  {author} {\bibinfo {author} {\bibfnamefont {G.}~\bibnamefont
  {Ithier}}, \bibinfo {author} {\bibfnamefont {E.}~\bibnamefont {Collin}},
  \bibinfo {author} {\bibfnamefont {P.}~\bibnamefont {Joyez}}, \bibinfo
  {author} {\bibfnamefont {P.~J.}\ \bibnamefont {Meeson}}, \bibinfo {author}
  {\bibfnamefont {D.}~\bibnamefont {Vion}}, \bibinfo {author} {\bibfnamefont
  {D.}~\bibnamefont {Esteve}}, \bibinfo {author} {\bibfnamefont
  {F.}~\bibnamefont {Chiarello}}, \bibinfo {author} {\bibfnamefont
  {A.}~\bibnamefont {Shnirman}}, \bibinfo {author} {\bibfnamefont
  {Y.}~\bibnamefont {Makhlin}}, \bibinfo {author} {\bibfnamefont
  {J.}~\bibnamefont {Schriefl}},\ and\ \bibinfo {author} {\bibfnamefont
  {G.}~\bibnamefont {Sch{\"o}n}},\ }\bibfield  {title} {\bibinfo {title}
  {Decoherence in a superconducting quantum bit circuit},\ }\href
  {https://doi.org/10.1103/PhysRevB.72.134519} {\bibfield  {journal} {\bibinfo
  {journal} {Phys. Rev. B}\ }\textbf {\bibinfo {volume} {72}},\ \bibinfo
  {pages} {134519} (\bibinfo {year} {2005})}\BibitemShut {NoStop}%
\bibitem [{\citenamefont {Gurvitz}(2019)}]{Gurvitz2019}%
  \BibitemOpen
  \bibfield  {author} {\bibinfo {author} {\bibfnamefont {S.}~\bibnamefont
  {Gurvitz}},\ }\bibfield  {title} {\bibinfo {title} {Generalized {{Landauer}}
  formula for time-dependent potentials and noise-induced zero-bias dc
  current},\ }\href {https://doi.org/10.1088/1751-8121/ab10ed} {\bibfield
  {journal} {\bibinfo  {journal} {J. Phys. A: Math. Theor.}\ }\textbf {\bibinfo
  {volume} {52}},\ \bibinfo {pages} {175301} (\bibinfo {year}
  {2019})}\BibitemShut {NoStop}%
\bibitem [{\citenamefont {Shapiro}\ and\ \citenamefont
  {Loginov}(1978)}]{Shapiro1978}%
  \BibitemOpen
  \bibfield  {author} {\bibinfo {author} {\bibfnamefont {V.~E.}\ \bibnamefont
  {Shapiro}}\ and\ \bibinfo {author} {\bibfnamefont {V.~M.}\ \bibnamefont
  {Loginov}},\ }\bibfield  {title} {\bibinfo {title} {``{{Formulae}} of
  differentiation'' and their use for solving stochastic equations},\ }\href
  {https://doi.org/10.1016/0378-4371(78)90198-X} {\bibfield  {journal}
  {\bibinfo  {journal} {Physica A: Statistical Mechanics and its Applications}\
  }\textbf {\bibinfo {volume} {91}},\ \bibinfo {pages} {563} (\bibinfo {year}
  {1978})}\BibitemShut {NoStop}%
\end{thebibliography}
\end{document}